\documentclass[twocolumn]{aastex63}
\usepackage{natbib}
\usepackage{txfonts}
\usepackage{color}
\usepackage{rotating}
\usepackage{subfigure}
\usepackage{graphics}
\usepackage{epstopdf}

\newcommand{\eqref}[1]{(\ref{#1})}
\newcommand\fmg{\mbox{$.\!\!^\prime$}}%
\newcommand{\tablenotemarknew}[1]{$^{#1}$}%
\newcommand{\tablenotetextnew}[2]{$^{#1}$#2}%

\newcommand{\textbi}[1]{\textbf{\textit{#1}}}
\newcommand{ \kms}{km s$^{-1}$}

\shorttitle{PGCCs at High Galactic Latitude}
\shortauthors{Xu et al.}

\begin{document}

\title{Planck Galactic Cold Clumps at High Galactic Latitude-A Study with CO Lines}


\author[0000-0001-5950-1932]{Fengwei Xu}
\affiliation{Department of Astronomy, School of Physics, Peking University, Beijing, 100871, People's Republic of China, \url{fengwei.astro@pku.edu.cn}, \url{ywu@pku.edu.cn}}
\affiliation{Kavli Institute for Astronomy and Astrophysics, Peking University, Haidian District, Beijing 100871, People’s Republic of China}

\author{Yuefang Wu}
\affiliation{Department of Astronomy, School of Physics, Peking University, Beijing, 100871, People's Republic of China, \url{fengwei.astro@pku.edu.cn}, \url{ywu@pku.edu.cn}}
\affiliation{Kavli Institute for Astronomy and Astrophysics, Peking University, Haidian District, Beijing 100871, People’s Republic of China}

\author{Tie Liu}
\affiliation{Shanghai Astronomical Observatory, Chinese Academy of Sciences, 80 Nandan Road, Shanghai 200030, Republic of China}
\affiliation{Korea Astronomy and Space Science Institute, 776 Daedeokdaero, Yuseong-gu, Daejeon 34055, Republic of Korea}

\author{Xunchuan Liu}
\affiliation{Department of Astronomy, School of Physics, Peking University, Beijing, 100871, People's Republic of China, \url{fengwei.astro@pku.edu.cn}, \url{ywu@pku.edu.cn}}
\affiliation{Kavli Institute for Astronomy and Astrophysics, Peking University, Haidian District, Beijing 100871, People’s Republic of China}

\author{Chao Zhang}
\affiliation{Department of Astronomy, Yunnan University, and Key Laboratory of Astroparticle Physics of Yunnan Province, Kunming 650091, China}
\affiliation{Department of Astronomy, School of Physics, Peking University, Beijing, 100871, People's Republic of China, \url{fengwei.astro@pku.edu.cn}, \url{ywu@pku.edu.cn}}

\author{Jarken Esimbek}
\affiliation{Xinjiang Astronomical Observatory, Chinese Academy of Sciences, Urumqi 830011, China}
\affiliation{Key Laboratory of Radio Astronomy, Chinese Academy of Sciences, Urumqi 830011, China}

\author{Sheng-Li Qin}
\affiliation{Department of Astronomy, Yunnan University, and Key Laboratory of Astroparticle Physics of Yunnan Province, Kunming 650091, China}

\author{Di Li}
\affiliation{CAS Key Laboratory of FAST, National Astronomical Observatories, CAS, Beijing 100012, China}
\affiliation{NAOC-UKZN Computational Astrophysics Centre (NUCAC), University of KwaZulu-Natal, Durban 4000, South Africa}

\author{Ke Wang}
\affiliation{Kavli Institute for Astronomy and Astrophysics, Peking University, Haidian District, Beijing 100871, People’s Republic of China}

\author{Jinghua Yuan}
\affiliation{National Astronomical Observatories, Chinese Academy of Sciences, 20A Datun Road, Chaoyang District, Beijing 100012, China}

\author{Fanyi Meng}
\affiliation{I. Physikalisches Institut, Universit\"{a}t z\"{u} K\"{o}ln, Z\"{u}lpicher Str. 77, D-50937 K\"{o}ln, Germany}
\affiliation{Department of Astronomy, School of Physics, Peking University, Beijing, 100871, People's Republic of China, \url{fengwei.astro@pku.edu.cn}, \url{ywu@pku.edu.cn}}

\author{Tianwei Zhang}
\affiliation{I. Physikalisches Institut, Universit\"{a}t z\"{u} K\"{o}ln, Z\"{u}lpicher Str. 77, D-50937 K\"{o}ln, Germany}
\affiliation{Department of Astronomy, School of Physics, Peking University, Beijing, 100871, People's Republic of China, \url{fengwei.astro@pku.edu.cn}, \url{ywu@pku.edu.cn}}

\author{David Eden}
\affiliation{Astrophysics Research Institute, Liverpool John Moores University, IC2, Liverpool Science Park, 146 Brownlow Hill, Liverpool L3 5RF, UK}

\author{K. Tatematsu}
\affiliation{Nobeyama Radio Observatory, National Astronomical Observatory of Japan, National Institutes of Natural Sciences, Nobeyama, Minamimaki, Minamisaku, Nagano 384-1305, Japan}

\author{Neal J. Evans}
\affiliation{Department of Astronomy The University of Texas at Austin 2515 Speedway, Stop C1400 Austin, TX 78712-1205, USA}
\affiliation{Korea Astronomy and Space Science Institute 776 Daedeokdae-ro, Yuseong-gu Daejeon, 34055, Republic of Korea}
\affiliation{Humanitas College, Global Campus, Kyung Hee University, Yongin-shi 17104, Republic of Korea}

\author{Paul. F. Goldsmith}
\affiliation{Jet Propulsion Laboratory, California Institute of Technology, 4800 Oak Grove Drive, Pasadena, CA 91109, USA}

\author{Qizhou Zhang}
\affiliation{Center for Astrophysics | Harvard \& Smithsonian, 60 Garden Street, Cambridge, MA 02138, USA}

\author{C. Henkel}
\affiliation{Max-Planck Institut f\"{u}r Radioastronomie, Auf Dem H\"{u}gel 69, 53121 Bonn, Germanyn}
\affiliation{Astronomy Department, Faculty of Science, King Abdulaziz University, PO Box 80203, Jeddah, 21589, Saudi Arabia}

\author{Hee-Weon Yi}
\affiliation{School of Space Research, Kyung Hee University, 1732, Deogyeong-daero, Giheung-gu, Yongin-si, Gyeonggi-do 17104, Republic of Korea}

\author{Jeong-Eun Lee}
\affiliation{School of Space Research, Kyung Hee University, 1732, Deogyeong-daero, Giheung-gu, Yongin-si, Gyeonggi-do 17104, Republic of Korea}

\author{Mika Saajasto}
\affiliation{Department of Physics, P.O.Box 64, FI-00014, University of Helsinki, Finland}

\author{Gwanjeong Kim}
\affiliation{Nobeyama Radio Observatory, National Astronomical Observatory of Japan, National Institutes of Natural Sciences, 462-2 Nobeyama, Minamimaki, Minamisaku, Nagano 384-1305, Japan}

\author{Mika Juvela}
\affiliation{Department of Physics, P.O.Box 64, FI-00014, University of Helsinki, Finland}

\author{Dipen Sahu}
\affiliation{Academia Sinica Institute of Astronomy and Astrophysics, 11F of AS/NTU Astronomy-Mathematics Building, No.1, Sec. 4, Roosevelt Rd, Taipei 10617, Taiwan, R.O.C.}

\author{Shih-Ying Hsu}
\affiliation{National Taiwan University (NTU), Taiwan, R.O.C.}

\author{Sheng-yuan Liu}
\affiliation{Academia Sinica Institute of Astronomy and Astrophysics, 11F of AS/NTU Astronomy-Mathematics Building, No.1, Sec. 4, Roosevelt Rd, Taipei 10617, Taiwan, R.O.C.}

\author{Somnath Dutta}
\affiliation{Academia Sinica Institute of Astronomy and Astrophysics, 11F of AS/NTU Astronomy-Mathematics Building, No.1, Sec. 4, Roosevelt Rd, Taipei 10617, Taiwan, R.O.C.}

\author{Chin-Fei Lee}
\affiliation{Academia Sinica Institute of Astronomy and Astrophysics, 11F of AS/NTU Astronomy-Mathematics Building, No.1, Sec. 4, Roosevelt Rd, Taipei 10617, Taiwan, R.O.C.}

\author{Chuan-Peng Zhang}
\affiliation{National Astronomical Observatories, Chinese Academy of Sciences, 100101 Beijing, China}

\author{Ye Xu}
\affiliation{Purple Mountain Observatory and Key Laboratory of Radio Astronomy, Chinese Academy of Sciences, 8 Yuanhua Road, Nanjing 210034, PR China}

\author{Binggang Ju}
\affiliation{Purple Mountain Observatory and Key Laboratory of Radio Astronomy, Chinese Academy of Sciences, 8 Yuanhua Road, Nanjing 210034, PR China}

\clearpage

\begin{abstract}
Gas at high Galactic latitude is a relatively little-noticed component of the interstellar medium. In an effort to address this, forty-one Planck Galactic Cold Clumps at high Galactic latitude (HGal; $|b|>25\degr$) were observed in $^{12}$CO, $^{13}$CO and C$^{18}$O J=1-0 lines, using the Purple Mountain Observatory 13.7-m telescope. $^{12}$CO (1-0) and $^{13}$CO (1-0) emission was detected in all clumps while C$^{18}$O (1-0) emission was only seen in sixteen clumps. The highest and average latitudes are $71.4\degr$ and $37.8\degr$, respectively. Fifty-one velocity components were obtained and then each was identified as a single clump. Thirty-three clumps were further mapped at 1$^\prime$ resolution and 54 dense cores were extracted. Among dense cores, the average excitation temperature $T_{\mathrm{ex}}$ of $^{12}$CO is 10.3 K. The average line widths of thermal and non-thermal velocity dispersions are $0.19$ \kms and $0.46$ \kms respectively, suggesting that these cores are dominated by turbulence. Distances of the HGal clumps given by Gaia dust reddening are about $120-360$ pc. The ratio of $X_{13}$/$X_{18}$ is significantly higher than that in the solar neighbourhood, implying that HGal gas has a different star formation history compared to the gas in the Galactic disk. HGal cores with sizes from $0.01-0.1$ pc show no notable Larson's relation and the turbulence remains supersonic down to a scale of slightly below $0.1$ pc. None of the HGal cores which bear masses from 0.01-1 $M_{\odot}$ are gravitationally bound and all appear to be confined by outer pressure.
\end{abstract}


\section{Introduction} \label{sec:intro}
Two directions in studying Galactic star formation nowadays are important: one is to study the initial condition of star formation and the other is to cover a more complete area. The former requires exploring the earliest stage. Therefore, astronomers are encouraged to search for those primordial molecular clumps which are close to the initial stage of forming stars and assumed to harbor prestellar cores \citep{1987ARA&A..25...23S}. The latter requires larger surveys, especially towards the regions beyond the Galactic disk or the low Galactic latitudes in traditional surveys \citep{1979PhDT.........9S,1978A&A....63....7B,1982ApJS...49..183B,1983ASSL..105....1R,1984A&A...134..396I,1984ApJ...276..182S,1987ApJ...322..706D}. The sky coverage at high Galatic latitude (hereafter HGal), the rest of the “iceberg”, provides an extensive view of gas distribution and star formation in our Galaxy. The infrared cirrus \citep{1984ApJ...278L..19L} found by the \textit{Infra-Red Astronomical Satellite} (hereafter (\textit{IRAS})) at HGal were thought to be an environment hostile to star formation. Also, previous works have shown that methanol masers which trace the massive star-forming regions are rare at high latitudes \citep{2017ApJ...846..160Y}. Besides, the gas at HGal as a variable gas reservoir probably plays an important role in the cycling of gas in and out of the Galactic plane \citep{2013ApJ...772..119L}. Therefore, the properties of the gas can be quite different from those in the Galactic plane, and this inspires the idea that latitude is a sensitive parameter for Galactic star formation and gas properties.

The molecular material at HGal had been poorly studied until the first CO survey searched for apparent optical obscuration on the Palomar Observatory Sky Survey (POSS) prints \citep[as well as the Whiteoak extension to the POSS]{1984ApJ...282L...9B} by the 5-m Millimeter Wave Observatory in Texas \citep{2012JAHH...15..232V}. 57 clouds were discovered by \citet*[hereafter MBM]{1985ApJ...295..402M} in 33 complexes at $|b|>20\degr$ and ten complexes at $|b| > 30\degr$ were mapped with a grid spacing 10$^\prime$-20$^\prime$ depending on the angular extent of the source. MBM provides the first glimpse of molecular clouds at high Galatic latitude (hereafter HGal clouds) and the number of identified HGal clouds has increased dramatically from then on (see \citet{1996ApJS..106..447M} for a review). Subsequently, unbiased CO surveys were brought up for more complete census of HGal clouds at northern Galactic hemisphere by \citet{1998ApJ...492..205H} and then southern one by \citet{2000ApJ...535..167M}.

As soon as \citet{1984ApJ...282L...9B} and \citet{1985ApJ...295..402M} detected CO molecular clouds in HGal, people began to explore the physical and chemical properties with the help of molecular lines. HGal clouds, for the most part, are thought to be gravitationally unbound \citep{1986ApJ...304..466K}, and may be in pressure equillibrium with the Interstellar medium (ISM). Besides, HGal clouds are reported to have lower densities and rare isotopic CO abundances than those encountered in dark clouds or GMCs \citep{1988ApJ...332..432P}. HGal clouds represent a distinct component of molecular material and may help to provide a new perspective of the ISM. This warrants further confirmed study.

During the flourishing age of the 8$^\prime$-resolution CO sky survey (with the help of the Columbia 1.2-m telescope twins), an important step toward discerning the nature of HGal clouds was made by \citet{1988ApJ...334..771V}. The HGal clouds were considered to be translucent clouds \citep[the total extinction $A^{\text{tot}}_{V}\simeq1-4$mag]{1991ApJ...366..141V} with CO column densities $\ge10^{15}$ cm$^{-2}$ and abundances from $10^{-6}$ to $10^{-4}$, which represent the transitional regime between traditional diffuse and dark clouds \citep{1996ApJS..106..447M}. Those HGal clouds appear to be extraordinarily young and may represent the earliest stages of molecular clouds condensing from the local ISM. Besides, the CO observations with sub-parsec resolution towards far-infrared emission in HGal clouds \citep{1990ApJ...352L..13B,1994ApJ...429..672R} have demonstrated that those clouds have smaller structures \citep{2000A&AS..144..123W}.
Therefore, higher resolution and spatial sampling rate are needed if we are to dissect and analyse the clouds on smaller scale, which also helps understand the initial condition of star formation better. Studies of the large-scale HGal cloud complexes with high spatial resolution have been rare, since the spatial extent is as large as tens of square degrees or more \citep{2003ApJ...592..217Y}. Composite CO observations were conducted towards the Polaris Flare with 620 $\deg^2$ by \citet{1990ApJ...353L..49H}, and towards Ursa Major clouds by \citet{1987ApJ...319..723D} and \citet{1997ApJ...482..334P}. Another observation toward the \textsc{Hi} filament regions including MBM 53, 54, and 55 with approximately 141 $\deg^2$ was conducted by \citet{2003ApJ...592..217Y}.

We summarize the large CO line surveys towards HGal regions as completely as we could in Table \ref{tab:surveys}, which are listed in chronological order. However, two main issues require further consideration. On one hand, the earliest stage of star formation buried deep inside the cold dust at about 10 K should be detected more sensitively at longer wavelengths, in millimeter or sub-millimeter bands. On the other hand, it is difficult to achieve both extensive coverage and high resolution (and also fine sampling): higher spatial accuracy requires more integration time so that an all-sky survey with high resolution is not feasible. Besides, signals may not be detected over a large fraction of the sky, leading to a low efficiency in the observations. Therefore, one possible strategy is to use a telescope with a relatively high resolution to map specific regions that are chosen from a previous survey with a larger coverage but lower resolution.

Fortunately, the Planck satellite provides an unprecedentedly complete spatial distribution of the HGal sources. The all-sky nature of this sample is particularly useful for studying the global properties of Galactic molecular gas and clouds. The Cold Clump Catalogue of Planck Objects (C3PO), also called Planck Galactic Cold Clumps (PGCCs), consists of 13,188 cold clumps derived from the fluxes in the three highest frequency Planck bands (353, 545, 857 GHz) as well as the 3000 GHz \textit{IRAS} band. PGCCs provides a rich catalog of sources with the a dust temperature between $10-15$ K, probably the coldest part of the ISM \citep{2011A&A...536A..22P} allowing us to probe the characteristics of the prestellar phase and the properties of starless clumps \citep{2012ApJ...756...76W}. A follow-up study was conducted with the 13.7-m telescope of the Purple Mountain Observatory (PMO) at Delingha, Qinghai Province. The main beam size of $\sim1^\prime$ and the grid spacing of 30$\arcsec$ are \textbf{the best among the surveys} listed in Table \ref{tab:surveys}, which satisfy the spatial accuracy to resolve the regions of several tenths of parsec in size at distance less than about 400 pc \citep{2012ApJS..202....4L}. Since Planck observations only provide the continuum emission, it is necessary to examine molecular lines to infer the dynamical information, which is essential for studying star formation among these samples. PMO observations provide all $^{12}$CO(1-0), $^{13}$CO(1-0), and C$^{18}$O(1-0) lines, guaranteeing fruitful information about gas dynamics. Besides, this work uses two other emission lines, N$_2$H$^+$ (1-0) and C$_2$H (1-0), to trace dense regions and to reveal the evolutionary state of the gas as supplementary information.

In this paper, we wish to answer the following questions: How do physical conditions in clouds away from the Galactic plane differ from those at low latitude? What phase are HGal gas cores in? We report an investigation of 41 PGCCs at high Galactic latitudes. We extract 54 cores condensed in these clumps and analyze their properties further. The data descriptions are in Section \ref{sec:data}. The results are shown in Section \ref{sec:results}. Further discussions follow in Section \ref{sec:discussion} while main conclusions are summarized in Section \ref{sec:summary}.

\section{Data} \label{sec:data}
\subsection{Sample Selection} \label{data:sample}
There are 41 clumps with absolute Galactic latitude higher than $25\degr$ ($|b|>25\degr$) from the Early Cold Core (ECC; \citet{2011A&A...536A...7P}) catalog, which were observed with the 13.7-m millimeter-wavelength telescope of the Purple Mountain Observatory (PMO). The ECC sources have high signal-to-noise ratios (SNR$>$15). Figure \ref{fig:COsurvey} shows the distribution of 41 clumps. The distribution of these 41 HGal clumps is not uniform along latitude or along longitude. As shown in Figure \ref{fig:COsurvey}, the clumps of belonging to the northern sky are located only in the latitude range $25\degr<b<42\degr$ and those in the southern sky mainly group around $-50\degr<b<-25\degr$. The longitudes of the detected clumps are mostly in the range of $90\degr<l<180\degr$ and $0\degr<l<45\degr$. Almost half, 46.3\% (19 out of 41) of these clumps are just \textbf{below} Taurus, Perseus, and California Complex at $b\le -20\degr$ and $l=155-180\degr$
The names of these clumps are listed in column (1), Galactic coordinates in columns (2)--(3) and equatorial coordinates in columns (4)--(7) of Table \ref{tab:samples}.

\subsection{PMO Observations} \label{data:obs}
The $3\times3$ beam sideband separation Superconduction Spectroscopic Array Receiver system was used as the front end \citep{6313968}. The half-power beam width is 56$\arcsec$ in the 115 GHz band. The mean beam efficiency is about $50\%$. The pointing accuracy is better than 5$\arcsec$. The Fast Fourier Transform Spectrometers (FFTSs) were used as the backend. Each FFTS with a total bandwidth of 1 GHz provides 16384 channels. 

The single point mode observations of $^{12}$CO, $^{13}$CO and C$^{18}$O J=1-0 towards 41 HGal clumps were conducted from April to May in 2011 and from December 2011 to January 2012. The velocity resolutions are 0.16 km s$^{-1}$ for $^{12}$CO and 0.17 km s$^{-1}$ for $^{13}$CO and C$^{18}$O. The $^{12}$CO emission was observed in the upper sideband (USB) with a system temperature (T$_{\mathrm{sys}}$) around 210 K, while the $^{13}$CO and C$^{18}$O emissions were observed simultaneously in the lower sideband (LSB) with T$_{\mathrm{sys}}$ around 120 K. The typical root mean square (rms) noise in T$_A^*$ was 0.2 K for $^{12}$CO but 0.1 K for $^{13}$CO and C$^{18}$O. Due to time limitations only 26 of these 41 HGal clumps were further mapped in the on-the-fly (OTF) mode in the three CO transitions. The antenna continuously scanned a region of $22\arcmin\times 22\arcmin$ centered on the PGCCs with a scanning rate of $20\arcsec$ per second. Because of the high rms noise level at the edges of the OTF maps, only the central $14\arcmin\times 14\arcmin$ regions were used for analyses in this work \citep{2020ApJS..247...29Z}. Whether the PGCC was mapped with OTF is shown in column (10) of Table \ref{tab:samples}. If mapped, it is marked by a `Y', if no, then `N'.

To trace the high-density region in these 26 mapped HGal clumps, we chose the center of cores defined in Section \ref{result:cores} to detect J=1-0 emissions of N$_2$H$^+$ and C$_2$H. We required that each candidate should have sufficient column density with an antenna temperature of $^{13}$CO higher than 1 K. Then, 18 of these HGal clumps with absolute latitude larger than 25\degr were chosen for further observations. To compare the detection rate between high and lower latitudes, we also observed an unbiased sample of 18 PGCCs with latitudes between 21\degr and 25\degr. Finally, 36 clumps in total were observed with N$_2$H$^+$ and C$_2$H in single-point mode from May to June in 2019. The velocity resolutions are 0.19 km s$^{-1}$ for N$_2$H$^+$ and 0.21 km s$^{-1}$ for C$_2$H. The N$_2$H$^+$ emission was observed in the USB with T$_{sys}$ also around 210 K and C$_2$H in LSB with T$_{sys}$ around 210 K. The typical noise is 0.2 K for both N$_2$H$^+$ and C$_2$H.

The raw data were reduced by CLASS and GREG in the GILDAS package \citep{2000ASPC..217..299G,2005sf2a.conf..721P}.

\subsection{Archived Data-Herschel FIR Continuum Emission} \label{data:archive}
To evaluate the definition of cores (see Section \ref{result:cores}) in the special clump G004.15+35.77, the northwestern edge of LDN 134 \citep{1962ApJS....7....1L}, the FIR continuum emission images were taken from the project Galactic Cold Cores \citep{2010A&A...518L..93J} which is a \textit{Herschel} satellite survey \citep{2010A&A...518L...1P} of the source populations revealed by Planck Cold Cores program. The data were obtained with the Photodetector Array Camera and Spectrometer \citep[PACS]{2010A&A...518L...2P} at 100 and 160 $\mu m$ wavelength and the Spectral and Photometric Imaging REceiver \citep[SPIRE]{2010A&A...518L...3G} at 250, 350, and 500 $\mu m$ wavelength. The original resolution of the maps, in order of increasing wavelengths, is $7\arcsec$, $12\arcsec$, $18\arcsec$, $25\arcsec$ and $36\arcsec$. PACS 100 $\mu m$ image shows no detection with a rms of 0.00144 $Jy/pixel$ and was therefore excluded from further study. We then performed pixel-by-pixel SED fitting following \citet{2017ApJS..231...11Y} with four infared bands (160, 250, 350, 500 $\mu m$) and obtained the temperature $T_{\text{dust}}$ and column density $N_{\text{H}_2}$ maps, which are then cut to the same size as our PMO's observation ($14'\times14'$). The method is described in detail in Appendix \ref{appendix:sed}.

\subsection{Distance} \label{data:distance}
Distance is always a puzzle in astronomy. Previous works estimated the distances of HGal molecular clouds to be 100 pc by the velocity dispersion and the scale height of the ensemble of clouds \citep{1984ApJ...282L...9B,1985ApJ...295..402M}. Follow-up star counts from POSS confirmed that HGal molecular clouds are indeed nearby objects and the upper limit ranges from 125 to 275 pc \citep{1986A&A...168..271M}. Besides, both small $V_{lsr}$ and the lack of a double-sine wave signature in the distribution on the $l-V_{lsr}$ plane demonstrate that HGal molecular gas is too close to the Sun for Galactic rotation to modulate the velocities \citep{1996ApJS..106..447M}. Therefore, we should not expect the co-rotation of HGal objects with the rotation curve, which means that the traditional method relying on the Milky Way rotation curve to derive distances may not be a good choice.


Recently, the Gaia satellite has provided new photometric measurement towards galactic stars \citep{2016A&A...595A...1G}. Together with 2MASS and Pan-STARRS 1 optical and near-infrared photometry, Gaia DR2 parallaxes can help to infer distances and reddenings of $\sim$ 800 million stars. These stars trace the reddening on a small patch of sky, both along the different lines of sight and different distance intervals, allowing us to build up a map of dust reddening in 3D \citep{2014ApJ...783..114G,2019ApJ...887...93G}. With the help of this 3D map, we could infer reddening as a function of distance in the given direction. We turn to this new method to derive the distances of HGal clumps for three reasons:
\begin{enumerate}
    \item Kinematic distances to nearby clouds are very unreliable due to the local Galactic velocity dispersion.
    \item HGal clumps are less affected by Galactic rotation so it will be better if the estimation of distances is independent from the rotation of the Milky Way. 
    \item Dust and gas are proved to be well coupled in PGCCs \citep{2012ApJ...756...76W}, so it is reasonable to derive the distance of molecular gas from the dust extinction.
\end{enumerate}
In the Gaia dust reddening algorithm, given Galactic longitude and latitude, different distances of dust are assigned with probabilities and we choose the distance with largest probability in our paper. This method is precise enough unless the distance is larger than 1 kpc where the extinction is too strong \citep{2019ApJ...887...93G}. But again, HGal clumps are assumed so close that such imprecise cases are avoided. 
Along with Galactic latitudes, the vertical distance of the clumps from the Galactic midplane (elevation Z) are also derived,
\begin{eqnarray}
	Z = d\times sin(|b|)
\end{eqnarray}
where Z, d, b are the elevation, distance and latitude, respectively. The distances and elevations of all 41 PGCCs from Gaia dust extinction are listed in columns (8) and (9) of Table \ref{tab:samples}.

\section{Results} \label{sec:results}
\subsection{Definition of Cores} \label{result:cores}
From single-point observations, we find several sources with multi-velocity components in our sample. Eight HGal PGCCs have double-velocity components and one has triple-velocity components. In total, 51 velocity components were obtained in HGal. We consider each velocity component as an individual clump, ie. 51 clumps are finally defined. Mapping observations are applied to 26 PGCCs (time is limited and the sample is unbiased) and due to multi-velocity components, we have 33 mapped clumps. The CO integrated intensity maps of three example clumps are shown in Figure \ref{fig:map_subsample} and the figures of total 33 clumps are shown in Appendeix \ref{appendix:map}. $^{12}$CO is shown as colored background and $^{13}$CO (left) and C$^{18}$O (right) are shown with blue contours from 50\% to 90 \% of the peak integrated intensity. The morphologies include isolated cores, filamentary structures and also diffuse or full-coverage features.

In this paper, we define all the local enhancements of $^{13}$CO intensity higher than 70\% of the peak intensity as individual dense cores. The reasons why we use 70\% rather than 50\% are two: one is that in some detections the SNR is limited especially on the edge of the field of view, and 70\% level helps to remove those mistaken cores by 50\% level; the other is that the $^{13}$CO contours are usually extensive and not concentrated, showing a flat density profile, and therefore we need a more strict criterion to distinguish intensity peaks. A typical example is the filamentary clump G101.62-28.84a, shown in the two medium panels of Figure \ref{fig:map_subsample}. A 70\% level contour helps to distinguish four cores while 50\% level cannot do this.

As shown in Figure \ref{fig:map_subsample}, we mark these cores with black crosses and names “C1”, “C2”, ... The 50\% isolines are further fitted with ellipses, providing measurements of the semi-major and semi-minor axes ($a$ and $b$). The ICRS coordinate offsets of each core from the observational center are listed in Columns (2)-(3) of Table \ref{tab:region-para}. The distances of cores are assumed to be the same as those of the host clumps and listed in Column (4) of Table \ref{tab:region-para}.
The sizes of the cores are given by $R=\sqrt{ab}D$ ($D$ is the distance). Also, due to the beam smearing effect, we only give upper limits for those cores with sizes smaller than the beam size of our telescope. The sizes of cores are listed in Column (5) of Table \ref{tab:region-para}. 

For three clumps G089.03-41.28a, G158.77-33.30b and G159.41-34.36b, no cores are found. For only one clump G004.15+35.77, the 50\% isolines significantly overstep the boundary of the 14-arcmin box. We check the column density map (blue background) given by pixel-by-pixel SED Fitting (detailed method in Appendix \ref{appendix:sed}), shown in panel (b) of Figure \ref{fig:L134}. Two dust cores are shown with deep blue color. The black and white contours give the $^{13}$CO and C$^{18}$O integrated intensity respectively and only one CO core appears in the map. A shallow yellow circle gives the location of ammonia dense core (L134-A) with kinetic temperature $T_k$ of 10 K and volume density of $10^{4.5}\ \text{cm}^{-3}$ \citep{1983ApJ...264..517M}. L134-A correlates well with the northern dust core with a low dust temperature of 10.3 K and C$^{18}$O core correlates well with southern dust core with slightly higher dust temperature of 11.1 K. We define the CO core “C1” traced by C$^{18}$O as well as dust column density. We also notice the reason why the northern dust core has lower CO emission. This may be caused by a higher CO depletion due to the low temperature \citep{2001ApJ...557..736G}. For consistency, we still choose $^{13}$CO emission to define the core size and neglect the part outside the boundary. This is safe as long as the core is nearly elliptical.

In total, 54 cores are identified among the 33 mapped clumps. Velocity components are denoted as “a, b, c, ...”, while the cores are labeled as “C1, C2, C3, ...” following \citet{2012ApJ...756...76W}.

\subsection{Line Parameters} \label{result:line_para}
We extract the 1-arcmin box towards the core center, averaging all the spectra inside the box. Then Gaussian fitting is obtained for the three CO lines. If the profiles arise from blended multi-velocity, multiple Gaussian components are fitted simultaneously. If the profile originates from self-absorption, only one Gaussian component is fitted. The centroid velocity ($V_{lsr}$), main beam brightness temperature ($T_b$) and FWHM ($\Delta$V) can be obtained from Gaussian fitting directly. For those cores without C$^{18}$O detection, the upper limit of $T_b$ is given by the noise of the baseline and the FWHM is given the same value as as that for $^{13}$CO detection, which will be useful in the calculation of the upper limits to the C$^{18}$O column density. Three fitting parameters are listed in Table \ref{tab:obs-para of cores}. Cores without C$^{18}$O detection have three parameters of C$^{18}$O only shown with '...'. The centroid velocity of the $^{13}$CO J=1-0 line of each core is adopted as the systemic velocity, because the $^{12}$CO J=1-0 line is optically thick and the C$^{18}$O J=1-0 line has only eight solid detections in our 54 cores.

\subsection{Line Profiles} \label{result:line_profile}
Sometimes, the line profile cannot be fitted as a single Gaussian function and we call them non-Gaussian. Generally, non-Gaussian profiles in optically thick molecular lines $^{12}$CO(1-0) fall into two main categories: multi-peak and asymmetry. 

The criterion set by \citet{1997ApJ...489..719M} is applied to check whether blue or red profiles indicating cloud contraction or expansion in certain cases. $\delta V$ = ($V_{\text{thick}}$-$V_{\text{thin}}$)/$\delta$ $V_{\text{thin}}$, where $V_{\text{thick}}$ is the peak velocity of $^{12}$CO(1-0), $V_{\text{thin}}$ and $\delta V_{\text{thin}}$ are the systemic velocity and the line width of $^{13}$CO(1-0). If $\delta V < $ -0.25, the line is classified as blue asymmetric; if $\delta V > $ 0.25, then red asymmetric. Among those asymmetric profiles, if the optically thick molecular line-$^{12}$CO has double peaks at the same time, then we have a lineshape termed blue or red profile, depending on the blue or red asymmetric profile defined above. The rest of the multi-peak profiles can be well described as multi-velocity components and can be fitted by composition of two or more Gaussian functions. 

In total, four non-Gaussian profiles are found by single-point observations (see Figure \ref{fig:special_profile}): G089.0-41.2 (blue profile), G182.5-25.3 (red profile), G210.6-36.7 (red profile) and G159.5-32.8 (red asymmetry).
However, single-point observations only provide one spectrum from the center (or maybe some little offset), and thus might not represent the real line profiles from the cores. Therefore, as mentioned before, we extract the molecular lines from the center of the 54 cores and then carefully examine the line profiles respectively. A more detailed discussion will be made in Section \ref{emission_line:line_profile}

\subsection{Parameters to Be Derived} \label{result:derpara}
Parameters from the 54 extracted dense cores are further calculated in this section.

The excitation temperatures ($T_{ex}$) and optical depths ($\tau$) can be derived from the solution of the radiation transfer equation:
\begin{eqnarray} \label{eq:radiation_transfer}
&T_b =& \frac{h\nu}{k}\left[\frac{1}{\exp(h\nu/kT_{ex})-1} - \frac{1}{\exp(h\nu/kT_{bg})-1}\right] \nonumber\\
&&\times [1-\exp(-\tau)]\ f,
\end{eqnarray}
where $T_{bg}$ is the background temperature (2.73 K). Under the assumptions that $^{12}$CO J=1-0 is optically thick ($\tau \gg$ 1), and the beam filling factor $f$ is equal to 1, according to Equation \eqref{eq:radiation_transfer} of $^{12}$CO, the excitation temperature can be expressed as 
\begin{eqnarray}
T_{ex} =& \frac{h\nu}{k} \ln^{-1}\left\{\left[\frac{kT_b(^{12}\mathrm{CO})}{h\nu}+\frac{1}{\exp(h\nu/kT_{bg})-1}\right]^{-1}+1\right\}.
\end{eqnarray}
Under the condition of local thermodynamic equilibrium (LTE), $^{12}$CO and $^{13}$CO have the same excitation temperature. We can obtain the optical depth of $^{13}$CO ($\tau_{13}$) by
\begin{eqnarray}
	&\tau_{13} =& -\ln\left[1-\frac{T_b(^{13}\mathrm{CO})}{T_b(^{12}\mathrm{CO})}\right].
\end{eqnarray}

The $T_{\mathrm{ex}}$ of the 54 cores ranges from 6 K to 13 K with a median value of $\sim$ 10.3 K. The values of T$_{\mathrm{ex}}$ and $\tau_{13}$ are listed in columns (2) and (3) of Table \ref{tab:deri-para of cores}.
The dust temperature (T$_{\mathrm{dust}}$) of all the PGCCs ranges from 5.8 to 20 K with a median value between 13 and 14.5 K \citep{2016A&A...594A..28P}.
The similarity of two temperatures indicates that the gas and dust are well coupled. Besides, the different median temperature (T$_{\mathrm{dust,median}}>$ T$_{\mathrm{ex,median}}$) implies that the gas might be heated by the dust \citep{1974ApJ...189..441G,1989ApJ...340..307W}.

The column density of $^{13}$CO (N$_{^{13}\mathrm{CO}}$) and C$^{18}$O (N$_{\mathrm{C}^{18}\mathrm{O}}$) can be derived from \citep{1978afcp.book.....L,1991ApJ...374..540G}
\begin{eqnarray} \label{eq:column_density}
N = && \frac{3k}{8\pi^3B\mu_d^2} \frac{\exp\left({h B J(J + 1)}/{kT_{ex}}\right)}{J + 1}  \nonumber \\
&&\times \frac{T_{ex} + h B/3k}{1 - \exp(-h\nu/kT_{ex})} \int\tau_{\nu}dV,
\end{eqnarray}
where $B$, $\mu_d$ and $J$ are the rotational constant, permanent dipole moment of the molecule and the rotational quantum number of the lower state in the observed transition. 

Adopting the typical abundance ratios, $[{\rm H_2}]/[{\rm ^{13}CO}] = 89\times10^4$ \citep{1980ApJ...237....9M} and $[{\rm H_2}]/[{\rm C^{18}O}] = 7\times10^6$ \citep{1982ApJ...262..590F} 
for the solar neighborhood, the column density of hydrogen (N$_{\mathrm{H}_2}$) can be calculated by Equation \eqref{eq:column_density}. 
Since the detection rate of C$^{18}$O is low, we use $^{13}$CO to calculate H$_2$ column densities, as listed in column (4) of Table. \ref{tab:deri-para of cores}.

The one-dimensional velocity dispersion of $^{13}$CO is given by
\begin{equation}\sigma_{^{13}\mathrm{CO}} = \frac{\Delta V_{^{13}\mathrm{CO}}}{\sqrt{8\ln 2}}\end{equation}
The thermal ($\sigma_{\text{Th}}$) and non-thermal ($\sigma_{\text{NT}}$) velocity dispersion can be calculated from
\begin{equation}\sigma_{\text{Th}} = \sqrt{\frac{k T_{ex}}{m_H\mu}},\end{equation}
\begin{equation}\sigma_{\text{NT}} = \left[\sigma_{^{13}\mathrm{CO}}^2 - \frac{k T_{ex}}{m_{^{13}\mathrm{CO}}}\right]^{1/2},\end{equation}
where m$_{\rm H}$ is the mass of atomic hydrogen, $\mu$ is the mean weight of molecule which equals 2.33 \citep{2008A&A...487..993K}, m$_{^{13}\mathrm{CO}}$ is the mass of the molecule $^{13}$CO and $k$ is the Boltzmann constant. Then the three-dimensional velocity dispersion $\sigma_{3\mathrm{D}}$ can be calculated from
\begin{equation}\sigma_{3\mathrm{D}} = \sqrt{3(\sigma_{\text{Th}}^2 + \sigma_{\text{NT}}^2)}.\end{equation}

All the derived parameters are listed in Table \ref{tab:deri-para of cores}.

\subsection{Parameters of Cores} \label{result:map_para}
The volume density can be derived through $n_{\mathrm{H}_2}$ = $N_{\mathrm{H}_2}/2R$ ($N$ is the column density of hydrogen molecules and $R$ is the size of the core) under the assumption of homogeneous spherical material distribution. The LTE mass can be obtained directly from
\begin{equation}\label{eq:Mlte} M_{\mathrm{LTE}} = \frac{4}{3}\pi R^3\mu n_{\mathrm{H}_2}m_{\mathrm{H}}.\end{equation}

Assuming that the cores are isothermal spheres and supported solely by random motions, we can calculate the virial mass $M_{\mathrm{vir}}$ as \citep{1994ApJ...428..693W}
\begin{equation}\label{eq:Mvir} M_{\mathrm{vir}} = \frac{5R\ \sigma^2_{3D}}{3\gamma \ G}, \end{equation}
where G is the gravitational constant. The density profile is assumed to be uniform (also corresponding to the isothermal distribution mentioned in Section \ref{stability}), and then $\gamma$ should be unity.

In molecular clouds, thermal pressure, turbulence and magnetic field can support the gas against gravity collapse. Taking thermal pressure and turbulence into account, the Jeans mass can be expressed as following \citep{2008ApJ...684..395H}
\begin{equation}\label{eq:Mj} \frac{M_{\mathrm{Jeans}}}{M_{\odot}} \approx 1.0 a_\mathrm{J} \left(\frac{T_{\mathrm{eff}}}{10K}\right)^{3/2}\left(\frac{\mu}{2.33}\right)^{-1/2}\left(\frac{n}{10^4cm^{-3}}\right)^{-1/2},\end{equation}
where $a_\mathrm{J}$ is a dimensionless parameter of the order unity, $n$ is the volume density of H$_2$, and the effective kinematic temperature $T_{\mathrm{eff}}$ is adopted as $\frac{\mu m_H C_\mathrm{s,eff}^2}{k_B}$. The effective sound speed $C_{\mathrm{s,eff}}$ is expressed by $[(\sigma_{\mathrm{NT}})^2+(\sigma_{\mathrm{Th}})^2]^{1/2}$ to account for combination of both thermal motion and turbulence.

According to \citet{2013ApJ...779..185K}, the virial parameter can be applied to gauge stability against gravitational collapse:
\begin{equation}\label{eq:alpha_vir}
    \alpha_{\mathrm{vir}} = \frac{M_{\mathrm{vir}}}{M_{\mathrm{LTE}}}
\end{equation}

The volume density, LTE mass, Jeans mass, virial mass and virial parameter are listed in column (6) -- (10) of Table \ref{tab:region-para}, respectively. Also notice that some cores only have upper limits for sizes, the size-related parameters above are then semi-constrained and presented with upper or lower limits.

\subsection{Dense Gas Tracers} \label{result:dense-gas}
We have observed two dense gas tracers N$_2$H$^+$ and C$_2$H in eighteen clumps with high enough $^{13}$CO antenna temperature among 26 mapped HGal PGCCs, which are thought to be the sub-sample with embedded dense cores. For comparison purpose, we also searched for N$_2$H$^+$ and C$_2$H in eighteen clumps at the latitude range between 21\degr and 25\degr. Only one HGal PGCC has obvious detections, of both two dense tracers, as shown in the lower right panel of Figure \ref{fig:dense_tracer}. For another eighteen low-latitude clumps, we find six of them having at least one kind of dense-gas tracer. G008.5+21.8 and G158.8-21.6 have both N$_2$H$^+$ and C$_2$H detection respectively in the upper left and upper right panel of Figure \ref{fig:dense_tracer} and clumps like G008.6+22.1 have only C$_2$H detection.

\section{Discussion} \label{sec:discussion}
\subsection{Galactic Distribution} \label{discuss:distribution}
\begin{figure*}
	\centering
	\includegraphics[width=1.0\linewidth,angle=0]{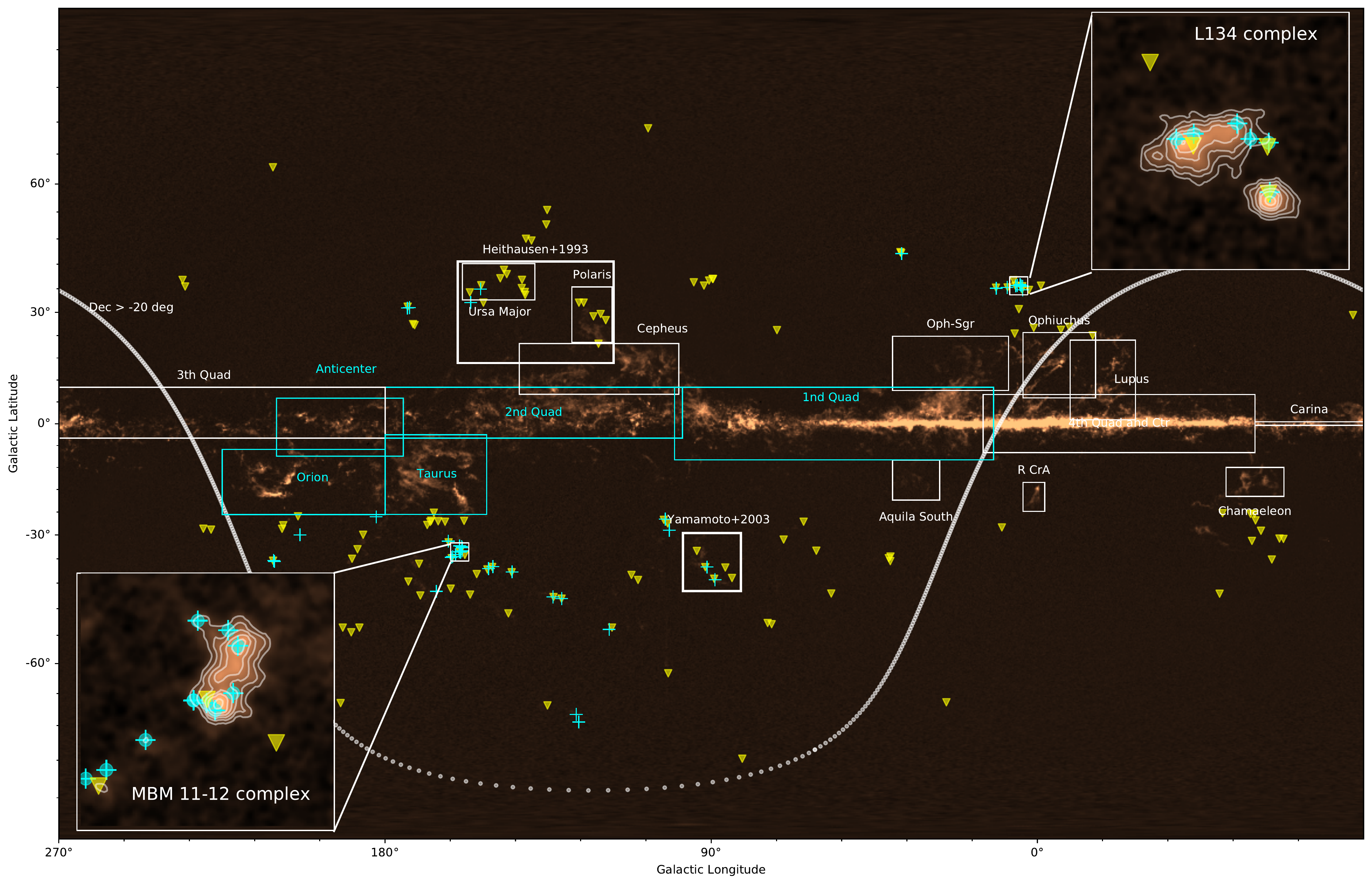} 
	 \caption{CO surveys are shown in this figure, including previous ones and our work. The all-sky map is projected onto a rectangle (cylinder projection by the package \textit{Montage}). The background is CO J=1-0 integrated intensity \citep{2014A&A...571A..13P}. The horizontal distortion at high latitude ($\ge60\degr$) is due to the effect of projection. Follow-up observations towards 41 PGCCs at HGal with the PMO 13.7-m telescope (PMO) are shown with blue crosses. The white point line \textbf{marks} a \textbf{declination} of -20\degr, above which the PMO can observe efficiently. Unequal spacing of the points is due the effect of projection. Rectangles with names were either defined by \citet{1987ApJ...322..706D} or \citet{2012ApJ...756...76W}. Those blue rectangles mark the published PMO data \citep{2012ApJS..202....4L,Meng_2013,Zhang_2016,2020ApJS..247...29Z}. Bold white rectangles mark the regions measured by large CO surveys \citep{1993AA...268..265H,2003ApJ...592..217Y}. The filled yellow inverted-triangles show 121 HGal objects \textbf{summarized} by \citet{1996ApJS..106..447M}. We zoom in two HGal cloud complexes, the L134 complex and the MBM 11--12 complex, where PGCCs are grouped in the northern and southern hemisphere, respectively. In the two zoomed-in figures, white contours give the CO integrated intensity and all the mentioned HGal objects inside the boxes are included. The concentric filled blue circles show the field of view (14$^\prime$) in our observation.}
	\label{fig:COsurvey}
\end{figure*}

\begin{figure*}
    \centering
    \includegraphics[width=0.8\linewidth,angle=0]{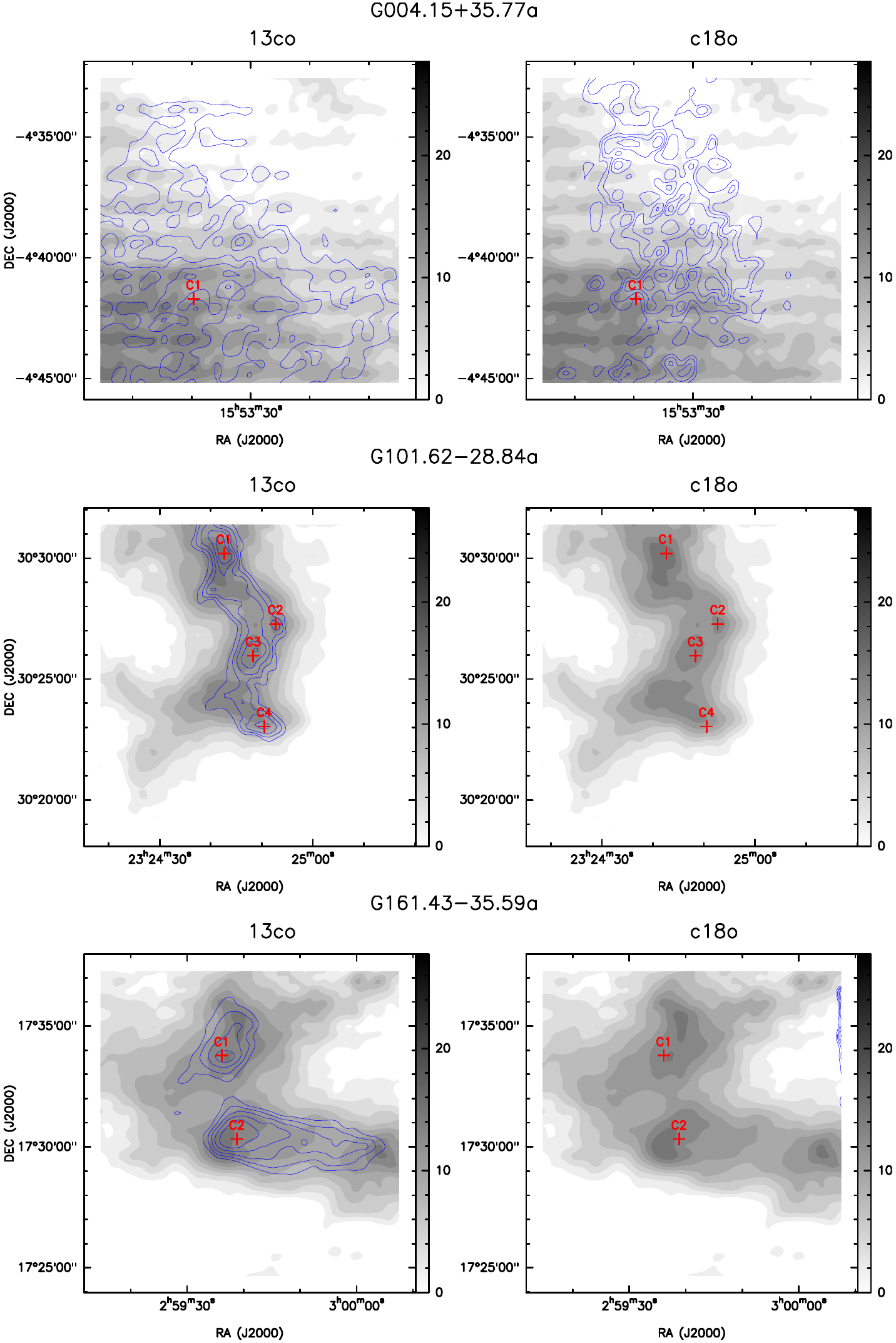}
	 \caption{Three examples of CO integrated intensity maps. Figures for the total sample are shown in Appendix \ref{appendix:map}. The background grey color maps show $^{12}$CO(1-0) emission and the colorbars show the value of W($^{12}$CO) in units of [K km s$^{-1}$]. The blue contours of $^{13}$CO(1-0) or C$^{18}$O(1-0) step from 50\% to 90\% by 10\% of the peak value. Defined cores are marked by red crosses and named `C1', `C2', and so on in the order of left-to-right and up-to-down. The specific line is labeled on the top of each panel. Source names are labeled above each pair of panels.}
	\label{fig:map_subsample}
\end{figure*}

The distribution of HGal clumps is shown in Figure \ref{fig:COsurvey}. The background colormap is the CO survey of the Planck Collaboration \citep{2014A&A...571A..13P}. HGal PGCCs we observed are marked with blue crosses. Rectangles with names were defined after \citet{1987ApJ...322..706D} as well as \citet{2012ApJ...756...76W}, marking the known CO molecular clouds or complexes. The blue rectangles mark the published PMO 13.7-m data \citep{2012ApJS..202....4L,Meng_2013,Zhang_2016,2020ApJS..247...29Z}. Two bold rectangles at high latitude mark the regions that have large and unbiased survey with higher resolution or spatial spacing \citep{1993AA...268..265H,2003ApJ...592..217Y}. However, the survey of \citet{1987ApJ...322..706D} was limited to the Galactic plane, while the surveys by \citet{1993AA...268..265H} and \citet{2003ApJ...592..217Y} are limited to certain regions (see Figure \ref{fig:COsurvey}). Compared to prior surveys, we have detected CO emission at higher latitude. The average latitude of 41 clumps is 36.87\degr. The highest elevation is 214 pc for G153.74+35.91 and the average value is 119 pc.

To examine the consistency between PGCCs and HGal clouds with CO detections, we cross-match the HGal PGCCs with molecular clouds that have been found and observed in CO before, including \citet{1993AA...268..265H,1996ApJS..106..447M,1998ApJ...492..205H,2000ApJ...535..167M}. The cross-matching result is shown in Column (11) of Table \ref{tab:samples}. Filled yellow circles in Figure \ref{fig:COsurvey} show 121 HGal objects summarized by \citet{1996ApJS..106..447M}. As we can see, most clumps are embedded in the known molecular clouds, and tend to group in several famous HGal cloud regions. In the Northern Hemisphere, 2 clumps come from the Ursa Major cloud complex, 2 come from MBM 25, 6 come from the L134 complex. In the southern hemisphere, most of clumps gather at 145\degr$\le l \le$195\degr, -46\degr$\le b\le$-30\degr, to the south of the Tarus-Aurgiae complex of dark clouds \citep{2000ApJ...535..167M}. For instance, MBM 11--12 (L1457-c) contains 7 clumps. There are also 2 clumps associated with MBM 53--55 complex surveyed by \citet{2003ApJ...592..217Y}, which has a significant extent 85\degr $\le l \le$ 96\degr and -43\degr $\le b \le$ -32\degr \citep{1988SpringerM,2000ApJ...535..167M}. We then zoom in two clump-crowded regions of the L134 complex (6 clumps included) and MBM 11--12 complex (7 clumps included) in the lower left and upper right panel of Figure \ref{fig:COsurvey}, respectively. Those clumps are mainly distributed at the outskirts of these HGal clouds/complexes.

There are 6 clumps G101.62-28.84, G102.72-25.98, G126.65-71.43, G127.31-70.09, G182.54-25.34 and G203.57-30.08 not associated with known HGal clouds, contributing to 14.6\% (6 in 41) of the sample. Those clumps, more diffuse in shape and smaller in size, might escape the search with coarse sampling grid of \textbf{about 1 degree} and beamwidth of $\sim8^\prime$ in the surveys of \citet{1998ApJ...492..205H} and \citet{2000ApJ...535..167M}. Besides, HGal PGCCs are generally embedded in translucent molecular clouds, so the MBM clouds selected by optical obscuration may also not include those diffuse clumps. Yet those clumps are cold enough to be detected by Planck.

\subsection{Emission Lines} \label{discuss:emission_line}
\subsubsection{Detection Rate} \label{emission_line:detection_rate}
All the HGal clumps were observed in $^{12}$CO and $^{13}$CO but only 16 clumps have C$^{18}$O detections. The detection rate of $^{12}$CO is 100\%, indicating that Planck Cold Galactic Clumps (PGCCs) are well-traced by CO emission. Such a result corresponds to similarly high detection rate of 98.4\% found in a surveyed 64 Lynds Dark Clouds sample where extinction exceeds one or two magnitudes \citep{1975ApJ...202...50D}. However, the detection rate of C$^{18}$O is only 40\%, compared to 68\% of the total sample-674 PGCCs \citep{2012ApJ...756...76W}, 77.5\% in Taurus\citep{2013ApJS..209...37M}, 81.5\% in the First Quadrant, 79.5\% in the Anti-centre Direction \citep{2020ApJS..247...29Z} and 84.4\% in the Second Quadrant \citep{Zhang_2016}.

Besides, two dense-gas tracers N$_2$H$^+$ and C$_2$H (see in Section \ref{result:dense-gas}) show that the detection rate of molecular lines with high critical density (molecular lines that trace dense gas) is different between the HGal region ($|b|>25\degr$) and lower region ($25\degr>|b|>21\degr$). Only one clump has both N$_2$H$^+$ J=1-0 and C$_2$H J=1-0 among the eighteen selected HGal clumps (see also Section \ref{result:dense-gas}), which are thought to contain dense cores. While for lower latitude regions, four of the eighteen dense cores show significant C$_2$H J=1-0 detections and three have N$_2$H$^+$ J=1-0. Such a result along with high detection rates of 58.7\% for C$_2$H and 47.9\% for N$_2$H$^+$ of 121 CO selected sample, mostly around $21\degr>|b|>0\degr$ \citep{2019A&A...622A..32L}, demonstrates a significant gradient of material density along latitude. Figure \ref{fig:detection_rate} shows the trend of detection rate decreasing along latitude. There are two explanations for this: one is that as Galactic latitude increases, gas becomes more diffuse and dense regions become rare; the other is that the abundances of N$_2$H$^+$ and C$_2$H are lower.

\subsubsection{CO Isotope Abundances}\label{emission_line:abundance}
The Galactic gradient of isotope ratio $^{12}$C/$^{13}$C is given by \citep{2020A&A...640A.125J}:
\begin{eqnarray}
    \frac{^{12}\mathrm{C}}{^{13}\mathrm{C}} = 5.87\frac{R_\mathrm{gal}}{\mathrm{kpc}} + 13.25
\end{eqnarray}
and $^{16}$O/$^{18}$O is given by \citep{1994ARA&A..32..191W}:
\begin{eqnarray}
    \frac{^{16}\mathrm{O}}{^{18}\mathrm{O}} = 58.8\frac{R_\mathrm{gal}}{\mathrm{kpc}} +37.1
\end{eqnarray} where $R_{\mathrm{gal}}$ is the distance towards the Galactic center.

The close distances of HGal clumps (see Section \ref{data:distance}) make it easier to calculate the CO abundances, simply by adopting $R_\mathrm{gal}$ = 8 kpc (solar neighbourhood). Considering CO and its isotopes $^{13}$CO and C$^{18}$O, we could derive the abundance $\frac{X(^{13}\mathrm{CO})}{X(\mathrm{C}^{18}\mathrm{O})}$ to be about 8, which is consistent with the single-point result \citep[Fig.11]{2012ApJ...756...76W} as well as the ratio we used in Section \ref{result:derpara}.
We then derived the column density of hydrogen molecules $N_{\mathrm{H}_2}$ for all 54 HGal cores by both $^{13}$CO and C$^{18}$O. Among these cores, 8 with C$^{18}$O (1-0) detections have been fitted with Gaussian profiles and the column densities can be derived directly. For the other 46 cores without C$^{18}$O, the upper limits are given by the noise $\sigma$ of baseline fitting (see Section \ref{result:derpara}). The cumulative distribution functions of the column densities derived by $^{13}$CO (black) and C$^{18}$O (blue) are shown in Figure \ref{fig:abundance} (a). Individual uncertainties are considered and 1000 runs of Monte Carlo simulate the data scatter. Besides, a KS test gives extremely low $p$-value, showing that the H$_2$ column densities derived from $^{13}$CO and C$^{18}$O are significantly different. 

We also check the possibility that low column density derived from C$^{18}$O is due to low excitation temperature of the C$^{18}$O molecule \citep[eq.10]{2003ApJ...585..823L}. Since the LTE condition is not always satisfied, especially when the volume density (for HGal, the mean value is $10^4$ cm$^{-3}$) is no higher than the critical density ($\sim 10^3$ cm$^3$) for pure collisional excitation by H$_2$ \citep{2013seg..book..491S}, the excitation temperature of C$^{18}$O cannot be as high as that for $^{12}$CO when the optical depth $\tau_{18}$ is low (mean value is 0.15). We assume that the excitation temperature of C$^{18}$O is as low as 5 K, then the column density shows only 13($\pm$4)\% enhancement. Such modest enhancement cannot explain the C$^{18}$O abundance we observed.

The Taurus region, reported to be close to us \citep{1987ApJ...322..706D,2009ApJ...703...52L,2010A&A...512A..67L}, is the best place to test whether HGal gas shows a significant deviation of CO isotopic abundances. This is because one would expect the closeness in location (and therefore a similar distance to the Galactic center) would produce similar abundances. We compare HGal (blue points with error bars) to Taurus as well as non-HGal PGCCs (red contours) in Figure \ref{fig:abundance} (b). The ratio $X_{13}/X_{18}$ in the solar neighborhood $\sim 8$ is satisfied along grey solid line. As shown in the Figure \ref{fig:abundance} (b), PGCCs in Taurus obey this ratio well and other regions have even lower ratios. However, HGal cores are mostly distributed much below the solid line, showing that they have a much higher $X_{13}$/$X_{18}$ ratio, which could be as high as 16. Above all, the ratio $X_{13}$/$X_{18}$ in HGal cores is systematically higher than those of previous results, which is consistent with only a 40\% detection rate of C$^{18}$O by single-point observed results (see in Section \ref{emission_line:detection_rate}).

High abundance ratios of $^{13}$CO to C$^{18}$O (16.47$\pm 0.10$) towards photo-dominated regions (PDR) are reported in the Orion-A giant molecular cloud and it suggests that FUV radiation penetrates the innermost part of clouds and the whole cloud experiences selective UV photo-dissociation \citep{2014A&A...564A..68S}. We examine the relation between the column density of C$^{18}$O and the abundance ratio $X_{13}$/$X_{18}$ in Figure \ref{fig:abundance} (c), shown with blue points (deep-colored points give solid abundance values and light-colored points give lower limits) as well as contours. The estimations of $A_v$ are shown at the top x-axis of the figure using the following relation derived in the Taurus region by \citet{1982ApJ...262..590F}:
\begin{equation}
    A_v = \frac{N_{\mathrm{C}^{18}\mathrm{O}}(\mathrm{cm}^{-2})}{2.4\times 10^{14}} + 2.9
\end{equation}
As the column density $N_{\mathrm{C}^{18}\mathrm{O}}$ increases (extinction $A_v$ increases from $\sim$3.0 to $\sim$6.0), the abundance ratio decreases. Such decreasing trend could be interpreted as lower penetrability of external photons into the clouds and therefore by a weaker isotopic selective photo-dissociation effect as the extinction becomes larger. HGal clouds, reported to be translucent \citep{1988ApJ...334..771V,1991ApJ...366..141V}, are in the astrochemical regime where most of the carbon becomes tied up in the form of CO \citep{1996ApJS..106..447M}. Hence, those clouds are more likely to be influenced by UV photons in the ISM though not as strong as the Orion-A giant molecular cloud mentioned above. Besides, the cold clumps we observed tend to be distributed in the outskirts of those clouds (see Figure \ref{fig:COsurvey} and therefore lower extinction), making the selective photo-dissociation easier.

We also notice that once column density becomes larger ($A_v > 7$), the ratio keeps about 14, which is still higher than the ratio of 5.5 in the solar neighborhood. Such high ratio could not be explained solely by selective photo-dissociation. A decreasing $X_{13}$/$X_{18}$ ratio as $\Sigma_{\mathrm{SFR}}$ increases is reported in nearby spiral galaxies and the possible explanation is that gas preferentially enriched by massive stars shows low $^{13}$CO/C$^{18}$O ratio \citep{2017ApJ...836L..29J}. In turn, in HGal regions, $\sim$ 100 pc above the disk, star formation history might not be the same as that in the disk. The scarce star formation activities (see two zoomed-in figures in Section \ref{discuss:core-status}) indicate a low star formation rate at HGal, which may be responsible for the high $X_{13}$/$X_{18}$ ratio.

\subsubsection{Line Profile} \label{emission_line:line_profile}
Core-extracted spectra show similar results as single-point observations did, except for the case of G159.5-32.8 (red asymmetry) which is later proved to have double velocity components bearing totally 5 cores. G210.6-36.7 (red profile) without mapping observation is excluded for further analyses.

\begin{figure*}
	\centering
	\includegraphics[width=0.7\linewidth,angle=0]{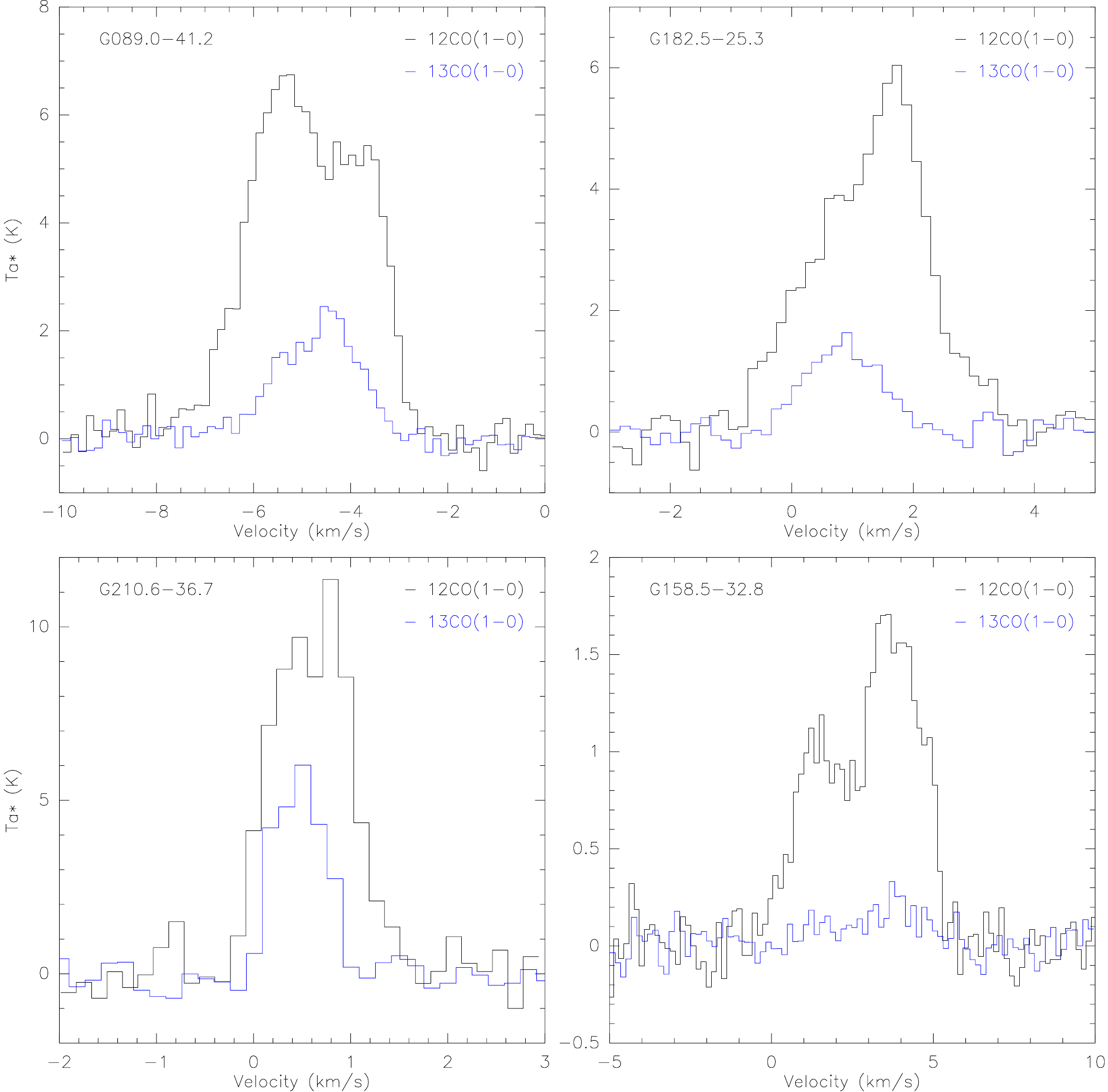}
	\caption{Non-Gaussian profiles defined by single-point observations. Upper left, right, lower left and right are respectively G089.0-41.2 (blue profile), G182.5-25.3 (red profile), G210.6-36.7 (red profile) and G159.5-32.8 (red asymmetry).}
	\label{fig:special_profile}
\end{figure*}

\begin{figure*}
	\centering
	\includegraphics[width=0.3\linewidth,angle=90]{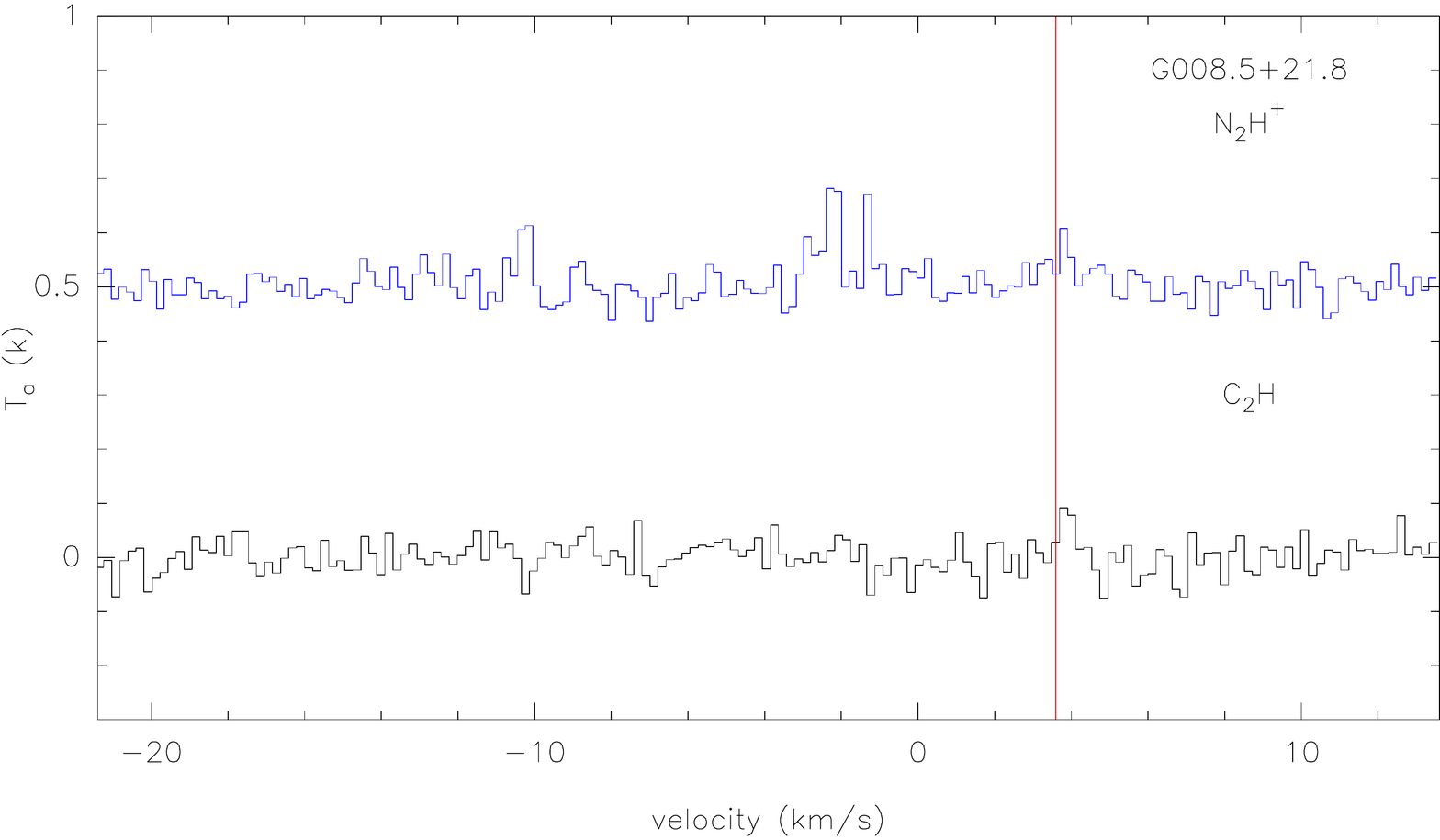}
	\includegraphics[width=0.3\linewidth,angle=90]{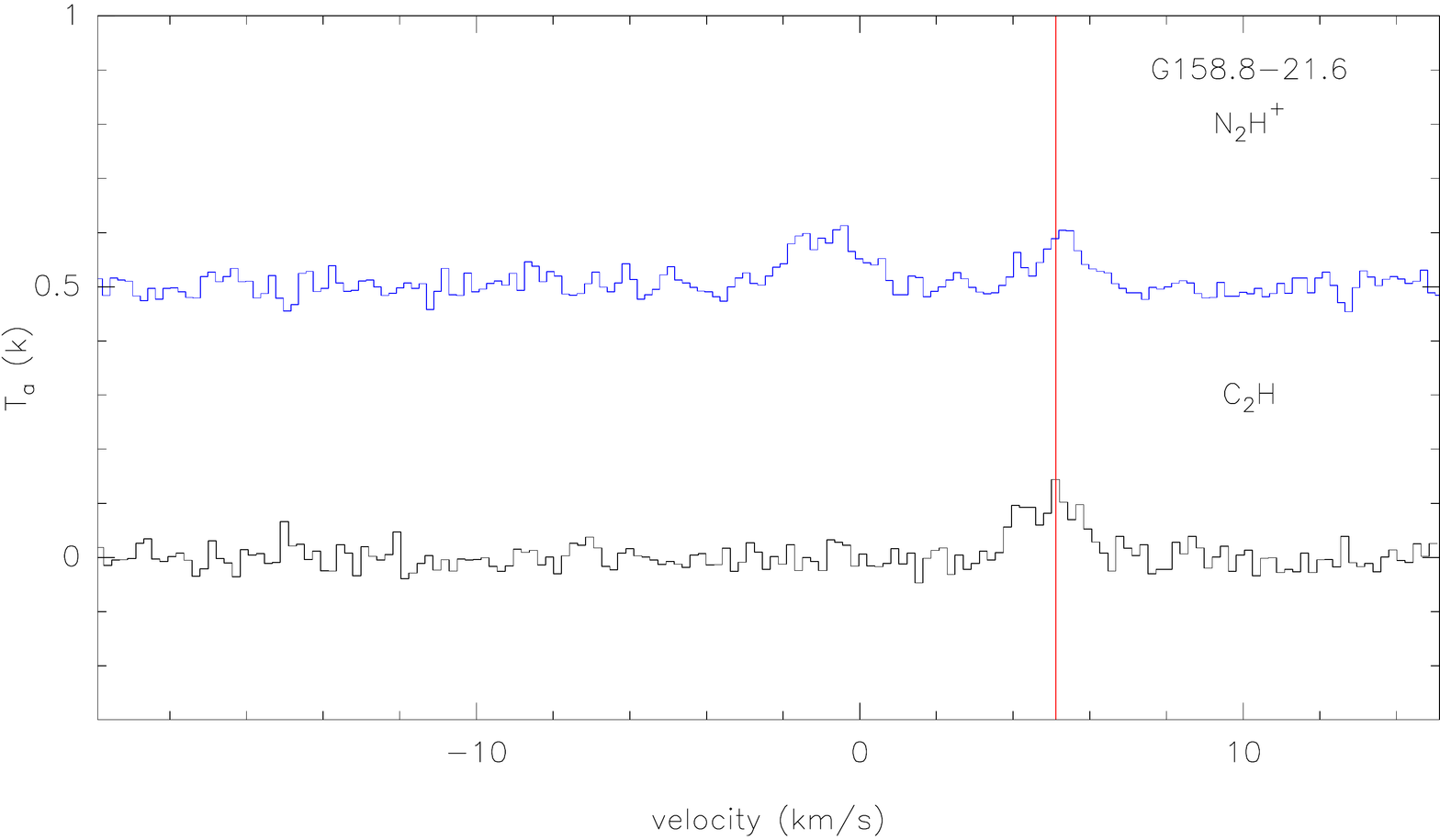}
	\includegraphics[width=0.3\linewidth,angle=90]{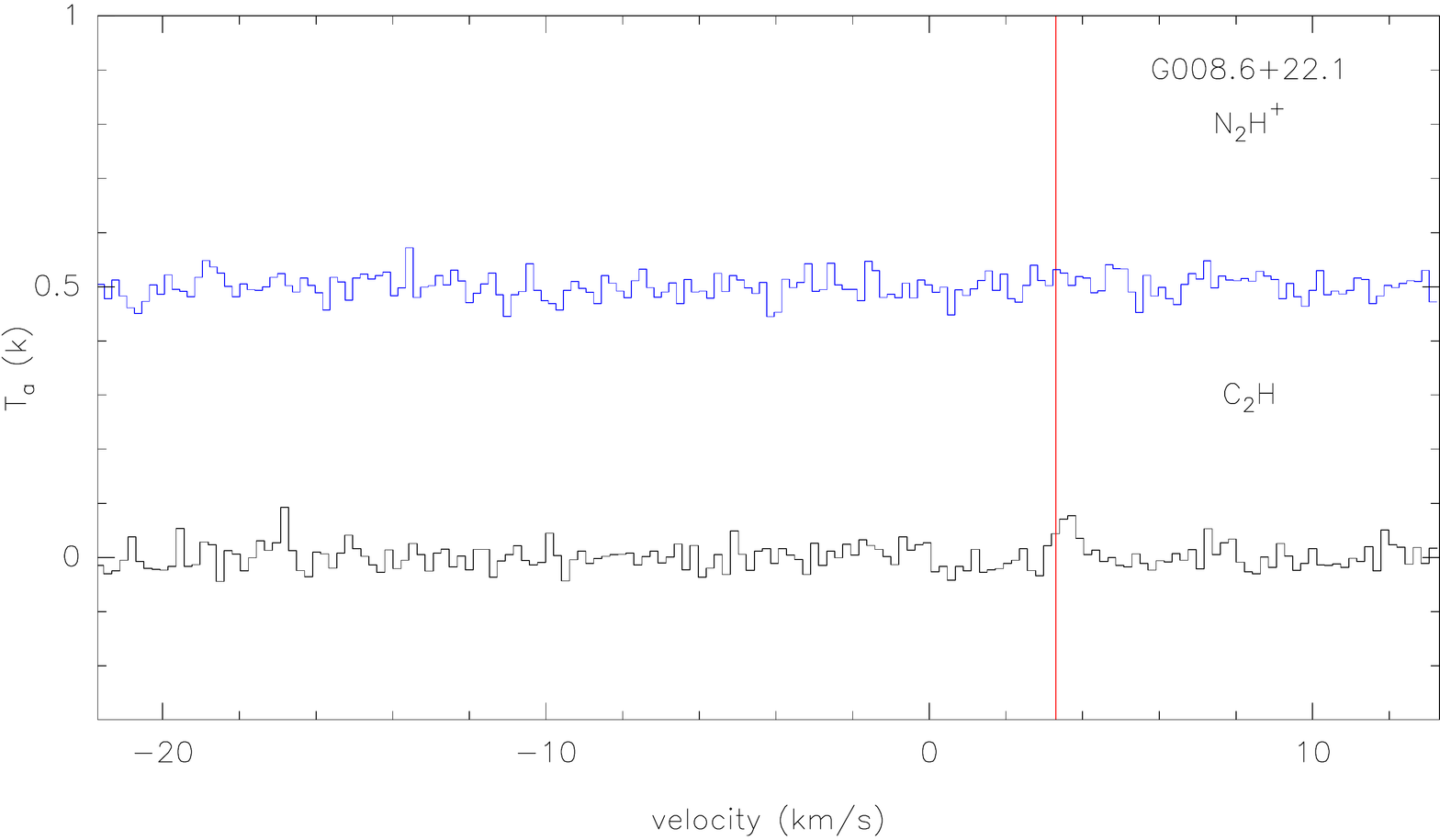}
	\includegraphics[width=0.3\linewidth,angle=90]{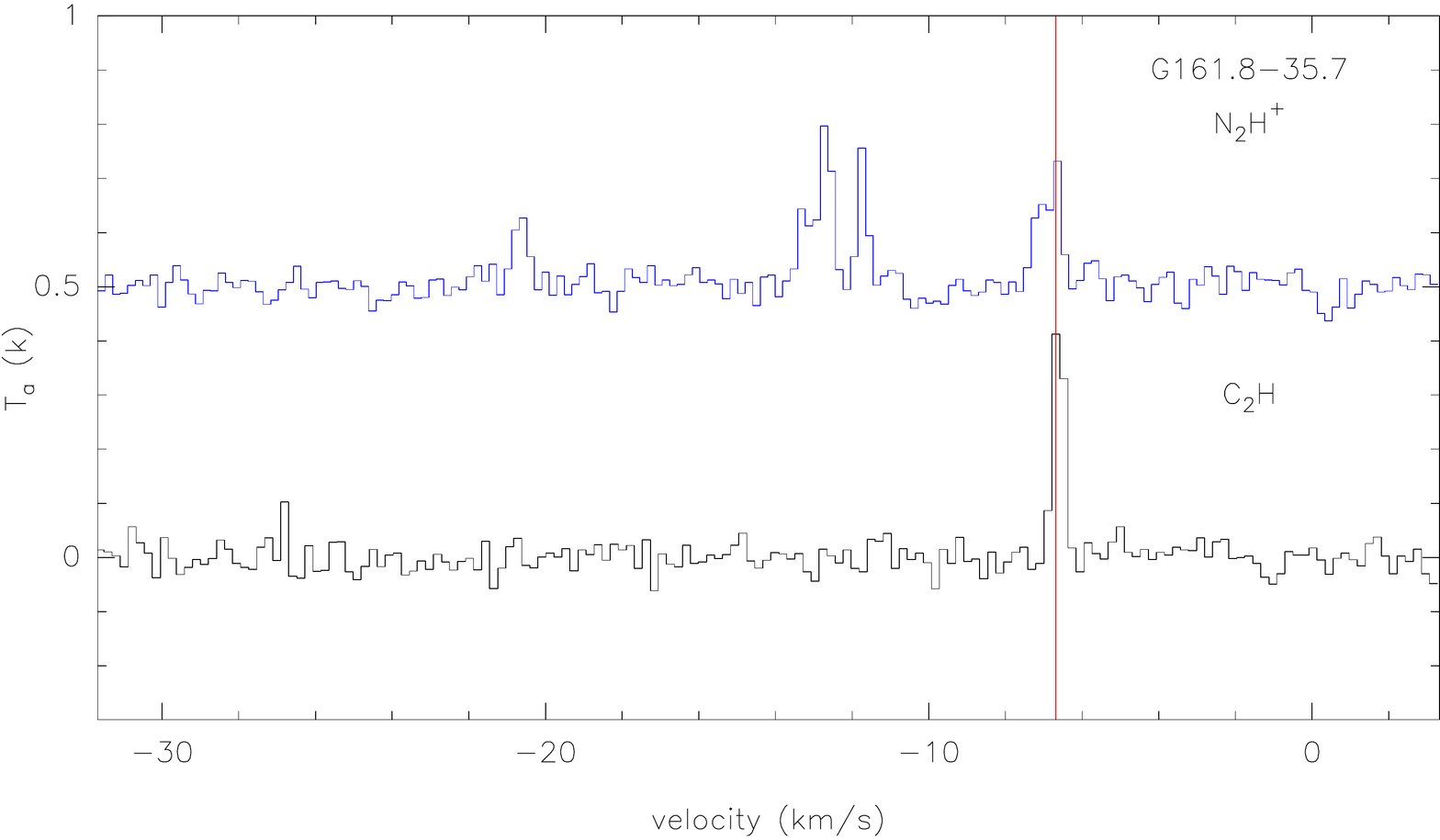}
	 \caption{Solid detection of N$_2$H$^+$ J=1-0 and C$_2$H J=1-0 at HGal and at lower latitude ($21\degr<|b|<25\degr$). Source names are labeled on the upper right of the panels. Three detections from lower latitude (upper left, upper right and lower left) and only one detection of both two molecules at HGal region (lower right).}
	\label{fig:dense_tracer}
\end{figure*}

\begin{figure*}
	\centering
	\includegraphics[width=0.5\linewidth,angle=0]{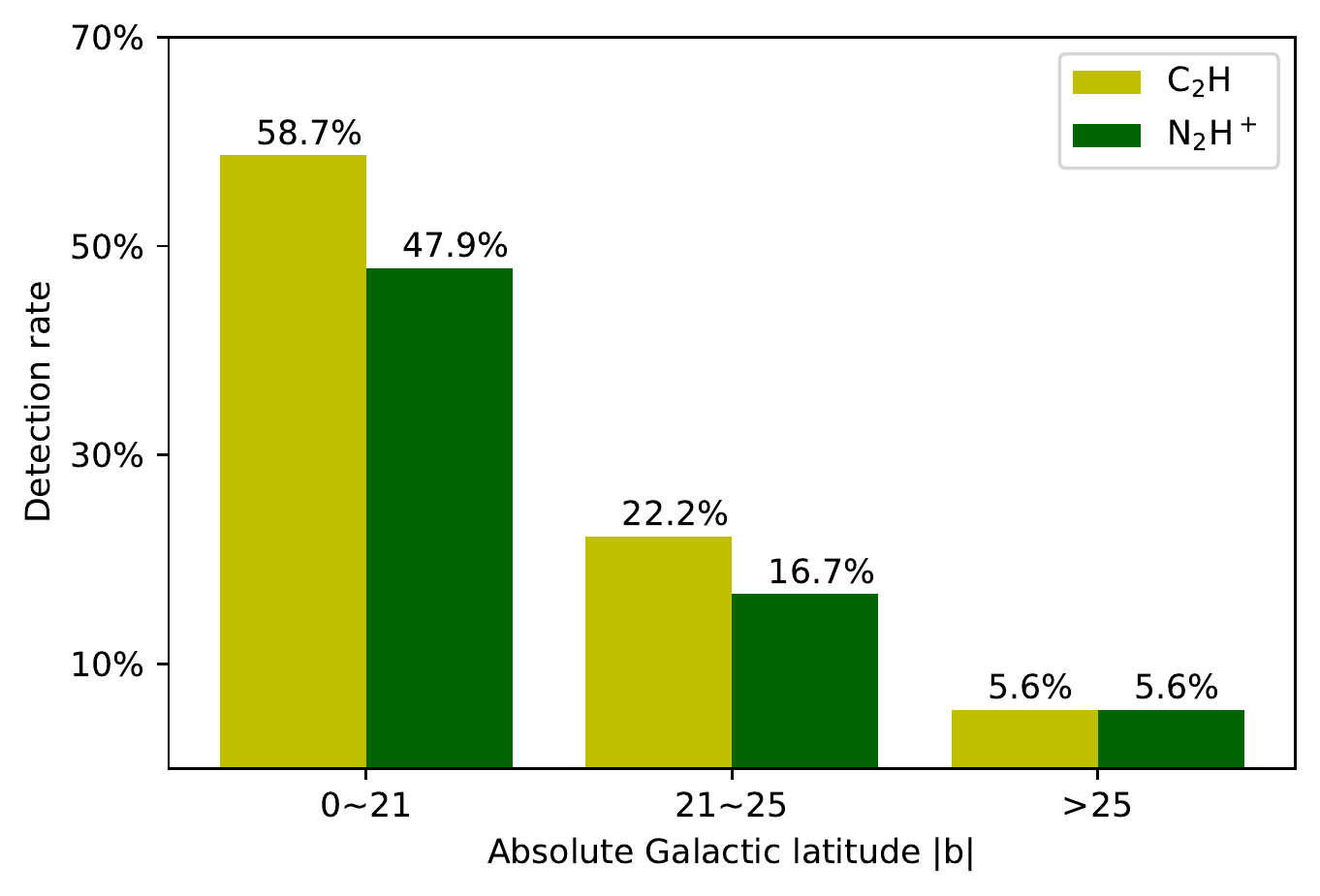}
	 \caption{The detection rate of both C$_2$H J=1-0 and N$_2$H$^+$ J=1-0 for three regions along the Galactic latitude: $0^\circ\sim21\degr$, $21\degr\sim25\degr$ and $>25\degr$. The negative gradient of detection rate along latitude is shown.}
	\label{fig:detection_rate}
\end{figure*}

\begin{figure*}
\centering
\subfigure[]{
    \begin{minipage}[b]{0.46\textwidth}
    \includegraphics[width=1\textwidth]{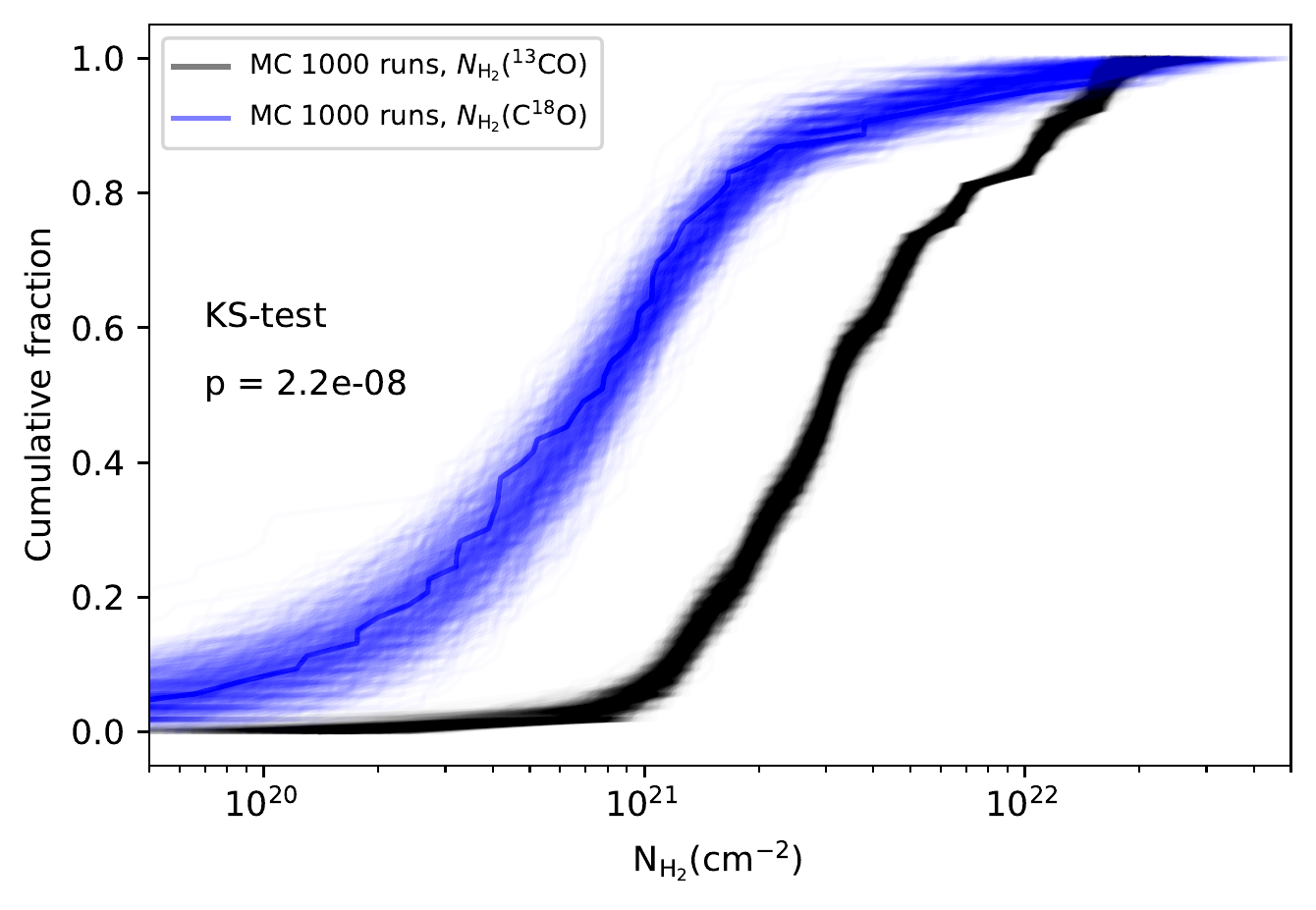}
    \end{minipage}
}%
\subfigure[]{
    \begin{minipage}[b]{0.46\textwidth}
    \includegraphics[width=1\textwidth]{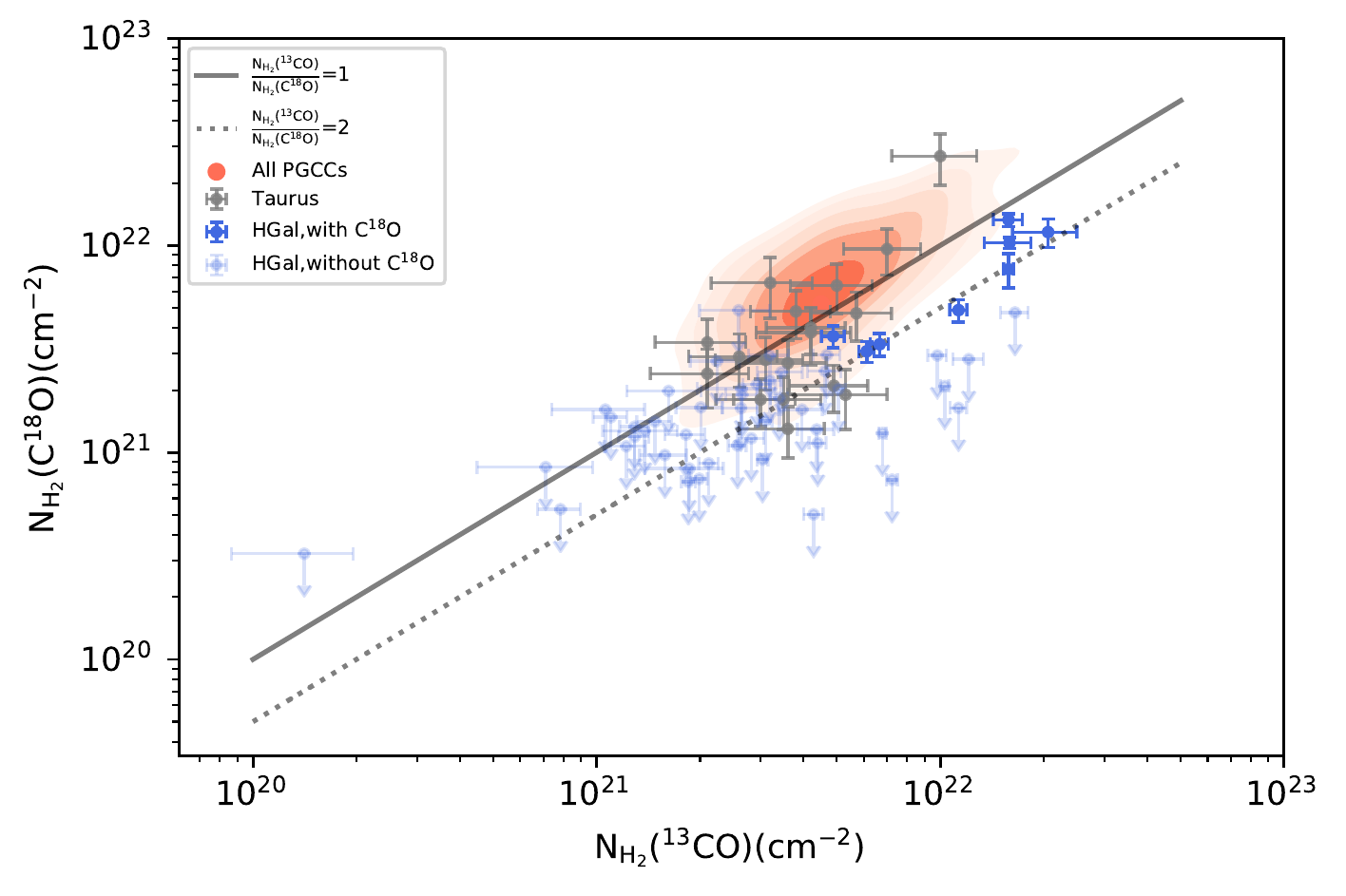}
    \end{minipage}
}\\
\subfigure[]{
    \begin{minipage}[b]{0.46\textwidth}
    \includegraphics[width=1\textwidth]{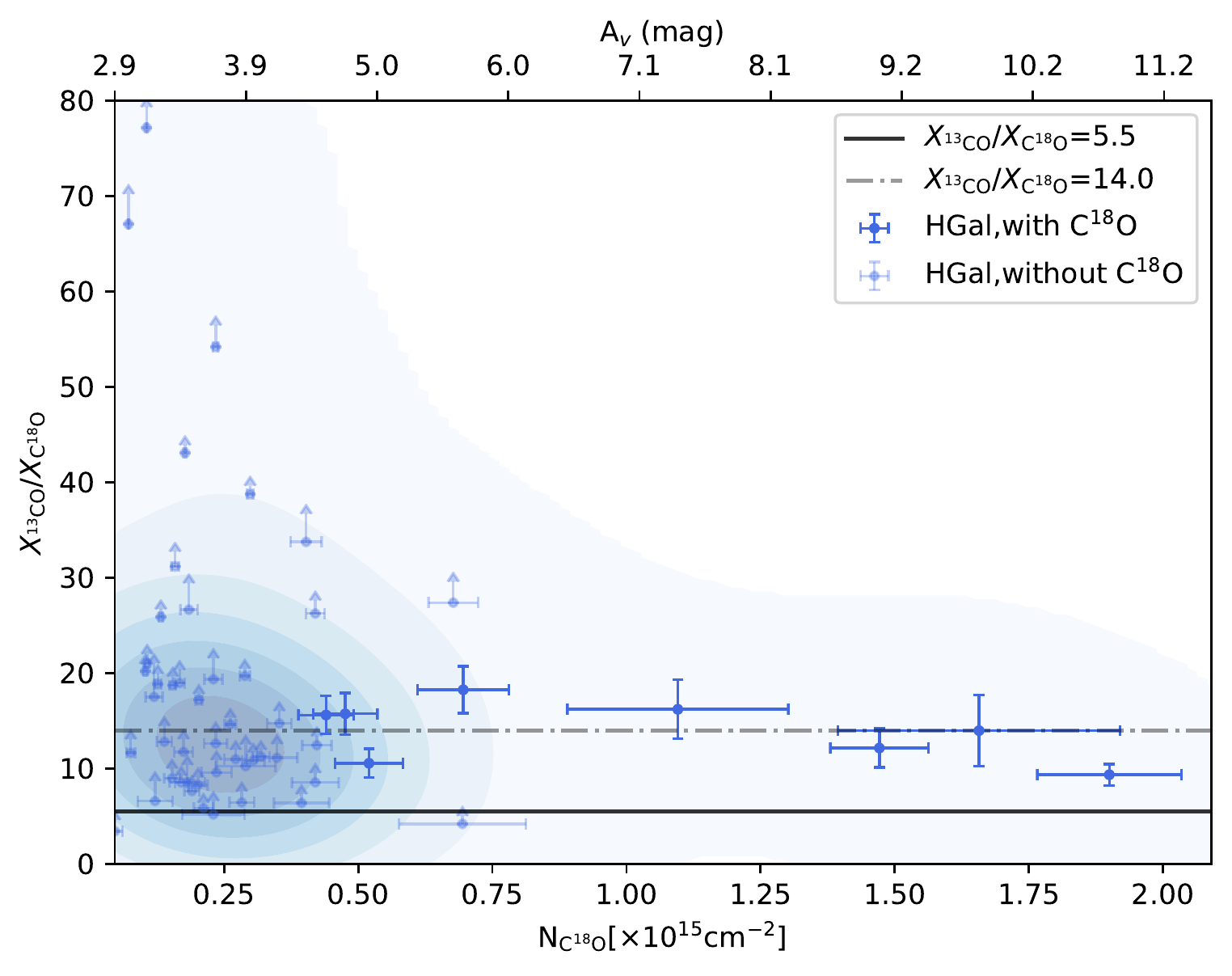}
    \end{minipage}
}%
\caption{(a) Cumulative distribution function of column density derived by two ways-$^{13}$CO (black) and C$^{18}$O (blue). Individual uncertainties are considered and 1000 runs of Monte Carlo simulates the data scatterings. The column density derived from two ways show significant difference. KS test also gives extremely low p value. (b) Assuming the abundance of isotope in the solar neighborhood, the column density of hydrogen molecules are derived. Along the black solid line, the column densities derived from $^{13}$CO and C$^{18}$O are consistent. Grey points are from Taurus \citep{2013ApJS..209...37M}. Orange contours show the distribution of all non-HGal PGCCs in previous work \citep{2012ApJS..202....4L,Meng_2013,Zhang_2016,2020ApJS..247...29Z}. Blue points are from HGal (this work). Deep blue points are 8 cores with solid C$^{18}$O detection while light blue points only give upper limits. Grey dotted line gives a much better estimation of abundance $X_{13}/X_{18}$ of 16. (c) Abundance ratio $X_{^{13}\mathrm{CO}}/X_{\mathrm{C}^{18}\mathrm{O}}$ versus the C$^{18}$O column density $N_{\mathrm{C}^{18}\mathrm{O}}$} \label{fig:abundance}
\end{figure*}

G089.03-41.28 and G182.54-25.34a (see Figure \ref{fig:special_profile}) are classified by single-point observations as blue and red line profiles, respectively. They both have mapping observations and further analyses find that the non-Gaussian line profiles are probable the result of double or complicated velocity components. These cases are discussed in the following.

\begin{figure*}
	\centering
	\includegraphics[width=0.5\linewidth,angle=0]{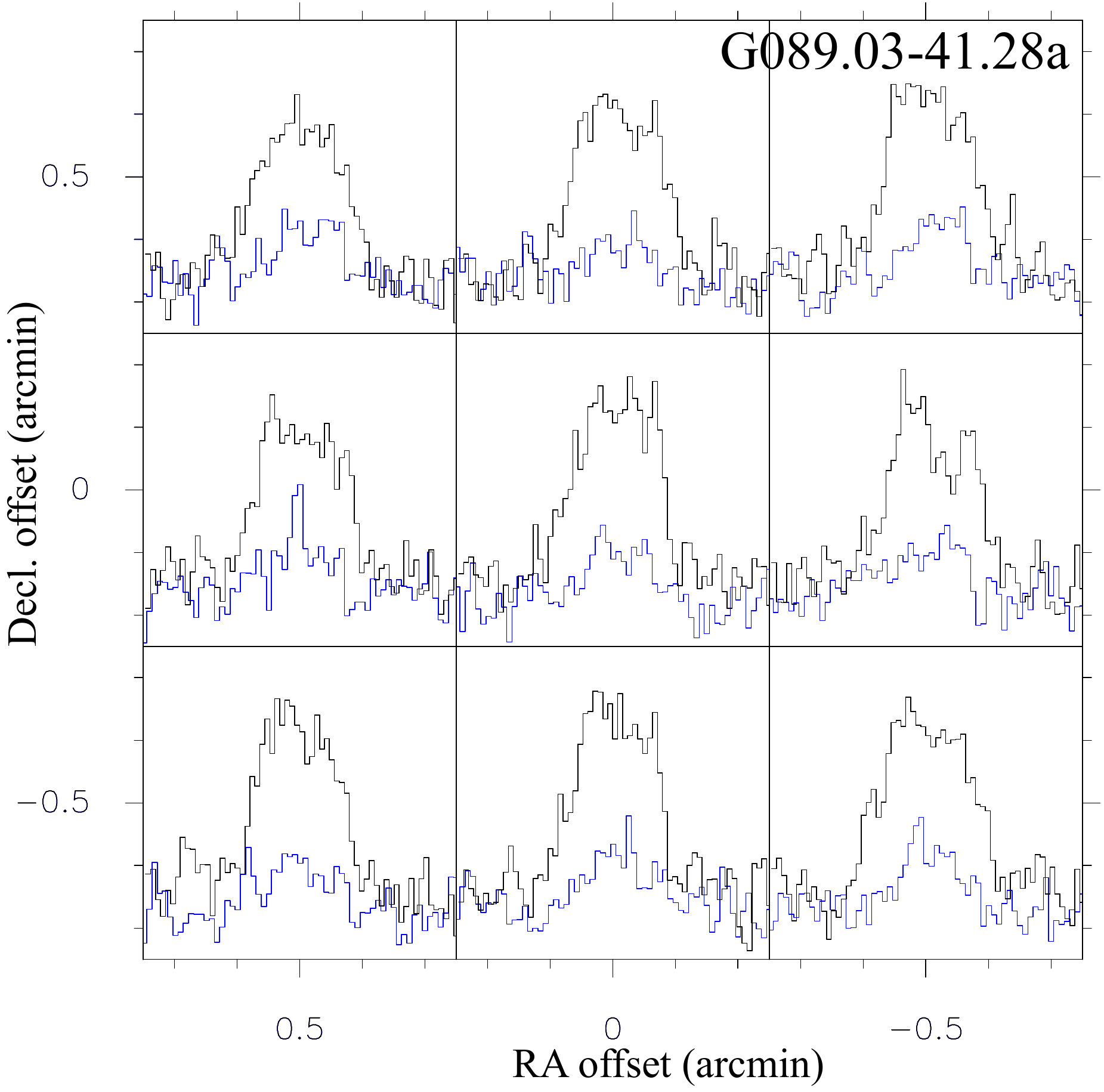}
	 \caption{3$\times$3 map grids of G089.03-41.28a are shown. The x and y coordinates are the offsets of two equatorial direction in unit of arcmin. Two CO molecular lines $^{12}$CO and $^{13}$CO are marked with black and blue, respectively.}
	\label{fig:G089.03-41.28}
\end{figure*}

G089.03-41.28 is the only blue profile detected in HGal region by single-point observations. We then check the central 3$\times$3 mapping grids in Figure \ref{fig:G089.03-41.28}. It is obvious that the source is dynamically complicated. Although single-point observation may display a potential blue profile, the line profiles from 3$\times$3 map grids (Figure \ref{fig:G089.03-41.28}) \textbf{show local differences at the size of 0.05-0.1 pc (0.5 arcmin per pixel at the distance of 0.183 kpc).}
Such a complex velocity structure has been studied in the molecular cloud MBM 55 containing the clump G089.03-41.28. The intrinsic velocity dispersion of MBM 55 is reported to be 2.09 km s$^{-1}$ \citep{1985ApJ...295..402M}, corresponding to the separation of our double peak $^{13}$CO emission lines.

\begin{figure*}
	\centering
	\includegraphics[width=0.8\linewidth,angle=0]{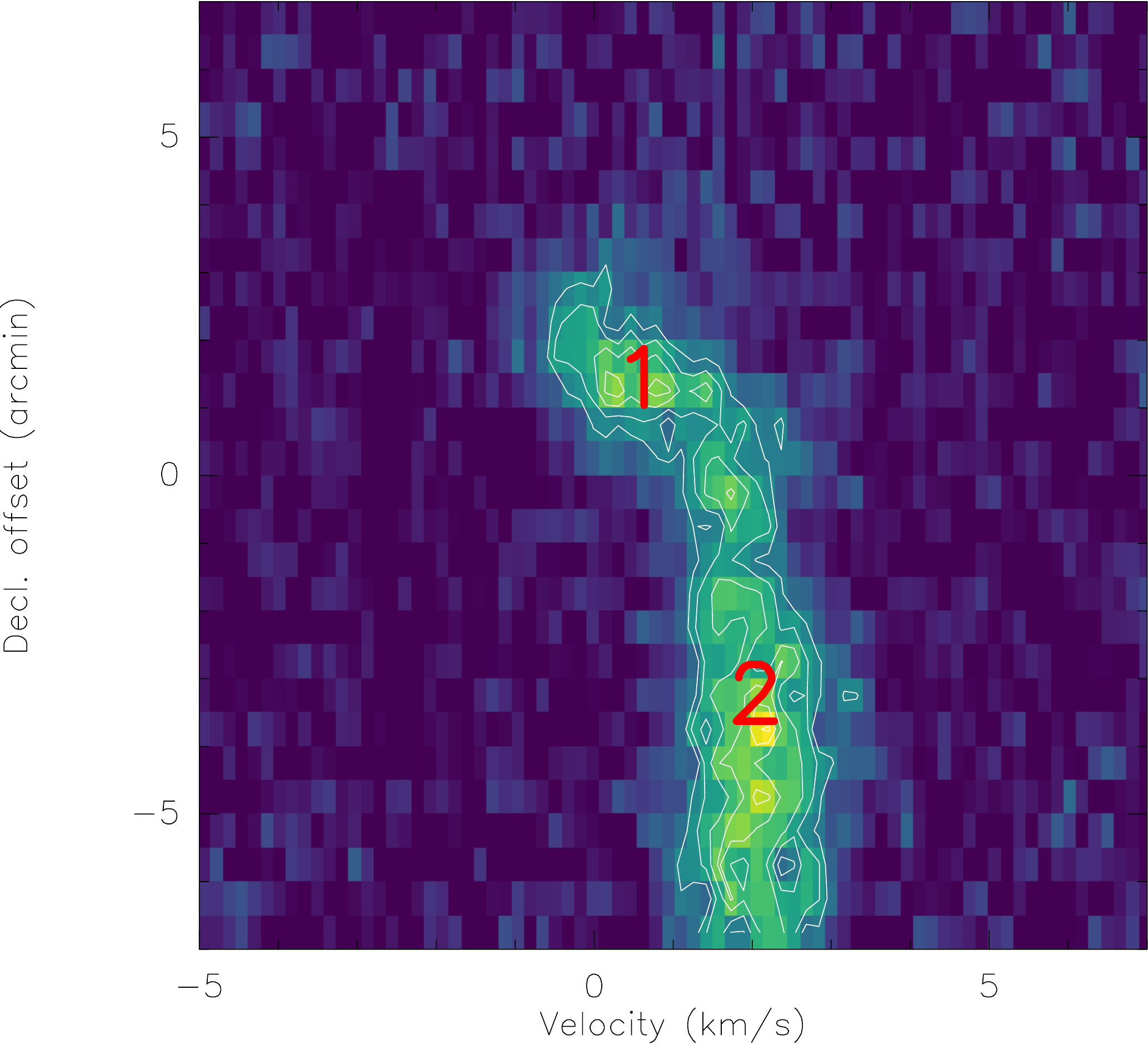}
	\caption{The position-velocity map along RA offset of -0.3 arcmin in the clump G182.54-25.34a. Red numbers '1' and '2' mark the velocity components we defined.}
	\label{fig:G182.54-25.34A}
\end{figure*}

G182.54-25.34 exhibits the only red profile observed with both single-point and mapping observations in HGal region.
To better understand the dynamical information, a position-velocity (PV) map is applied to the central area of G182.54-25.34. We extract the molecular lines along the strip with the same RA offset of -0.343 arcmin in Figure \ref{fig:G182.54-25.34A}. Two velocity components are shown: one is located in the center of the map (marked red “1”) and the other with uniform velocity dispersion extends to the south (marked red “2”). Therefore, the central line profile of G182.54-25.34a.C1 is considered to result from two velocity components with 1.3 \kms shift rather than from an expanding motion.

\subsection{Non-thermal Motion} \label{discuss:nonthermal}
\begin{figure*}
	\centering
	\includegraphics[width=0.8\linewidth,angle=0]{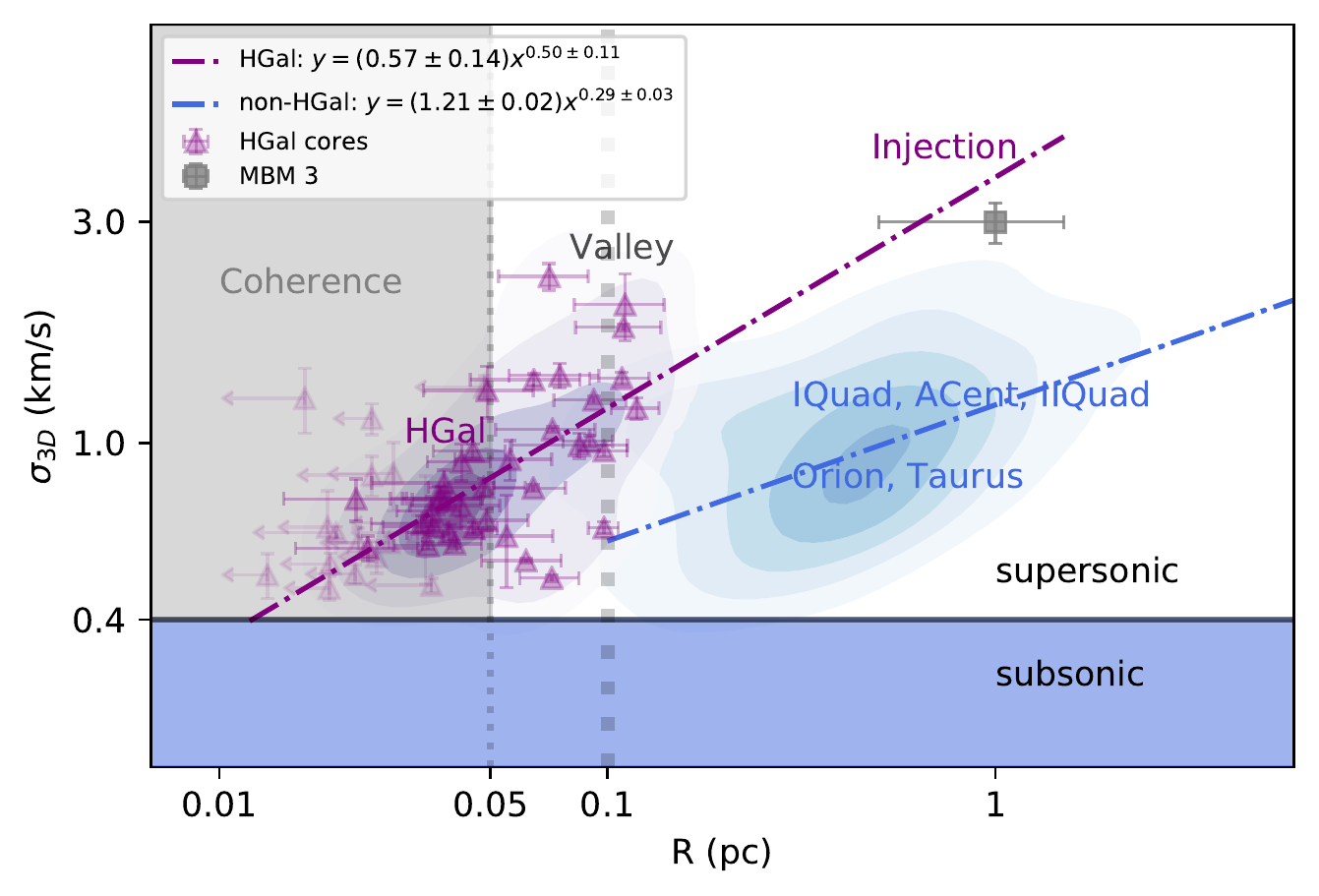}
	\caption{The relationship between of core size and internal velocity dispersion is shown. HGal cores (purple points with error bars, also filled contours) gives a power-law index 0.50. Cores with only upper limits of sizes are shown with lighter purple errorbars. The linear fitting is shown with purple dash-dotted line. The scattering is large with $R^2$=0.36. Grey dash-dotted line from IQuad \citep{2020ApJS..247...29Z} gives a better Larson's relation with index of 0.33 at large scale (0.1-3 pc). Blue contours represent other PGCCs cores observed in previous works \citep{2012ApJS..202....4L,Meng_2013,Zhang_2016,2020ApJS..247...29Z}. There is a valley (0.1 pc) for bimodal distribution of total observed PGCCs sample, shown with black vertical dotted line. Different turbulence regime is separated by grey solid horizontal line and sub-sonic turbulence regime is filled with royal blue color, assuming the temperature is 10 K and sound speed is 0.17 km s$^{-1}$.}
	\label{fig:Larson}
\end{figure*}

PGCCs are relatively cold and quiescent on the whole compared to other typical star forming regions \citep{2012ApJ...756...76W} and the non-thermal motions in these clumps are therefore mainly contributed by turbulence \citep{2020ApJS..247...29Z}. In most of the PGCC cores we observed, non-thermal velocity dispersion is larger than thermal one ($\sigma_{NT}\ge\sigma_{Th}$). In other words, the turbulence is super-sonic (above blue-shaded region in Figure \ref{fig:Larson}) in most of the samples \citep[Table.2]{2012ApJ...756...76W}. These cores might be still at an early stage along the time sequence of condensations: confined by outer \textsc{Hi} gas pressure \citep{1996A&A...314..251H} or even just in a transient phase \citep{1993ApJ...402..226M}.

First suggested by \citet{1981MNRAS.194..809L}, a cascade in an incompressible fluid should be responsible for observed $\sigma\sim R^{0.38}$ relation. Further refinements have changed this simple picture and the power law index differs depending on environment \citep{2007IAUS..237...17S}. The grey dash-dotted line (in Figure \ref{fig:Larson}) from the First Quadrant (IQuad) gives a correlation similar to Larson's relation \citep{2020ApJS..247...29Z}.
For HGal cores (purple points with error bars, also purple filled contours), the relation is also checked as:
\begin{eqnarray}\label{eq:larson}
    \sigma_{3D} = (0.57\pm0.14)\times R^{0.50\pm0.11}
\end{eqnarray}
with a poor correlation coefficient of $R^2=0.36$. We exclude the cores with only upper limits of sizes in this linear regression and mark them as light purple with semi-errorbars. Blue contours represent other PGCC cores (hereafter non-HGal) also observed with 13.7-m PMO telescope \citep{2012ApJS..202....4L,Meng_2013,Zhang_2016,2020ApJS..247...29Z}. These non-HGal cores also obey the Larson's relation derived from grey dash-dotted line on the whole.

HGal clumps are closer in distance and enable us to observe at smaller scale, and on the other hand we tend to observe a farther clump at larger scale. Such observational differences in size enable us to study the turbulent property at the different scales. The conceptual foundation of turbulence is a monotonic cascade from large injection scales to small scales where it is dissipated \citep{1972fct..book.....T,1992PhRvA..46.4797M}. As shown in Figure \ref{fig:Larson}, along the blue dash-dotted line from upper right to lower left, $\sigma_{3D}$ decreases with core size R. However, a large-scale (0.1-3 pc) Larson's relation cannot be extended to the small-scale (0.01-0.1 pc) continuously, which is dramatically shown by the bimodal distribution with purple and blue filled contours. HGal cores (purple), with the smallest size, have significantly higher $\sigma$ than a continuous cascade from the large scale expected, even if the error bars are taken into account.

Since energy can be injected on many scales ranging from Galactic dynamics to individual stellar processes, the turbulent mechanism in ISM can vary due to \textbf{local} differences. Many of the HGal clouds are reported to be part of large \textsc{Hi} shells or fragments of shells in the local ISM \citep{1994ApJ...434..162G,2000A&A...362..715B} where the shock compression at the cloud boundary plays the dominant role. However, HGal clouds like MBM 3, MBM 16 and MBM 40 are thought to be immersed in strong shear flows from the \textsc{Hi} medium, powering the turbulence inside the clouds \citep{2003ApJ...593..413S,2006A&A...457..197S}. We notice that whichever mechanism is operating, HGal cores with extra energy injections will restart a new cascade “path” along the purple dash-dotted line (see Figure \ref{fig:Larson}). We then plot the grey point as MBM 3, showing the typical size $\sim$ 1 pc where turbulent energy is injected by shear flow \citep{2006A&A...457..197S}.

However, the scatter in the data is large and the light purple points can't be simply explained by a monotonic turbulent cascade along the new “path”. Instead, once the self-similar turbulence scales down to a typical size $R_0$, velocity dispersion $\sigma$ doesn't obey the decreasing trend but seems to be \textbf{independent of} size.
Such a turning-point scale $R_{\text{coh}}$ corresponds to the transition to “coherence” \citep{1998ApJ...504..223G}, which can be the case of HGal clouds MBM 16 \citep{1993ApJ...402..226M}. One of the accepted efxplanations for “coherence” is Alfv\'en wave cutoff \citep{1998ApJ...504..223G}. If the Alfv\'en waves support the turbulent cascade, then $R_{\text{coh}}$ may be very close to the scale $R_{\text{cut}}$ below which no Alfv\'en wave can propagate in the medium. $R_{\text{cut}}$ can be simply calculated from \citep{1995ApJ...440..686M}:

\begin{eqnarray}
R_{\text{cut}} = \frac{\lambda_{\text{cut}}}{2} = \frac{\pi B}{4\pi^{\frac{1}{2}}\rho^{\frac{3}{2}}x_i\gamma}
\end{eqnarray}
where $B$, $\rho$, $x_i$ and $\gamma$ are magnetic field, density, ionization fraction and drag coefficient, respectively. If we take the typical value of number density $n=5\times10^3$ cm$^{-3}$, $B=20\mu G$, $x_i=10^{-7}$ and $\gamma = 3.5\times10^{13}$ cm$^{3}$g$^{-1}$s$^{-1}$, then $R_{\text{cut}}\simeq 0.05$ pc. As shown with grey-shaded area in Figure \ref{fig:Larson}, the turbulent dissipation is suppressed below 0.05 pc.



\subsection{Core Status} \label{discuss:core-status}
\subsubsection{Starless Cores}\label{starless}
Matching these HGal cores with stellar objects is very useful for understanding the environment and evolutionary status of the cloud cores. Based on the Wide field Infrared Survey Explorer (WISE) mission, a catalog of 133,980 identified Class I/II YSO candidates (CIs) as well as Class III YSO candidates (CIIIs) have been presented by \cite{2016MNRAS.458.3479M}. The CIs and CIIIs sources are believed to be associated with cores inside the clumps if they are located in the mapped area of 14'$\times$14' as well as offset to the core center less than a beam size ($\sim 1'$, corresponding to 0.05 pc with the typical distance of 200 pc). As a result, none of HGal cores have associated sources mentioned above and they are all starless cores. \textbf{The non-detection} of YSO candidates indicates that the HGal gas traced by PGCCs remains to be at early stage with barely active star formation. We also notice that the advantage of studying associated objects of HGal PGCCs is to avoid most of the contamination at line of sight for those distant sources in the direction of the Galactic plane. Therefore, no associated YSO candidates detected in HGal cores reflects the starless status.

However, we note that this result does not mean HGal regions bear no active star formation activities. Considering PGCCs are cold and early \citep{2011A&A...536A..22P,2012ApJ...756...76W}, we \textbf{unfortunately} miss the gas or dust with higher temperature which traces but also is heated by young stellar objects. Actually, although rare, HGal regions still have a handful of ongoing star formation. A typical example is the famous HGal cloud L1642 showing definitely active star-formation \citep{2014A&A...563A.125M}, The HGal cloud consists of dense regions A1, A2, B, and C \citep{2004A&A...423..975L}, in which L1642-B ($\alpha_{2000} = 4^\mathrm{h}35^\mathrm{m}13^\mathrm{s}.1$, $\delta_{2000}=-14\degr15\arcmin43\arcsec$) contains two well-known binary T Tauri stars \citep{1987A&A...181..283S,2014A&A...563A.125M}, while L1642-A1 ($\alpha_{2000} = 4^\mathrm{h}34^\mathrm{m}00^\mathrm{s}.3$, $\delta_{2000}=-14\degr10\arcmin12\arcsec$) corresponds to the PGCC G210.67-36.77 that we observed. These two dense regions are $18\fmg5$ away from each other and therefore would not be identified to be associated by the YSO matching. The other example is the MBM12 cloud (also known as L1453, L1454, L1457 and L1458), which is identified to contain the young association MBM 12A \citep{2001ApJ...560..287L}. Towards MBM12, PGCCs have 6 detections of G158.77-33.30, G158.88-34.18, G158.97-33.01, G159.23-34.49, G159.41-34.36 and G159.67-34.31, and we all adopted mapping observations. Nevertheless, dense cores inside these clumps are all starless.

\subsubsection{Gravitational Stability}\label{stability}
Based on the criterion of virial parameters derived in Section \ref{result:derpara}, we can determine the gravitational stability in each core.
More specifically, if the virial mass $M_{vir}$ (eq.\ref{eq:Mvir}) is smaller than the $M_{\mathrm{LTE}}$ (eq.\ref{eq:Mlte}), the system has enough gravitational potential to restrain the material inside and would probably collapse in the future. In turn, if $M_{vir}$ is much larger than $M_{\mathrm{LTE}}$, the core is probably confined by external pressure or is in a transient phase. Whichever the status is, the core would not currently go into gravitational collapse.

\begin{figure*}
    \centering
    \includegraphics[width=0.8\linewidth,angle=0]{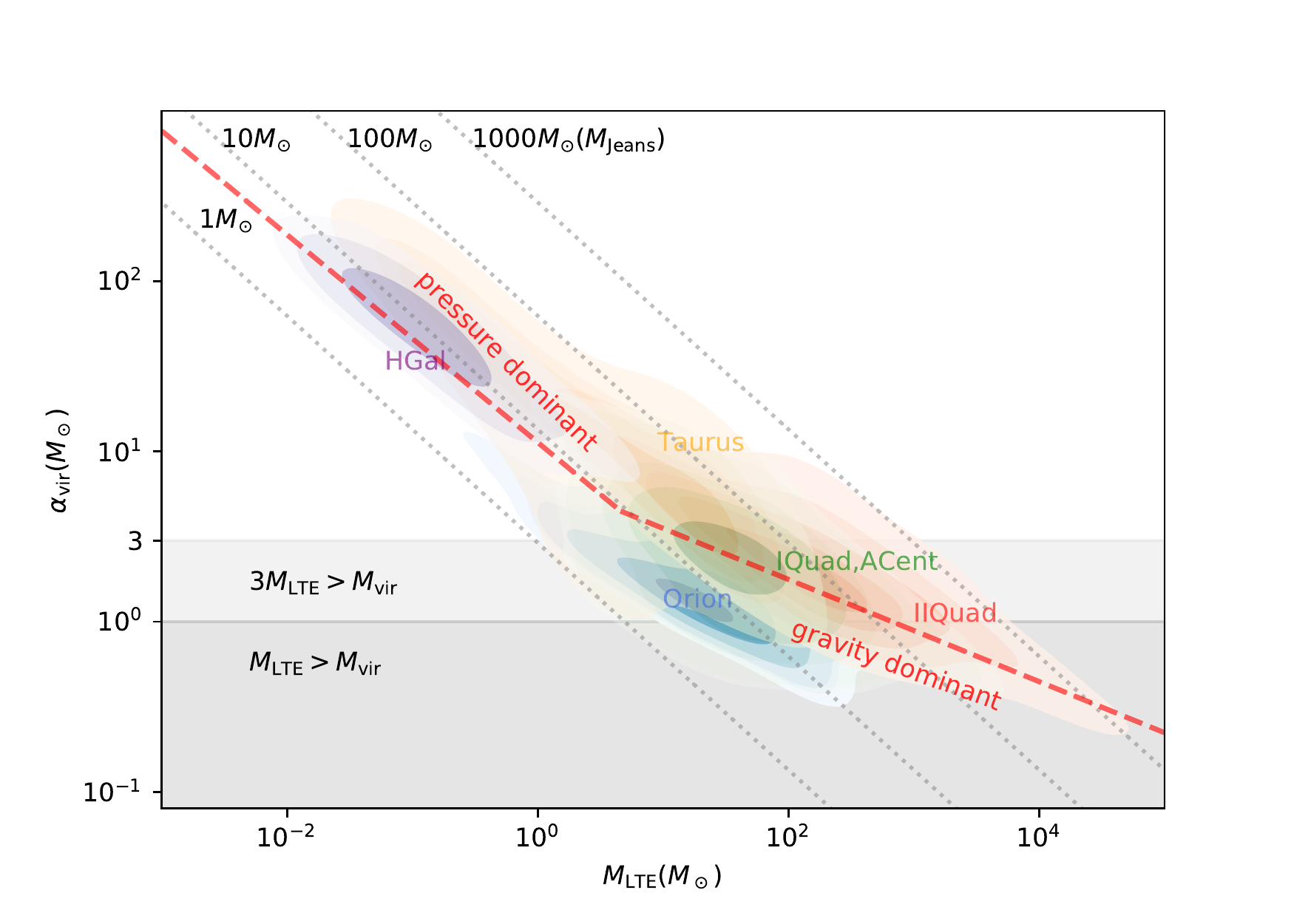}
    \caption{IQuad, ACent (green), IIQuad (red), Orion (blue), Taurus (orange) and HGal (purple) cores are shown in $M_{\mathrm{LTE}}-\alpha_{\mathrm{vir}}$ plane. Grey and light grey-shaded region are the gravitational bound and marginally gravitational bound, respectively. Pressure-confined cores obey paralleled dashed grey lines with different Jeans mass. The dashed red line is the fitted broken power-law with two slopes -0.61 and -0.3 as well as the 'ankle' (4.6$M_{\odot}$, 4.5).}
    \label{fig:alpha_vir}
\end{figure*}

The $M_{\text{LTE}}$-$\alpha_{\mathrm{vir}}$ relation is checked in all observed cores as shown in Figure \ref{fig:alpha_vir}. Grey and light grey-shaded area gives $M_{\mathrm{LTE}}>M_{\mathrm{vir}}$ and $3M_{\mathrm{LTE}}>M_{\mathrm{vir}}$ \citep{2012ApJS..202....4L} regime as criteria for gravitational bound and marginally gravitational bound, respectively. The results of four previous studies (Orion, Taurus, IIQuad and IQuad\&ACent) as well as this work are shown with blue, orange, red, green and purple filled contours. Most cores in Orion, IQuad, ACent and part of cores in IIQuad satisfy $3M_{\mathrm{LTE}}>M_{\mathrm{vir}}$, let alone some satisfy $M_{\mathrm{LTE}}>M_{\mathrm{vir}}$. However, none of HGal cores are gravitational bound and 11\% of the cores are close to a highly unbound phase with $\alpha\ge 100$. Only 21\% of the Taurus cores are marginally gravitational bound and 13\% of the cores are in a highly unbound phase as HGal cores are. The similar position in $M_{\mathrm{LTE}}$-$\alpha_{\mathrm{vir}}$ plane is assumed to be related to spatial proximity: almost half of HGal cores are the spatial extension of Taurus region (see in Section \ref{discuss:distribution}).

\begin{figure*}
\centering
\subfigure[]{
    \begin{minipage}[b]{0.46\textwidth}
    \includegraphics[width=1\textwidth]{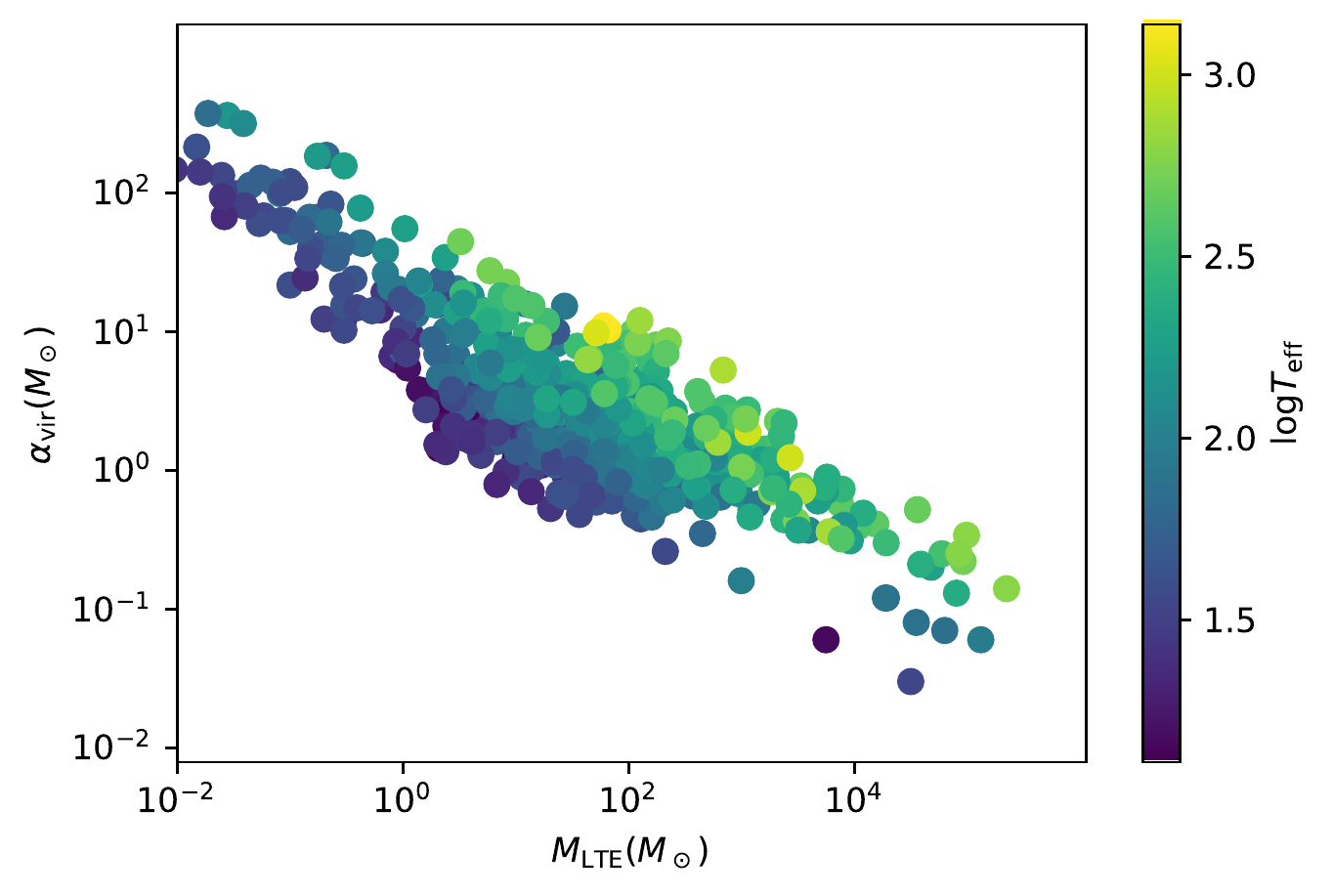}
    \end{minipage}
}%
\subfigure[]{
    \begin{minipage}[b]{0.46\textwidth}
    \includegraphics[width=1\textwidth]{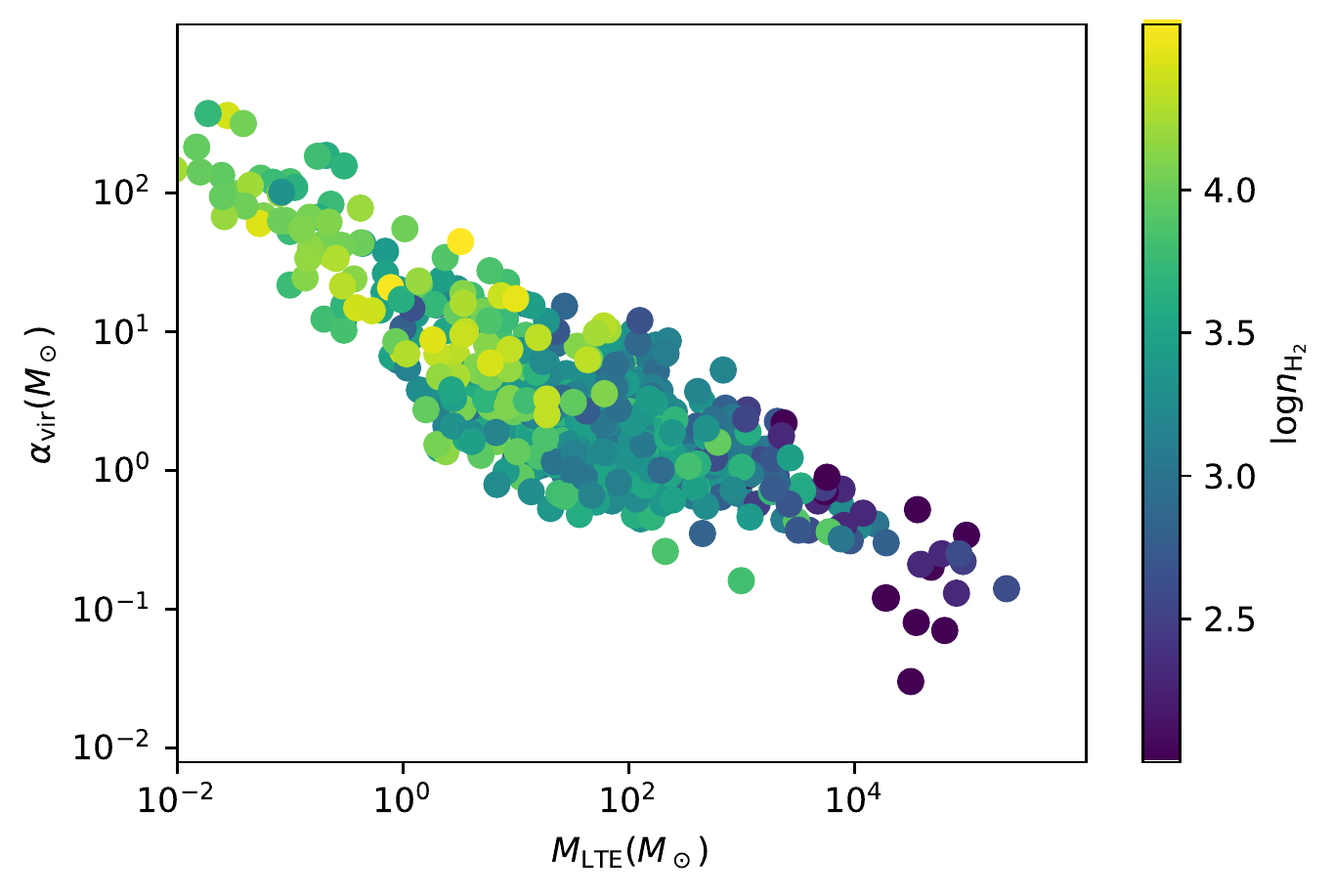}
    \end{minipage}
}%
\caption{(a) $M_{\mathrm{LTE}}-\alpha$ relation with color coding for the effective temperature $T_{\mathrm{eff}}$ ($=\frac{\mu m_H \left(\sigma_{\mathrm{NT}}^2+\sigma_{\mathrm{Th}}^2\right)}{k_B}$) in log scale. (b) $M_{\mathrm{LTE}}-\alpha$ relation with color coding for the volume density $n_{\mathrm{H}_2}$ in log scale.}
\label{fig:alpha_vir_2}
\end{figure*}

A broken power-law is fitted to the total sample (including this work and Taurus, Orion, IQuad, Acent and IIQuad) with the ankle point (4.6$M_{\odot}$, 4.5) and two power-law indices of $k_1 = -0.61$ and $k_2 = -0.30$ (see red dotted line in Figure \ref{fig:alpha_vir}). In the lower LTE mass regime, HGal as well as part of Taurus cores dominate and produce a steep power-law index which is consistent with -2/3 for pressure-confined cores \citep{1992ApJ...395..140B}.
But in the high mass regime, IQuad, ACent, Orion and IIQuad cores dominate and produce a much flatter slope of -0.3. Those grey dotted lines shown with different $M_{\mathrm{Jeans}}$ of 1, 10, 100, 1000 $M_{\odot}$ give the $M_{LTE}$-$\alpha_{\mathrm{vir}}$ relation under the assumption of pressure-confined cores. For these cores (along grey dotted lines), self-gravity is unimportant and there is barely a gradient of density or temperature (isothermal). Therefore the Jeans mass does not change with LTE mass. As indicated by the left end of red dotted line, HGal and a part of the Taurus cores are dominated by external pressure with Jeans mass of 1-100 $M_{\odot}$. As $M_{\mathrm{LTE}}$ increases, self-gravity cannot be ignored. Jeans mass gets larger as velocity dispersion (effective temperature $T_{\mathrm{eff}}$; see (a) in Figure \ref{fig:alpha_vir_2}) increases and volume density decreases (see (b) in Figure \ref{fig:alpha_vir_2}), when the LTE mass becomes larger (from the ankle to the right end of the red dotted line).

\subsection{Unusual Characteristics of HGal Gas}
The idea that HGal gas is different from that in the Galactic disk has been reported already \citep{1984ApJ...282L...9B,1985ApJ...295..402M} and our follow-up work gives \textbf{further evidence} and makes the idea much clearer. Section \ref{emission_line:detection_rate} shows a significant gradient of the dense-molecule detection rate along latitude. According to the result of 54 HGal cores, the abundance of CO isotopes bears a large deviation from that of Galactic disk (see Section \ref{emission_line:detection_rate}). And no blue or red profiles are found (see Section \ref{emission_line:line_profile}). Besides, there are no associated YSO candidates found in the host clumps (see Section \ref{starless}). As shown in Section \ref{stability}, none of the cores are gravitational bound even with a looser criterion. These results demonstrate that at that high latitude, gas is totally different from that of Galactic disk, although they are \textbf{part of the local} interstellar medium within 200 pc or so. Above all, HGal gas traced by PGCCs is at an extremely early stage and transient phase still confined with pressure, \textbf{without clear indications of star formation at the current time}.

\section{Summary} \label{sec:summary}
We have performed a single-point survey in the J=1-0 transition of $^{12}$CO, $^{13}$CO and C$^{18}$O with the PMO 13.7-m telescope toward 41 Planck Galactic cold clumps (PGCCs) at high Galactic latitude ($|b|>25\degr$; HGal). In total 51 velocity components are identified and 33 of them have CO mapping observations. $^{12}$CO and $^{13}$CO lines were detected in all of them, while C$^{18}$O has been detected in only 16 of HGal clumps. We identified 54 dense cores from the $^{13}$CO velocity integrated intensity maps. The parameters of detected lines and identified cores were derived. Further single-point observations towards 18 HGal cores and 18 cores with lower latitude $25\degr>|b|>21\degr$ were adopted. We also include other previously observed PGCCs from Orion, Taurus, the Second Quadrant (IIQuad), the First Quadrant (IQuad) and Anti-center (ACent) regions \citep{2012ApJS..202....4L,Meng_2013,Zhang_2016,2020ApJS..247...29Z}. We call these previous observed PGCCs as `non-HGal PGCCs' and some related detailed statistic analyses are done. The main findings are summarized as follows.

\begin{enumerate}
\item The highest latitude clump we detected by single point observation is G126.65-71.43 with $b=-71.43\degr$, and the highest one we mapped is G089.03-41.28 with $b=-41.28\degr$. 14.6\% of the observed clumps are not included in previous CO surveys. All the clumps in our work are now mapped with the high angular resolution of 1\arcmin. The one with the highest elevation is located 214 pc above the Galactic plane. The sample from PGCCs is unbiased and facilitates an overall knowledge of gas in our Galaxy.

\item The ratio $X_{13}$/$X_{18}(>\sim 14)$ in HGal cores is systematically higher than that found in the solar neighborhood or the Galactic disk, which is consistent with our low C$^{18}$O detection rate (40\%) by single-point observed results. Such a high ratio suggests that the HGal ISM is unique and might be explained by a different star forming history compared to that in the Galactic disk. Extremely low detection rates of two dense core tracers illustrates the decreasing gradient of N$_2$H$^+$ and C$_2$H along latitude. Such a low detection rate shows that HGal gas is relatively diffuse and has not yet form dense regions.

\item The $^{12}$CO excitation temperature ($T_{\rm ex}$) of HGal cores ranges from 6 to 13 K respectively. The excitation temperatures are similar to the dust temperatures (T$_{\rm dust}$) \citep{2016A&A...594A..28P}, which might suggest that gas and dust are well coupled in these cores. We improve the calculation of clump distances by adopting Gaia dust extinction maps. The typical distance is about 200 pc.

\item No blue or red profiles are detected in HGal cores. There seem to be no signs of star forming activities. Besides, no cores have associated YSO candidates. The environment of these cores is proved to be quiescent and inactive.

\item HGal cores are distinctly separated from other observed PGCCs (non-HGal) cores on the $\sigma-R$ plane. \textbf{This could indicate another turbulence energy injection mechanism}, like shear flows at the scale of clouds $\sim1$ pc. Besides, the monotonic turbulent cascade seems to be suppressed below 0.05 pc where the transition to coherence takes place.

\item All the HGal cores are not gravitationally bound but instead appear to be pressure-confined, while for massive non-HGal cores, the gravity dominates. These two different behaviours are shown by a broken power law in the $M_{\mathrm{LTE}}-\alpha_{\mathrm{vir}}$ plane with an ankle (M$_{\mathrm{LTE}}$,$\alpha_{\mathrm{vir}}$) = (4.6$M_{\odot}$,4.5)
\end{enumerate}

\begin{acknowledgements}

This project was supported by NSFC No.11433008 and the China Ministry of Science and Technology under State Key Development Program for Basic Research (No. 2012CB821800), and the National Key R\&D Program of China No.2017YFA0402600, and the NSFC No.12033005, 11988101, 11373009, 11503035 and 11573036. Natalia Inostroza acknowledges CONICYT/PCI/REDI170243. We thank Lianghao Lin's helpful reminder of GAIA \textsc{DRii} and related reference, which contributes a lot to our distance calculator. We thank the staff at PMO 13.7-m for their support during the observations. This research was performed in part at the Jet Propulsion Laboratory, California Institute of Technology, under a contract with the National Aeronautics and Space Administration (80NM0018D0004). This research made use of Montage which is funded by the National Science Foundation under Grant Number ACI-1440620.
\end{acknowledgements}

\software{GILDAS/CLASS \url{https://www.iram.fr/IRAMFR/GILDAS/}, Astropy \citep{price2018astropy}, Matplotlib \citep{Hunter2007}, Aplpy \citep{2012ascl.soft08017R}}

\clearpage
\bibliographystyle{aasjournal}

      \bibliography{Paper_fengwei}

\begin{thebibliography}{}
\expandafter\ifx\csname natexlab\endcsname\relax\def\natexlab#1{#1}\fi
\providecommand{\url}[1]{\href{#1}{#1}}
\providecommand{\dodoi}[1]{doi:~\href{http://doi.org/#1}{\nolinkurl{#1}}}
\providecommand{\doeprint}[1]{\href{http://ascl.net/#1}{\nolinkurl{http://ascl.net/#1}}}
\providecommand{\doarXiv}[1]{\href{https://arxiv.org/abs/#1}{\nolinkurl{https://arxiv.org/abs/#1}}}

\bibitem[{{Bertoldi} \& {McKee}(1992)}]{1992ApJ...395..140B}
{Bertoldi}, F., \& {McKee}, C.~F. 1992, \apj, 395, 140, \dodoi{10.1086/171638}

\bibitem[{{Bhatt}(2000)}]{2000A&A...362..715B}
{Bhatt}, H.~C. 2000, \aap, 362, 715.
\newblock \doarXiv{astro-ph/0009486}

\bibitem[{{Blitz} {et~al.}(1990){Blitz}, {Bazell}, \&
  {Desert}}]{1990ApJ...352L..13B}
{Blitz}, L., {Bazell}, D., \& {Desert}, F.~X. 1990, \apjl, 352, L13,
  \dodoi{10.1086/185682}

\bibitem[{{Blitz} {et~al.}(1982){Blitz}, {Fich}, \&
  {Stark}}]{1982ApJS...49..183B}
{Blitz}, L., {Fich}, M., \& {Stark}, A.~A. 1982, \apjs, 49, 183,
  \dodoi{10.1086/190795}

\bibitem[{Blitz \& Lockman(1988)}]{1988SpringerM}
Blitz, L., \& Lockman, F.~J., eds. 1988, The Outer Galaxy (Springer Berlin
  Heidelberg), \dodoi{10.1007/3-540-19484-3}

\bibitem[{{Blitz} {et~al.}(1984){Blitz}, {Magnani}, \&
  {Mundy}}]{1984ApJ...282L...9B}
{Blitz}, L., {Magnani}, L., \& {Mundy}, L. 1984, \apjl, 282, L9,
  \dodoi{10.1086/184293}

\bibitem[{{Burton} \& {Gordon}(1978)}]{1978A&A....63....7B}
{Burton}, W.~B., \& {Gordon}, M.~A. 1978, \aap, 63, 7

\bibitem[{{Clemens} \& {Barvainis}(1988)}]{1988ApJS...68..257C}
{Clemens}, D.~P., \& {Barvainis}, R. 1988, \apjs, 68, 257,
  \dodoi{10.1086/191288}

\bibitem[{{Dame} {et~al.}(1987){Dame}, {Ungerechts}, {Cohen}, {de Geus},
  {Grenier}, {May}, {Murphy}, {Nyman}, \& {Thaddeus}}]{1987ApJ...322..706D}
{Dame}, T.~M., {Ungerechts}, H., {Cohen}, R.~S., {et~al.} 1987, \apj, 322, 706,
  \dodoi{10.1086/165766}

\bibitem[{{de Vries} {et~al.}(1987){de Vries}, {Heithausen}, \&
  {Thaddeus}}]{1987ApJ...319..723D}
{de Vries}, H.~W., {Heithausen}, A., \& {Thaddeus}, P. 1987, \apj, 319, 723,
  \dodoi{10.1086/165492}

\bibitem[{{Dickman}(1975)}]{1975ApJ...202...50D}
{Dickman}, R.~L. 1975, \apj, 202, 50, \dodoi{10.1086/153951}

\bibitem[{{Frerking} {et~al.}(1982){Frerking}, {Langer}, \&
  {Wilson}}]{1982ApJ...262..590F}
{Frerking}, M.~A., {Langer}, W.~D., \& {Wilson}, R.~W. 1982, \apj, 262, 590,
  \dodoi{10.1086/160451}

\bibitem[{{Gaia Collaboration} {et~al.}(2016){Gaia Collaboration}, {Prusti},
  {de Bruijne}, {Brown}, {Vallenari}, {Babusiaux}, {Bailer-Jones}, {Bastian},
  {Biermann}, {Evans}, {Eyer}, {Jansen}, {Jordi}, {Klioner}, {Lammers},
  {Lindegren}, {Luri}, {Mignard}, {Milligan}, {Panem}, {Poinsignon},
  {Pourbaix}, {Randich}, {Sarri}, {Sartoretti}, {Siddiqui}, {Soubiran},
  {Valette}, {van Leeuwen}, {Walton}, {Aerts}, {Arenou}, {Cropper}, {Drimmel},
  {H{\o}g}, {Katz}, {Lattanzi}, {O'Mullane}, {Grebel}, {Holland}, {Huc},
  {Passot}, {Bramante}, {Cacciari}, {Casta{\~n}eda}, {Chaoul}, {Cheek}, {De
  Angeli}, {Fabricius}, {Guerra}, {Hern{\'a}ndez}, {Jean-Antoine-Piccolo},
  {Masana}, {Messineo}, {Mowlavi}, {Nienartowicz}, {Ord{\'o}{\~n}ez-Blanco},
  {Panuzzo}, {Portell}, {Richards}, {Riello}, {Seabroke}, {Tanga},
  {Th{\'e}venin}, {Torra}, {Els}, {Gracia-Abril}, {Comoretto},
  {Garcia-Reinaldos}, {Lock}, {Mercier}, {Altmann}, {Andrae}, {Astraatmadja},
  {Bellas-Velidis}, {Benson}, {Berthier}, {Blomme}, {Busso}, {Carry},
  {Cellino}, {Clementini}, {Cowell}, {Creevey}, {Cuypers}, {Davidson}, {De
  Ridder}, {de Torres}, {Delchambre}, {Dell'Oro}, {Ducourant}, {Fr{\'e}mat},
  {Garc{\'\i}a-Torres}, {Gosset}, {Halbwachs}, {Hambly}, {Harrison}, {Hauser},
  {Hestroffer}, {Hodgkin}, {Huckle}, {Hutton}, {Jasniewicz}, {Jordan},
  {Kontizas}, {Korn}, {Lanzafame}, {Manteiga}, {Moitinho}, {Muinonen},
  {Osinde}, {Pancino}, {Pauwels}, {Petit}, {Recio-Blanco}, {Robin}, {Sarro},
  {Siopis}, {Smith}, {Smith}, {Sozzetti}, {Thuillot}, {van Reeven}, {Viala},
  {Abbas}, {Abreu Aramburu}, {Accart}, {Aguado}, {Allan}, {Allasia},
  {Altavilla}, {{\'A}lvarez}, {Alves}, {Anderson}, {Andrei}, {Anglada Varela},
  {Antiche}, {Antoja}, {Ant{\'o}n}, {Arcay}, {Atzei}, {Ayache}, {Bach},
  {Baker}, {Balaguer-N{\'u}{\~n}ez}, {Barache}, {Barata}, {Barbier}, {Barblan},
  {Baroni}, {Barrado y Navascu{\'e}s}, {Barros}, {Barstow}, {Becciani},
  {Bellazzini}, {Bellei}, {Bello Garc{\'\i}a}, {Belokurov}, {Bendjoya},
  {Berihuete}, {Bianchi}, {Bienaym{\'e}}, {Billebaud}, {Blagorodnova},
  {Blanco-Cuaresma}, {Boch}, {Bombrun}, {Borrachero}, {Bouquillon}, {Bourda},
  {Bouy}, {Bragaglia}, {Breddels}, {Brouillet}, {Br{\"u}semeister},
  {Bucciarelli}, {Budnik}, {Burgess}, {Burgon}, {Burlacu}, {Busonero}, {Buzzi},
  {Caffau}, {Cambras}, {Campbell}, {Cancelliere}, {Cantat-Gaudin}, {Carlucci},
  {Carrasco}, {Castellani}, {Charlot}, {Charnas}, {Charvet}, {Chassat},
  {Chiavassa}, {Clotet}, {Cocozza}, {Collins}, {Collins}, {Costigan}, {Crifo},
  {Cross}, {Crosta}, {Crowley}, {Dafonte}, {Damerdji}, {Dapergolas}, {David},
  {David}, {De Cat}, {de Felice}, {de Laverny}, {De Luise}, {De March}, {de
  Martino}, {de Souza}, {Debosscher}, {del Pozo}, {Delbo}, {Delgado},
  {Delgado}, {di Marco}, {Di Matteo}, {Diakite}, {Distefano}, {Dolding}, {Dos
  Anjos}, {Drazinos}, {Dur{\'a}n}, {Dzigan}, {Ecale}, {Edvardsson}, {Enke},
  {Erdmann}, {Escolar}, {Espina}, {Evans}, {Eynard Bontemps}, {Fabre},
  {Fabrizio}, {Faigler}, {Falc{\~a}o}, {Farr{\`a}s Casas}, {Faye}, {Federici},
  {Fedorets}, {Fern{\'a}ndez-Hern{\'a}ndez}, {Fernique}, {Fienga}, {Figueras},
  {Filippi}, {Findeisen}, {Fonti}, {Fouesneau}, {Fraile}, {Fraser}, {Fuchs},
  {Furnell}, {Gai}, {Galleti}, {Galluccio}, {Garabato}, {Garc{\'\i}a-Sedano},
  {Gar{\'e}}, {Garofalo}, {Garralda}, {Gavras}, {Gerssen}, {Geyer}, {Gilmore},
  {Girona}, {Giuffrida}, {Gomes}, {Gonz{\'a}lez-Marcos},
  {Gonz{\'a}lez-N{\'u}{\~n}ez}, {Gonz{\'a}lez-Vidal}, {Granvik}, {Guerrier},
  {Guillout}, {Guiraud}, {G{\'u}rpide}, {Guti{\'e}rrez-S{\'a}nchez}, {Guy},
  {Haigron}, {Hatzidimitriou}, {Haywood}, {Heiter}, {Helmi}, {Hobbs},
  {Hofmann}, {Holl}, {Holland}, {Hunt}, {Hypki}, {Icardi}, {Irwin}, {Jevardat
  de Fombelle}, {Jofr{\'e}}, {Jonker}, {Jorissen}, {Julbe}, {Karampelas},
  {Kochoska}, {Kohley}, {Kolenberg}, {Kontizas}, {Koposov}, {Kordopatis},
  {Koubsky}, {Kowalczyk}, {Krone-Martins}, {Kudryashova}, {Kull}, {Bachchan},
  {Lacoste-Seris}, {Lanza}, {Lavigne}, {Le Poncin-Lafitte}, {Lebreton},
  {Lebzelter}, {Leccia}, {Leclerc}, {Lecoeur-Taibi}, {Lemaitre}, {Lenhardt},
  {Leroux}, {Liao}, {Licata}, {Lindstr{\o}m}, {Lister}, {Livanou}, {Lobel},
  {L{\"o}ffler}, {L{\'o}pez}, {Lopez-Lozano}, {Lorenz}, {Loureiro},
  {MacDonald}, {Magalh{\~a}es Fernandes}, {Managau}, {Mann}, {Mantelet},
  {Marchal}, {Marchant}, {Marconi}, {Marie}, {Marinoni}, {Marrese},
  {Marschalk{\'o}}, {Marshall}, {Mart{\'\i}n-Fleitas}, {Martino}, {Mary},
  {Matijevi{\v{c}}}, {Mazeh}, {McMillan}, {Messina}, {Mestre}, {Michalik},
  {Millar}, {Miranda}, {Molina}, {Molinaro}, {Molinaro}, {Moln{\'a}r},
  {Moniez}, {Montegriffo}, {Monteiro}, {Mor}, {Mora}, {Morbidelli}, {Morel},
  {Morgenthaler}, {Morley}, {Morris}, {Mulone}, {Muraveva}, {Musella},
  {Narbonne}, {Nelemans}, {Nicastro}, {Noval}, {Ord{\'e}novic},
  {Ordieres-Mer{\'e}}, {Osborne}, {Pagani}, {Pagano}, {Pailler}, {Palacin},
  {Palaversa}, {Parsons}, {Paulsen}, {Pecoraro}, {Pedrosa}, {Pentik{\"a}inen},
  {Pereira}, {Pichon}, {Piersimoni}, {Pineau}, {Plachy}, {Plum}, {Poujoulet},
  {Pr{\v{s}}a}, {Pulone}, {Ragaini}, {Rago}, {Rambaux}, {Ramos-Lerate},
  {Ranalli}, {Rauw}, {Read}, {Regibo}, {Renk}, {Reyl{\'e}}, {Ribeiro},
  {Rimoldini}, {Ripepi}, {Riva}, {Rixon}, {Roelens}, {Romero-G{\'o}mez},
  {Rowell}, {Royer}, {Rudolph}, {Ruiz-Dern}, {Sadowski}, {Sagrist{\`a}
  Sell{\'e}s}, {Sahlmann}, {Salgado}, {Salguero}, {Sarasso}, {Savietto},
  {Schnorhk}, {Schultheis}, {Sciacca}, {Segol}, {Segovia}, {Segransan},
  {Serpell}, {Shih}, {Smareglia}, {Smart}, {Smith}, {Solano}, {Solitro},
  {Sordo}, {Soria Nieto}, {Souchay}, {Spagna}, {Spoto}, {Stampa}, {Steele},
  {Steidelm{\"u}ller}, {Stephenson}, {Stoev}, {Suess}, {S{\"u}veges}, {Surdej},
  {Szabados}, {Szegedi-Elek}, {Tapiador}, {Taris}, {Tauran}, {Taylor},
  {Teixeira}, {Terrett}, {Tingley}, {Trager}, {Turon}, {Ulla}, {Utrilla},
  {Valentini}, {van Elteren}, {Van Hemelryck}, {van Leeuwen}, {Varadi},
  {Vecchiato}, {Veljanoski}, {Via}, {Vicente}, {Vogt}, {Voss}, {Votruba},
  {Voutsinas}, {Walmsley}, {Weiler}, {Weingrill}, {Werner}, {Wevers},
  {Whitehead}, {Wyrzykowski}, {Yoldas}, {{\v{Z}}erjal}, {Zucker}, {Zurbach},
  {Zwitter}, {Alecu}, {Allen}, {Allende Prieto}, {Amorim},
  {Anglada-Escud{\'e}}, {Arsenijevic}, {Azaz}, {Balm}, {Beck}, {Bernstein},
  {Bigot}, {Bijaoui}, {Blasco}, {Bonfigli}, {Bono}, {Boudreault}, {Bressan},
  {Brown}, {Brunet}, {Bunclark}, {Buonanno}, {Butkevich}, {Carret}, {Carrion},
  {Chemin}, {Ch{\'e}reau}, {Corcione}, {Darmigny}, {de Boer}, {de Teodoro}, {de
  Zeeuw}, {Delle Luche}, {Domingues}, {Dubath}, {Fodor}, {Fr{\'e}zouls},
  {Fries}, {Fustes}, {Fyfe}, {Gallardo}, {Gallegos}, {Gardiol}, {Gebran},
  {Gomboc}, {G{\'o}mez}, {Grux}, {Gueguen}, {Heyrovsky}, {Hoar}, {Iannicola},
  {Isasi Parache}, {Janotto}, {Joliet}, {Jonckheere}, {Keil}, {Kim},
  {Klagyivik}, {Klar}, {Knude}, {Kochukhov}, {Kolka}, {Kos}, {Kutka}, {Lainey},
  {LeBouquin}, {Liu}, {Loreggia}, {Makarov}, {Marseille}, {Martayan},
  {Martinez-Rubi}, {Massart}, {Meynadier}, {Mignot}, {Munari}, {Nguyen},
  {Nordlander}, {Ocvirk}, {O'Flaherty}, {Olias Sanz}, {Ortiz}, {Osorio},
  {Oszkiewicz}, {Ouzounis}, {Palmer}, {Park}, {Pasquato}, {Peltzer}, {Peralta},
  {P{\'e}turaud}, {Pieniluoma}, {Pigozzi}, {Poels}, {Prat}, {Prod'homme},
  {Raison}, {Rebordao}, {Risquez}, {Rocca-Volmerange}, {Rosen}, {Ruiz-Fuertes},
  {Russo}, {Sembay}, {Serraller Vizcaino}, {Short}, {Siebert}, {Silva},
  {Sinachopoulos}, {Slezak}, {Soffel}, {Sosnowska}, {Strai{\v{z}}ys}, {ter
  Linden}, {Terrell}, {Theil}, {Tiede}, {Troisi}, {Tsalmantza}, {Tur},
  {Vaccari}, {Vachier}, {Valles}, {Van Hamme}, {Veltz}, {Virtanen}, {Wallut},
  {Wichmann}, {Wilkinson}, {Ziaeepour}, \& {Zschocke}}]{2016A&A...595A...1G}
{Gaia Collaboration}, {Prusti}, T., {de Bruijne}, J.~H.~J., {et~al.} 2016,
  \aap, 595, A1, \dodoi{10.1051/0004-6361/201629272}

\bibitem[{{Garden} {et~al.}(1991){Garden}, {Hayashi}, {Hasegawa}, {Gatley}, \&
  {Kaifu}}]{1991ApJ...374..540G}
{Garden}, R.~P., {Hayashi}, M., {Hasegawa}, T., {Gatley}, I., \& {Kaifu}, N.
  1991, \apj, 374, 540, \dodoi{10.1086/170143}

\bibitem[{{Gir} {et~al.}(1994){Gir}, {Blitz}, \&
  {Magnani}}]{1994ApJ...434..162G}
{Gir}, B.-Y., {Blitz}, L., \& {Magnani}, L. 1994, \apj, 434, 162,
  \dodoi{10.1086/174713}

\bibitem[{{Goldreich} \& {Kwan}(1974)}]{1974ApJ...189..441G}
{Goldreich}, P., \& {Kwan}, J. 1974, \apj, 189, 441, \dodoi{10.1086/152821}

\bibitem[{{Goldsmith}(2001)}]{2001ApJ...557..736G}
{Goldsmith}, P.~F. 2001, \apj, 557, 736, \dodoi{10.1086/322255}

\bibitem[{{Goodman} {et~al.}(1998){Goodman}, {Barranco}, {Wilner}, \&
  {Heyer}}]{1998ApJ...504..223G}
{Goodman}, A.~A., {Barranco}, J.~A., {Wilner}, D.~J., \& {Heyer}, M.~H. 1998,
  \apj, 504, 223, \dodoi{10.1086/306045}

\bibitem[{{Green} {et~al.}(2019){Green}, {Schlafly}, {Zucker}, {Speagle}, \&
  {Finkbeiner}}]{2019ApJ...887...93G}
{Green}, G.~M., {Schlafly}, E., {Zucker}, C., {Speagle}, J.~S., \&
  {Finkbeiner}, D. 2019, \apj, 887, 93, \dodoi{10.3847/1538-4357/ab5362}

\bibitem[{{Green} {et~al.}(2014){Green}, {Schlafly}, {Finkbeiner}, {Juri{\'c}},
  {Rix}, {Burgett}, {Chambers}, {Draper}, {Flewelling}, {Kudritzki}, {Magnier},
  {Martin}, {Metcalfe}, {Tonry}, {Wainscoat}, \&
  {Waters}}]{2014ApJ...783..114G}
{Green}, G.~M., {Schlafly}, E.~F., {Finkbeiner}, D.~P., {et~al.} 2014, \apj,
  783, 114, \dodoi{10.1088/0004-637X/783/2/114}

\bibitem[{{Griffin} {et~al.}(2010){Griffin}, {Abergel}, {Abreu}, {Ade},
  {Andr{\'e}}, {Augueres}, {Babbedge}, {Bae}, {Baillie}, {Baluteau}, {Barlow},
  {Bendo}, {Benielli}, {Bock}, {Bonhomme}, {Brisbin}, {Brockley-Blatt},
  {Caldwell}, {Cara}, {Castro-Rodriguez}, {Cerulli}, {Chanial}, {Chen},
  {Clark}, {Clements}, {Clerc}, {Coker}, {Communal}, {Conversi}, {Cox},
  {Crumb}, {Cunningham}, {Daly}, {Davis}, {de Antoni}, {Delderfield}, {Devin},
  {di Giorgio}, {Didschuns}, {Dohlen}, {Donati}, {Dowell}, {Dowell}, {Duband},
  {Dumaye}, {Emery}, {Ferlet}, {Ferrand}, {Fontignie}, {Fox}, {Franceschini},
  {Frerking}, {Fulton}, {Garcia}, {Gastaud}, {Gear}, {Glenn}, {Goizel},
  {Griffin}, {Grundy}, {Guest}, {Guillemet}, {Hargrave}, {Harwit}, {Hastings},
  {Hatziminaoglou}, {Herman}, {Hinde}, {Hristov}, {Huang}, {Imhof}, {Isaak},
  {Israelsson}, {Ivison}, {Jennings}, {Kiernan}, {King}, {Lange}, {Latter},
  {Laurent}, {Laurent}, {Leeks}, {Lellouch}, {Levenson}, {Li}, {Li},
  {Lilienthal}, {Lim}, {Liu}, {Lu}, {Madden}, {Mainetti}, {Marliani}, {McKay},
  {Mercier}, {Molinari}, {Morris}, {Moseley}, {Mulder}, {Mur}, {Naylor},
  {Nguyen}, {O'Halloran}, {Oliver}, {Olofsson}, {Olofsson}, {Orfei}, {Page},
  {Pain}, {Panuzzo}, {Papageorgiou}, {Parks}, {Parr-Burman}, {Pearce},
  {Pearson}, {P{\'e}rez-Fournon}, {Pinsard}, {Pisano}, {Podosek}, {Pohlen},
  {Polehampton}, {Pouliquen}, {Rigopoulou}, {Rizzo}, {Roseboom}, {Roussel},
  {Rowan-Robinson}, {Rownd}, {Saraceno}, {Sauvage}, {Savage}, {Savini},
  {Sawyer}, {Scharmberg}, {Schmitt}, {Schneider}, {Schulz}, {Schwartz},
  {Shafer}, {Shupe}, {Sibthorpe}, {Sidher}, {Smith}, {Smith}, {Smith},
  {Spencer}, {Stobie}, {Sudiwala}, {Sukhatme}, {Surace}, {Stevens}, {Swinyard},
  {Trichas}, {Tourette}, {Triou}, {Tseng}, {Tucker}, {Turner}, {Vaccari},
  {Valtchanov}, {Vigroux}, {Virique}, {Voellmer}, {Walker}, {Ward}, {Waskett},
  {Weilert}, {Wesson}, {White}, {Whitehouse}, {Wilson}, {Winter}, {Woodcraft},
  {Wright}, {Xu}, {Zavagno}, {Zemcov}, {Zhang}, \&
  {Zonca}}]{2010A&A...518L...3G}
{Griffin}, M.~J., {Abergel}, A., {Abreu}, A., {et~al.} 2010, \aap, 518, L3,
  \dodoi{10.1051/0004-6361/201014519}

\bibitem[{{Guilloteau} \& {Lucas}(2000)}]{2000ASPC..217..299G}
{Guilloteau}, S., \& {Lucas}, R. 2000, in Astronomical Society of the Pacific
  Conference Series, Vol. 217, Imaging at Radio through Submillimeter
  Wavelengths, ed. J.~G. {Mangum} \& S.~J.~E. {Radford}, 299

\bibitem[{{Hartmann} {et~al.}(1998){Hartmann}, {Magnani}, \&
  {Thaddeus}}]{1998ApJ...492..205H}
{Hartmann}, D., {Magnani}, L., \& {Thaddeus}, P. 1998, \apj, 492, 205,
  \dodoi{10.1086/305019}

\bibitem[{{Heithausen}(1996)}]{1996A&A...314..251H}
{Heithausen}, A. 1996, \aap, 314, 251

\bibitem[{{Heithausen} {et~al.}(1993){Heithausen}, {Stacy}, {de Vries},
  {Mebold}, \& {Thaddeus}}]{1993AA...268..265H}
{Heithausen}, A., {Stacy}, J.~G., {de Vries}, H.~W., {Mebold}, U., \&
  {Thaddeus}, P. 1993, \aap, 268, 265

\bibitem[{{Heithausen} \& {Thaddeus}(1990)}]{1990ApJ...353L..49H}
{Heithausen}, A., \& {Thaddeus}, P. 1990, \apjl, 353, L49,
  \dodoi{10.1086/185705}

\bibitem[{{Hennebelle} \& {Chabrier}(2008)}]{2008ApJ...684..395H}
{Hennebelle}, P., \& {Chabrier}, G. 2008, \apj, 684, 395,
  \dodoi{10.1086/589916}

\bibitem[{{Hildebrand}(1983)}]{1983QJRAS..24..267H}
{Hildebrand}, R.~H. 1983, \qjras, 24, 267

\bibitem[{Hunter(2007)}]{Hunter2007}
Hunter, J.~D. 2007, Computing in Science {\&} Engineering, 9, 90,
  \dodoi{10.1109/mcse.2007.55}

\bibitem[{{Israel} {et~al.}(1984){Israel}, {de Graauw}, {van der Biezen}, {de
  Vries}, {Brand}, {Habing}, {Leene}, {van de Stadt}, {van Amerongen},
  {Wouterloot}, \& {Selman}}]{1984A&A...134..396I}
{Israel}, F.~P., {de Graauw}, T., {van der Biezen}, J., {et~al.} 1984, \aap,
  134, 396

\bibitem[{{Jacob} {et~al.}(2020){Jacob}, {Menten}, {Wiesemeyer}, {G{\"u}sten},
  {Wyrowski}, \& {Klein}}]{2020A&A...640A.125J}
{Jacob}, A.~M., {Menten}, K.~M., {Wiesemeyer}, H., {et~al.} 2020, \aap, 640,
  A125, \dodoi{10.1051/0004-6361/201937385}

\bibitem[{{Jim{\'e}nez-Donaire} {et~al.}(2017){Jim{\'e}nez-Donaire}, {Cormier},
  {Bigiel}, {Leroy}, {Gallagher}, {Krumholz}, {Usero}, {Hughes}, {Kramer},
  {Meier}, {Murphy}, {Pety}, {Schinnerer}, {Schruba}, {Schuster}, {Sliwa}, \&
  {Tomicic}}]{2017ApJ...836L..29J}
{Jim{\'e}nez-Donaire}, M.~J., {Cormier}, D., {Bigiel}, F., {et~al.} 2017,
  \apjl, 836, L29, \dodoi{10.3847/2041-8213/836/2/L29}

\bibitem[{{Juvela} {et~al.}(2010){Juvela}, {Ristorcelli}, {Montier},
  {Marshall}, {Pelkonen}, {Malinen}, {Ysard}, {T{\'o}th}, {Harju}, {Bernard},
  {Schneider}, {Vereb{\'e}lyi}, {Anderson}, {Andr{\'e}}, {Giard}, {Krause},
  {Lehtinen}, {Macias-Perez}, {Martin}, {McGehee}, {Meny}, {Motte}, {Pagani},
  {Paladini}, {Reach}, {Valenziano}, {Ward-Thompson}, \&
  {Zavagno}}]{2010A&A...518L..93J}
{Juvela}, M., {Ristorcelli}, I., {Montier}, L.~A., {et~al.} 2010, \aap, 518,
  L93, \dodoi{10.1051/0004-6361/201014619}

\bibitem[{{Kauffmann} {et~al.}(2008){Kauffmann}, {Bertoldi}, {Bourke}, {Evans},
  \& {Lee}}]{2008A&A...487..993K}
{Kauffmann}, J., {Bertoldi}, F., {Bourke}, T.~L., {Evans}, N.~J., I., \& {Lee},
  C.~W. 2008, \aap, 487, 993, \dodoi{10.1051/0004-6361:200809481}

\bibitem[{{Kauffmann} {et~al.}(2013){Kauffmann}, {Pillai}, \&
  {Goldsmith}}]{2013ApJ...779..185K}
{Kauffmann}, J., {Pillai}, T., \& {Goldsmith}, P.~F. 2013, \apj, 779, 185,
  \dodoi{10.1088/0004-637X/779/2/185}

\bibitem[{{Keto} \& {Myers}(1986)}]{1986ApJ...304..466K}
{Keto}, E.~R., \& {Myers}, P.~C. 1986, \apj, 304, 466, \dodoi{10.1086/164181}

\bibitem[{{Lada} {et~al.}(2009){Lada}, {Lombardi}, \&
  {Alves}}]{2009ApJ...703...52L}
{Lada}, C.~J., {Lombardi}, M., \& {Alves}, J.~F. 2009, \apj, 703, 52,
  \dodoi{10.1088/0004-637X/703/1/52}

\bibitem[{{Lang}(1978)}]{1978afcp.book.....L}
{Lang}, K.~R. 1978, {Astrophysical formulae. A compendium for the physicist and
  astrophysicist.}

\bibitem[{{Larson}(1981)}]{1981MNRAS.194..809L}
{Larson}, R.~B. 1981, \mnras, 194, 809, \dodoi{10.1093/mnras/194.4.809}

\bibitem[{{Lehtinen} {et~al.}(2004){Lehtinen}, {Russeil}, {Juvela}, {Mattila},
  \& {Lemke}}]{2004A&A...423..975L}
{Lehtinen}, K., {Russeil}, D., {Juvela}, M., {Mattila}, K., \& {Lemke}, D.
  2004, \aap, 423, 975, \dodoi{10.1051/0004-6361:20047087}

\bibitem[{{Li} \& {Goldsmith}(2003)}]{2003ApJ...585..823L}
{Li}, D., \& {Goldsmith}, P.~F. 2003, \apj, 585, 823, \dodoi{10.1086/346227}

\bibitem[{{Lilly} {et~al.}(2013){Lilly}, {Carollo}, {Pipino}, {Renzini}, \&
  {Peng}}]{2013ApJ...772..119L}
{Lilly}, S.~J., {Carollo}, C.~M., {Pipino}, A., {Renzini}, A., \& {Peng}, Y.
  2013, \apj, 772, 119, \dodoi{10.1088/0004-637X/772/2/119}

\bibitem[{{Liu} {et~al.}(2012){Liu}, {Wu}, \& {Zhang}}]{2012ApJS..202....4L}
{Liu}, T., {Wu}, Y., \& {Zhang}, H. 2012, \apjs, 202, 4,
  \dodoi{10.1088/0067-0049/202/1/4}

\bibitem[{{Liu} {et~al.}(2019){Liu}, {Wu}, {Zhang}, {Liu}, {Yuan}, {Qin}, {Ju},
  \& {Li}}]{2019A&A...622A..32L}
{Liu}, X.~C., {Wu}, Y., {Zhang}, C., {et~al.} 2019, \aap, 622, A32,
  \dodoi{10.1051/0004-6361/201834411}

\bibitem[{{Lombardi} {et~al.}(2010){Lombardi}, {Lada}, \&
  {Alves}}]{2010A&A...512A..67L}
{Lombardi}, M., {Lada}, C.~J., \& {Alves}, J. 2010, \aap, 512, A67,
  \dodoi{10.1051/0004-6361/200912670}

\bibitem[{{Low} {et~al.}(1984){Low}, {Beintema}, {Gautier}, {Gillett},
  {Beichman}, {Neugebauer}, {Young}, {Aumann}, {Boggess}, {Emerson}, {Habing},
  {Hauser}, {Houck}, {Rowan-Robinson}, {Soifer}, {Walker}, \&
  {Wesselius}}]{1984ApJ...278L..19L}
{Low}, F.~J., {Beintema}, D.~A., {Gautier}, T.~N., {et~al.} 1984, \apjl, 278,
  L19, \dodoi{10.1086/184213}

\bibitem[{{Luhman}(2001)}]{2001ApJ...560..287L}
{Luhman}, K.~L. 2001, \apj, 560, 287, \dodoi{10.1086/322386}

\bibitem[{{Lynds}(1962)}]{1962ApJS....7....1L}
{Lynds}, B.~T. 1962, \apjs, 7, 1, \dodoi{10.1086/190072}

\bibitem[{{Lynds}(1965)}]{1965ApJS...12..163L}
---. 1965, \apjs, 12, 163, \dodoi{10.1086/190123}

\bibitem[{{Magnani} {et~al.}(1985){Magnani}, {Blitz}, \&
  {Mundy}}]{1985ApJ...295..402M}
{Magnani}, L., {Blitz}, L., \& {Mundy}, L. 1985, \apj, 295, 402,
  \dodoi{10.1086/163385}

\bibitem[{{Magnani} \& {de Vries}(1986)}]{1986A&A...168..271M}
{Magnani}, L., \& {de Vries}, C.~P. 1986, \aap, 168, 271

\bibitem[{{Magnani} {et~al.}(2000){Magnani}, {Hartmann}, {Holcomb}, {Smith}, \&
  {Thaddeus}}]{2000ApJ...535..167M}
{Magnani}, L., {Hartmann}, D., {Holcomb}, S.~L., {Smith}, L.~E., \& {Thaddeus},
  P. 2000, \apj, 535, 167, \dodoi{10.1086/308841}

\bibitem[{{Magnani} {et~al.}(1996){Magnani}, {Hartmann}, \&
  {Speck}}]{1996ApJS..106..447M}
{Magnani}, L., {Hartmann}, D., \& {Speck}, B.~G. 1996, \apjs, 106, 447,
  \dodoi{10.1086/192344}

\bibitem[{{Magnani} {et~al.}(1993){Magnani}, {Larosa}, \&
  {Shore}}]{1993ApJ...402..226M}
{Magnani}, L., {Larosa}, T.~N., \& {Shore}, S.~N. 1993, \apj, 402, 226,
  \dodoi{10.1086/172125}

\bibitem[{{Malinen} {et~al.}(2014){Malinen}, {Juvela}, {Zahorecz},
  {Rivera-Ingraham}, {Montillaud}, {Arimatsu}, {Bernard}, {Doi}, {Haikala},
  {Kawabe}, {Marton}, {McGehee}, {Pelkonen}, {Ristorcelli}, {Shimajiri},
  {Takita}, {T{\'o}th}, {Tsukagoshi}, \& {Ysard}}]{2014A&A...563A.125M}
{Malinen}, J., {Juvela}, M., {Zahorecz}, S., {et~al.} 2014, \aap, 563, A125,
  \dodoi{10.1051/0004-6361/201323026}

\bibitem[{{Mardones} {et~al.}(1997){Mardones}, {Myers}, {Tafalla}, {Wilner},
  {Bachiller}, \& {Garay}}]{1997ApJ...489..719M}
{Mardones}, D., {Myers}, P.~C., {Tafalla}, M., {et~al.} 1997, \apj, 489, 719,
  \dodoi{10.1086/304812}

\bibitem[{{Marton} {et~al.}(2016){Marton}, {T{\'o}th}, {Paladini}, {Kun},
  {Zahorecz}, {McGehee}, \& {Kiss}}]{2016MNRAS.458.3479M}
{Marton}, G., {T{\'o}th}, L.~V., {Paladini}, R., {et~al.} 2016, \mnras, 458,
  3479, \dodoi{10.1093/mnras/stw398}

\bibitem[{{McComb} \& {Watt}(1992)}]{1992PhRvA..46.4797M}
{McComb}, W.~D., \& {Watt}, A.~G. 1992, \pra, 46, 4797,
  \dodoi{10.1103/PhysRevA.46.4797}

\bibitem[{{McCutcheon} {et~al.}(1980){McCutcheon}, {Shuter}, {Dickman}, \&
  {Roger}}]{1980ApJ...237....9M}
{McCutcheon}, W.~H., {Shuter}, W.~L.~H., {Dickman}, R.~L., \& {Roger}, R.~S.
  1980, \apj, 237, 9, \dodoi{10.1086/157838}

\bibitem[{{McKee} \& {Zweibel}(1995)}]{1995ApJ...440..686M}
{McKee}, C.~F., \& {Zweibel}, E.~G. 1995, \apj, 440, 686,
  \dodoi{10.1086/175306}

\bibitem[{Meng {et~al.}(2013)Meng, Wu, \& Liu}]{Meng_2013}
Meng, F., Wu, Y., \& Liu, T. 2013, The Astrophysical Journal Supplement Series,
  209, 37, \dodoi{10.1088/0067-0049/209/2/37}

\bibitem[{{Meng} {et~al.}(2013){Meng}, {Wu}, \& {Liu}}]{2013ApJS..209...37M}
{Meng}, F., {Wu}, Y., \& {Liu}, T. 2013, \apjs, 209, 37,
  \dodoi{10.1088/0067-0049/209/2/37}

\bibitem[{{Moriarty-Schieven} {et~al.}(1997){Moriarty-Schieven}, {Wannier}, \&
  {P.~G.}}]{1997ApJ...475..642M}
{Moriarty-Schieven}, G.~H., {Wannier}, \& {P.~G.} 1997, \apj, 475, 642,
  \dodoi{10.1086/303543}

\bibitem[{{Myers} {et~al.}(1983){Myers}, {Linke}, \&
  {Benson}}]{1983ApJ...264..517M}
{Myers}, P.~C., {Linke}, R.~A., \& {Benson}, P.~J. 1983, \apj, 264, 517,
  \dodoi{10.1086/160619}

\bibitem[{{Ossenkopf} \& {Henning}(1994)}]{1994A&A...291..943O}
{Ossenkopf}, V., \& {Henning}, T. 1994, \aap, 291, 943

\bibitem[{{Pety}(2005)}]{2005sf2a.conf..721P}
{Pety}, J. 2005, in SF2A-2005: Semaine de l'Astrophysique Francaise, ed.
  F.~{Casoli}, T.~{Contini}, J.~M. {Hameury}, \& L.~{Pagani}, 721

\bibitem[{{Pilbratt} {et~al.}(2010){Pilbratt}, {Riedinger}, {Passvogel},
  {Crone}, {Doyle}, {Gageur}, {Heras}, {Jewell}, {Metcalfe}, {Ott}, \&
  {Schmidt}}]{2010A&A...518L...1P}
{Pilbratt}, G.~L., {Riedinger}, J.~R., {Passvogel}, T., {et~al.} 2010, \aap,
  518, L1, \dodoi{10.1051/0004-6361/201014759}

\bibitem[{{Planck Collaboration} {et~al.}(2011{\natexlab{a}}){Planck
  Collaboration}, {Ade}, {Aghanim}, {Arnaud}, {Ashdown}, {Aumont},
  {Baccigalupi}, {Balbi}, {Banday}, {Barreiro}, {Bartlett}, {Battaner},
  {Benabed}, {Beno{\^\i}t}, {Bernard}, {Bersanelli}, {Bhatia}, {Bock},
  {Bonaldi}, {Bond}, {Borrill}, {Bouchet}, {Boulanger}, {Bucher}, {Burigana},
  {Cabella}, {Cantalupo}, {Cardoso}, {Catalano}, {Cay{\'o}n}, {Challinor},
  {Chamballu}, {Chiang}, {Christensen}, {Clements}, {Colombi}, {Couchot},
  {Coulais}, {Crill}, {Cuttaia}, {Danese}, {Davies}, {de Bernardis}, {de
  Gasperis}, {de Rosa}, {de Zotti}, {Delabrouille}, {Delouis}, {D{\'e}sert},
  {Dickinson}, {Doi}, {Donzelli}, {Dor{\'e}}, {D{\"o}rl}, {Douspis}, {Dupac},
  {Efstathiou}, {En{\ss}lin}, {Falgarone}, {Finelli}, {Forni}, {Frailis},
  {Franceschi}, {Galeotta}, {Ganga}, {Giard}, {Giardino}, {Giraud-H{\'e}raud},
  {Gonz{\'a}lez-Nuevo}, {G{\'o}rski}, {Gratton}, {Gregorio}, {Gruppuso},
  {Hansen}, {Harrison}, {Helou}, {Henrot-Versill{\'e}}, {Herranz},
  {Hildebrandt}, {Hivon}, {Hobson}, {Holmes}, {Hovest}, {Hoyland},
  {Huffenberger}, {Ikeda}, {Jaffe}, {Jones}, {Juvela}, {Keih{\"a}nen},
  {Keskitalo}, {Kisner}, {Kitamura}, {Kneissl}, {Knox}, {Kurki-Suonio},
  {Lagache}, {Lamarre}, {Lasenby}, {Laureijs}, {Lawrence}, {Leach}, {Leonardi},
  {Leroy}, {Linden-V{\o}rnle}, {L{\'o}pez-Caniego}, {Lubin},
  {Mac{\'\i}as-P{\'e}rez}, {MacTavish}, {Maffei}, {Malinen}, {Mandolesi},
  {Mann}, {Maris}, {Marshall}, {Martin}, {Mart{\'\i}nez-Gonz{\'a}lez}, {Masi},
  {Matarrese}, {Matthai}, {Mazzotta}, {McGehee}, {Melchiorri}, {Mendes},
  {Mennella}, {Meny}, {Mitra}, {Miville-Desch{\^e}nes}, {Moneti}, {Montier},
  {Morgante}, {Mortlock}, {Munshi}, {Murphy}, {Naselsky}, {Nati}, {Natoli},
  {Netterfield}, {N{\o}rgaard-Nielsen}, {Noviello}, {Novikov}, {Novikov},
  {Osborne}, {Pagani}, {Pajot}, {Paladini}, {Pasian}, {Patanchon}, {Pelkonen},
  {Perdereau}, {Perotto}, {Perrotta}, {Piacentini}, {Piat}, {Plaszczynski},
  {Pointecouteau}, {Polenta}, {Ponthieu}, {Poutanen}, {Pr{\'e}zeau}, {Prunet},
  {Puget}, {Reach}, {Rebolo}, {Reinecke}, {Renault}, {Ricciardi}, {Riller},
  {Ristorcelli}, {Rocha}, {Rosset}, {Rowan-Robinson}, {Rubi{\~n}o-Mart{\'\i}n},
  {Rusholme}, {Sandri}, {Santos}, {Savini}, {Scott}, {Seiffert}, {Smoot},
  {Starck}, {Stivoli}, {Stolyarov}, {Sudiwala}, {Sygnet}, {Tauber}, {Terenzi},
  {Toffolatti}, {Tomasi}, {Torre}, {Toth}, {Tristram}, {Tuovinen}, {Umana},
  {Valenziano}, {Vielva}, {Villa}, {Vittorio}, {Wade}, {Wandelt}, {Ysard},
  {Yvon}, {Zacchei}, \& {Zonca}}]{2011A&A...536A..22P}
{Planck Collaboration}, {Ade}, P.~A.~R., {Aghanim}, N., {et~al.}
  2011{\natexlab{a}}, \aap, 536, A22, \dodoi{10.1051/0004-6361/201116481}

\bibitem[{{Planck Collaboration} {et~al.}(2011{\natexlab{b}}){Planck
  Collaboration}, {Ade}, {Aghanim}, {Arnaud}, {Ashdown}, {Aumont},
  {Baccigalupi}, {Balbi}, {Banday}, {Barreiro}, {Bartlett}, {Battaner},
  {Benabed}, {Beno{\^\i}t}, {Bernard}, {Bersanelli}, {Bhatia}, {Bonaldi},
  {Bonavera}, {Bond}, {Borrill}, {Bouchet}, {Bucher}, {Burigana}, {Butler},
  {Cabella}, {Cantalupo}, {Cappellini}, {Cardoso}, {Carvalho}, {Catalano},
  {Cay{\'o}n}, {Challinor}, {Chamballu}, {Chary}, {Chen}, {Chiang}, {Chiang},
  {Christensen}, {Clements}, {Colombi}, {Couchot}, {Coulais}, {Crill},
  {Cuttaia}, {Danese}, {Davis}, {de Bernardis}, {de Rosa}, {de Zotti},
  {Delabrouille}, {Delouis}, {D{\'e}sert}, {Dickinson}, {Diego}, {Dolag},
  {Dole}, {Donzelli}, {Dor{\'e}}, {D{\"o}rl}, {Douspis}, {Dupac}, {Efstathiou},
  {En{\ss}lin}, {Eriksen}, {Finelli}, {Forni}, {Fosalba}, {Frailis},
  {Franceschi}, {Galeotta}, {Ganga}, {Giard}, {Giraud-H{\'e}raud},
  {Gonz{\'a}lez-Nuevo}, {G{\'o}rski}, {Gratton}, {Gregorio}, {Gruppuso},
  {Haissinski}, {Hansen}, {Harrison}, {Helou}, {Henrot-Versill{\'e}},
  {Hern{\'a}ndez-Monteagudo}, {Herranz}, {Hildebrand t}, {Hivon}, {Hobson},
  {Holmes}, {Hornstrup}, {Hovest}, {Hoyland}, {Huffenberger}, {Huynh}, {Jaffe},
  {Jones}, {Juvela}, {Keih{\"a}nen}, {Keskitalo}, {Kisner}, {Kneissl}, {Knox},
  {Kurki-Suonio}, {Lagache}, {L{\"a}hteenm{\"a}ki}, {Lamarre}, {Lasenby},
  {Laureijs}, {Lawrence}, {Leach}, {Leahy}, {Leonardi}, {Le{\'o}n-Tavares},
  {Leroy}, {Lilje}, {Linden-V{\o}rnle}, {L{\'o}pez-Caniego}, {Lubin},
  {Mac{\'\i}as-P{\'e}rez}, {MacTavish}, {Maffei}, {Maggio}, {Maino},
  {Mandolesi}, {Mann}, {Maris}, {Marleau}, {Marshall},
  {Mart{\'\i}nez-Gonz{\'a}lez}, {Masi}, {Massardi}, {Matarrese}, {Matthai},
  {Mazzotta}, {McGehee}, {Meinhold}, {Melchiorri}, {Melin}, {Mendes},
  {Mennella}, {Mitra}, {Miville-Desch{\^e}nes}, {Moneti}, {Montier},
  {Morgante}, {Mortlock}, {Munshi}, {Murphy}, {Naselsky}, {Natoli},
  {Netterfield}, {N{\o}rgaard-Nielsen}, {Noviello}, {Novikov}, {Novikov},
  {O'Dwyer}, {Osborne}, {Pajot}, {Paladini}, {Partridge}, {Pasian},
  {Patanchon}, {Pearson}, {Perdereau}, {Perotto}, {Perrotta}, {Piacentini},
  {Piat}, {Piffaretti}, {Plaszczynski}, {Platania}, {Pointecouteau}, {Polenta},
  {Ponthieu}, {Poutanen}, {Pratt}, {Pr{\'e}zeau}, {Prunet}, {Puget}, {Rachen},
  {Reach}, {Rebolo}, {Reinecke}, {Renault}, {Ricciardi}, {Riller},
  {Ristorcelli}, {Rocha}, {Rosset}, {Rowan-Robinson}, {Rubi{\~n}o-Mart{\'\i}n},
  {Rusholme}, {Sajina}, {Sandri}, {Santos}, {Savini}, {Schaefer}, {Scott},
  {Seiffert}, {Shellard}, {Smoot}, {Starck}, {Stivoli}, {Stolyarov},
  {Sudiwala}, {Sunyaev}, {Sygnet}, {Tauber}, {Tavagnacco}, {Terenzi},
  {Toffolatti}, {Tomasi}, {Torre}, {Tristram}, {Tuovinen}, {T{\"u}rler},
  {Umana}, {Valenziano}, {Valiviita}, {Varis}, {Vielva}, {Villa}, {Vittorio},
  {Wade}, {Wandelt}, {White}, {Wilkinson}, {Yvon}, {Zacchei}, \&
  {Zonca}}]{2011A&A...536A...7P}
---. 2011{\natexlab{b}}, \aap, 536, A7, \dodoi{10.1051/0004-6361/201116474}

\bibitem[{{Planck Collaboration} {et~al.}(2011{\natexlab{c}}){Planck
  Collaboration}, {Ade}, {Aghanim}, {Arnaud}, {Ashdown}, {Aumont},
  {Baccigalupi}, {Balbi}, {Banday}, {Barreiro}, {Bartlett}, {Battaner},
  {Benabed}, {Beno{\^\i}t}, {Bernard}, {Bersanelli}, {Bhatia}, {Bock},
  {Bonaldi}, {Bond}, {Borrill}, {Bouchet}, {Boulanger}, {Bucher}, {Burigana},
  {Cabella}, {Cantalupo}, {Cardoso}, {Catalano}, {Cay{\'o}n}, {Challinor},
  {Chamballu}, {Chary}, {Chiang}, {Christensen}, {Clements}, {Colombi},
  {Couchot}, {Coulais}, {Crill}, {Cuttaia}, {Danese}, {Davies}, {Davis}, {de
  Bernardis}, {de Gasperis}, {de Rosa}, {de Zotti}, {Delabrouille}, {Delouis},
  {D{\'e}sert}, {Dickinson}, {Dobashi}, {Donzelli}, {Dor{\'e}}, {D{\"o}rl},
  {Douspis}, {Dupac}, {Efstathiou}, {En{\ss}lin}, {Falgarone}, {Finelli},
  {Forni}, {Frailis}, {Franceschi}, {Galeotta}, {Ganga}, {Giard}, {Giardino},
  {Giraud-H{\'e}raud}, {Gonz{\'a}lez-Nuevo}, {G{\'o}rski}, {Gratton},
  {Gregorio}, {Gruppuso}, {Hansen}, {Harrison}, {Helou}, {Henrot-Versill{\'e}},
  {Herranz}, {Hildebrandt}, {Hivon}, {Hobson}, {Holmes}, {Hovest}, {Hoyland},
  {Huffenberger}, {Jaffe}, {Joncas}, {Jones}, {Juvela}, {Keih{\"a}nen},
  {Keskitalo}, {Kisner}, {Kneissl}, {Knox}, {Kurki-Suonio}, {Lagache},
  {Lamarre}, {Lasenby}, {Laureijs}, {Lawrence}, {Leach}, {Leonardi}, {Leroy},
  {Linden-V{\o}rnle}, {L{\'o}pez-Caniego}, {Lubin}, {Mac{\'\i}as-P{\'e}rez},
  {MacTavish}, {Maffei}, {Mandolesi}, {Mann}, {Maris}, {Marshall}, {Martin},
  {Mart{\'\i}nez-Gonz{\'a}lez}, {Marton}, {Masi}, {Matarrese}, {Matthai},
  {Mazzotta}, {McGehee}, {Melchiorri}, {Mendes}, {Mennella}, {Mitra},
  {Miville-Desch{\^e}nes}, {Moneti}, {Montier}, {Morgante}, {Mortlock},
  {Munshi}, {Murphy}, {Naselsky}, {Nati}, {Natoli}, {Netterfield},
  {N{\o}rgaard-Nielsen}, {Noviello}, {Novikov}, {Novikov}, {Osborne}, {Pajot},
  {Paladini}, {Pasian}, {Patanchon}, {Pearson}, {Pelkonen}, {Perdereau},
  {Perotto}, {Perrotta}, {Piacentini}, {Piat}, {Plaszczynski}, {Pointecouteau},
  {Polenta}, {Ponthieu}, {Poutanen}, {Pr{\'e}zeau}, {Prunet}, {Puget}, {Reach},
  {Rebolo}, {Reinecke}, {Renault}, {Ricciardi}, {Riller}, {Ristorcelli},
  {Rocha}, {Rosset}, {Rowan-Robinson}, {Rubi{\~n}o-Mart{\'\i}n}, {Rusholme},
  {Sandri}, {Santos}, {Savini}, {Scott}, {Seiffert}, {Smoot}, {Starck},
  {Stivoli}, {Stolyarov}, {Sudiwala}, {Sygnet}, {Tauber}, {Terenzi},
  {Toffolatti}, {Tomasi}, {Torre}, {Toth}, {Tristram}, {Tuovinen}, {Umana},
  {Valenziano}, {Vielva}, {Villa}, {Vittorio}, {Wade}, {Wandelt}, {Ysard},
  {Yvon}, {Zacchei}, {Zahorecz}, \& {Zonca}}]{2011A&A...536A..23P}
---. 2011{\natexlab{c}}, \aap, 536, A23, \dodoi{10.1051/0004-6361/201116472}

\bibitem[{{Planck Collaboration} {et~al.}(2014){Planck Collaboration}, {Ade},
  {Aghanim}, {Alves}, {Armitage-Caplan}, {Arnaud}, {Ashdown},
  {Atrio-Barandela}, {Aumont}, {Baccigalupi}, {Banday}, {Barreiro}, {Bartlett},
  {Battaner}, {Benabed}, {Beno{\^\i}t}, {Benoit-L{\'e}vy}, {Bernard},
  {Bersanelli}, {Bielewicz}, {Bobin}, {Bock}, {Bonaldi}, {Bond}, {Borrill},
  {Bouchet}, {Boulanger}, {Bridges}, {Bucher}, {Burigana}, {Butler}, {Cardoso},
  {Catalano}, {Chamballu}, {Chary}, {Chen}, {Chiang}, {Chiang}, {Christensen},
  {Church}, {Clements}, {Colombi}, {Colombo}, {Combet}, {Couchot}, {Coulais},
  {Crill}, {Curto}, {Cuttaia}, {Danese}, {Davies}, {de Bernardis}, {de Rosa},
  {de Zotti}, {Delabrouille}, {Delouis}, {Dempsey}, {D{\'e}sert}, {Dickinson},
  {Diego}, {Dole}, {Donzelli}, {Dor{\'e}}, {Douspis}, {Dupac}, {Efstathiou},
  {En{\ss}lin}, {Eriksen}, {Falgarone}, {Finelli}, {Forni}, {Frailis},
  {Franceschi}, {Fukui}, {Galeotta}, {Ganga}, {Giard}, {Giraud-H{\'e}raud},
  {Gonz{\'a}lez-Nuevo}, {G{\'o}rski}, {Gratton}, {Gregorio}, {Gruppuso},
  {Handa}, {Hansen}, {Hanson}, {Harrison}, {Henrot-Versill{\'e}},
  {Hern{\'a}ndez-Monteagudo}, {Herranz}, {Hildebrandt}, {Hily-Blant}, {Hivon},
  {Hobson}, {Holmes}, {Hornstrup}, {Hovest}, {Huffenberger}, {Hurier}, {Jaffe},
  {Jaffe}, {Jewell}, {Jones}, {Juvela}, {Keih{\"a}nen}, {Keskitalo}, {Kisner},
  {Knoche}, {Knox}, {Kunz}, {Kurki-Suonio}, {Lagache}, {L{\"a}hteenm{\"a}ki},
  {Lamarre}, {Lasenby}, {Laureijs}, {Lawrence}, {Leonardi}, {Le{\'o}n-Tavares},
  {Lesgourgues}, {Liguori}, {Lilje}, {Linden-V{\o}rnle}, {L{\'o}pez-Caniego},
  {Lubin}, {Mac{\'\i}as-P{\'e}rez}, {Maffei}, {Mandolesi}, {Maris}, {Marshall},
  {Martin}, {Mart{\'\i}nez-Gonz{\'a}lez}, {Masi}, {Massardi}, {Matarrese},
  {Matthai}, {Mazzotta}, {McGehee}, {Melchiorri}, {Mendes}, {Mennella},
  {Migliaccio}, {Mitra}, {Miville-Desch{\^e}nes}, {Moneti}, {Montier}, {Moore},
  {Morgante}, {Morino}, {Mortlock}, {Munshi}, {Murphy}, {Nakajima}, {Naselsky},
  {Nati}, {Natoli}, {Netterfield}, {N{\o}rgaard-Nielsen}, {Noviello},
  {Novikov}, {Novikov}, {Okuda}, {Osborne}, {Oxborrow}, {Paci}, {Pagano},
  {Pajot}, {Paladini}, {Paoletti}, {Pasian}, {Patanchon}, {Perdereau},
  {Perotto}, {Perrotta}, {Piacentini}, {Piat}, {Pierpaoli}, {Pietrobon},
  {Plaszczynski}, {Pointecouteau}, {Polenta}, {Ponthieu}, {Popa}, {Poutanen},
  {Pratt}, {Pr{\'e}zeau}, {Prunet}, {Puget}, {Rachen}, {Reach}, {Rebolo},
  {Reinecke}, {Remazeilles}, {Renault}, {Ricciardi}, {Riller}, {Ristorcelli},
  {Rocha}, {Rosset}, {Roudier}, {Rowan-Robinson}, {Rubi{\~n}o-Mart{\'\i}n},
  {Rusholme}, {Sandri}, {Santos}, {Savini}, {Scott}, {Seiffert}, {Shellard},
  {Spencer}, {Starck}, {Stolyarov}, {Stompor}, {Sudiwala}, {Sunyaev}, {Sureau},
  {Sutton}, {Suur-Uski}, {Sygnet}, {Tauber}, {Tavagnacco}, {Terenzi}, {Thomas},
  {Toffolatti}, {Tomasi}, {Torii}, {Tristram}, {Tucci}, {Tuovinen}, {Umana},
  {Valenziano}, {Valiviita}, {Van Tent}, {Vielva}, {Villa}, {Vittorio}, {Wade},
  {Wandelt}, {Wehus}, {Yamamoto}, {Yoda}, {Yvon}, {Zacchei}, \&
  {Zonca}}]{2014A&A...571A..13P}
---. 2014, \aap, 571, A13, \dodoi{10.1051/0004-6361/201321553}

\bibitem[{{Planck Collaboration} {et~al.}(2016){Planck Collaboration}, {Ade},
  {Aghanim}, {Arnaud}, {Ashdown}, {Aumont}, {Baccigalupi}, {Banday},
  {Barreiro}, {Bartolo}, \& et~al.}]{2016A&A...594A..28P}
---. 2016, \aap, 594, A28, \dodoi{10.1051/0004-6361/201525819}

\bibitem[{{Poglitsch} {et~al.}(2010){Poglitsch}, {Waelkens}, {Geis},
  {Feuchtgruber}, {Vandenbussche}, {Rodriguez}, {Krause}, {Renotte}, {van
  Hoof}, {Saraceno}, {Cepa}, {Kerschbaum}, {Agn{\`e}se}, {Ali}, {Altieri},
  {Andreani}, {Augueres}, {Balog}, {Barl}, {Bauer}, {Belbachir}, {Benedettini},
  {Billot}, {Boulade}, {Bischof}, {Blommaert}, {Callut}, {Cara}, {Cerulli},
  {Cesarsky}, {Contursi}, {Creten}, {De Meester}, {Doublier}, {Doumayrou},
  {Duband}, {Exter}, {Genzel}, {Gillis}, {Gr{\"o}zinger}, {Henning},
  {Herreros}, {Huygen}, {Inguscio}, {Jakob}, {Jamar}, {Jean}, {de Jong},
  {Katterloher}, {Kiss}, {Klaas}, {Lemke}, {Lutz}, {Madden}, {Marquet},
  {Martignac}, {Mazy}, {Merken}, {Montfort}, {Morbidelli}, {M{\"u}ller},
  {Nielbock}, {Okumura}, {Orfei}, {Ottensamer}, {Pezzuto}, {Popesso},
  {Putzeys}, {Regibo}, {Reveret}, {Royer}, {Sauvage}, {Schreiber}, {Stegmaier},
  {Schmitt}, {Schubert}, {Sturm}, {Thiel}, {Tofani}, {Vavrek}, {Wetzstein},
  {Wieprecht}, \& {Wiezorrek}}]{2010A&A...518L...2P}
{Poglitsch}, A., {Waelkens}, C., {Geis}, N., {et~al.} 2010, \aap, 518, L2,
  \dodoi{10.1051/0004-6361/201014535}

\bibitem[{{Polk} {et~al.}(1988){Polk}, {Knapp}, {Stark}, \&
  {Wilson}}]{1988ApJ...332..432P}
{Polk}, K.~S., {Knapp}, G.~R., {Stark}, A.~A., \& {Wilson}, R.~W. 1988, \apj,
  332, 432, \dodoi{10.1086/166667}

\bibitem[{{Pound} \& {Goodman}(1997)}]{1997ApJ...482..334P}
{Pound}, M.~W., \& {Goodman}, A.~A. 1997, \apj, 482, 334,
  \dodoi{10.1086/304136}

\bibitem[{Price-Whelan {et~al.}(2018)Price-Whelan, Sip{\H{o}}cz, G{\"u}nther,
  Lim, Crawford, Conseil, Shupe, Craig, Dencheva, Ginsburg,
  {et~al.}}]{price2018astropy}
Price-Whelan, A.~M., Sip{\H{o}}cz, B., G{\"u}nther, H., {et~al.} 2018, The
  Astronomical Journal, 156, 123

\bibitem[{{Reach} {et~al.}(1994){Reach}, {Koo}, \&
  {Heiles}}]{1994ApJ...429..672R}
{Reach}, W.~T., {Koo}, B.-C., \& {Heiles}, C. 1994, \apj, 429, 672,
  \dodoi{10.1086/174353}

\bibitem[{{Reach} {et~al.}(1998){Reach}, {Wall}, \&
  {Odegard}}]{1998ApJ...507..507R}
{Reach}, W.~T., {Wall}, W.~F., \& {Odegard}, N. 1998, \apj, 507, 507,
  \dodoi{10.1086/306357}

\bibitem[{{Robinson} {et~al.}(1983){Robinson}, {Manchester}, {Whiteoak}, \&
  {McCutcheon}}]{1983ASSL..105....1R}
{Robinson}, B.~J., {Manchester}, R.~N., {Whiteoak}, J.~B., \& {McCutcheon},
  W.~H. 1983, Astrophysics and Space Science Library, Vol. 105, {CO
  distribution along the southern galactic plane}, ed. W.~B. {Burton} \& F.~P.
  {Israel}, 1--15, \dodoi{10.1007/978-94-009-7217-9_1}

\bibitem[{{Robitaille} \& {Bressert}(2012)}]{2012ascl.soft08017R}
{Robitaille}, T., \& {Bressert}, E. 2012, {APLpy: Astronomical Plotting Library
  in Python}.
\newblock \doeprint{1208.017}

\bibitem[{Robitaille {et~al.}(2013)Robitaille, Tollerud, Greenfield,
  Droettboom, Bray, Aldcroft, Davis, Ginsburg, Price-Whelan, Kerzendorf,
  Conley, Crighton, Barbary, Muna, Ferguson, Grollier, Parikh, Nair,
  G\"{u}nther, Deil, Woillez, Conseil, Kramer, Turner, Singer, Fox, Weaver,
  Zabalza, Edwards, Bostroem, Burke, Casey, Crawford, Dencheva, Ely, Jenness,
  Labrie, Lim, Pierfederici, Pontzen, Ptak, Refsdal, Servillat, \&
  Streicher}]{2013Robitaille}
Robitaille, T.~P., Tollerud, E.~J., Greenfield, P., {et~al.} 2013, Astronomy
  {\&} Astrophysics, 558, A33, \dodoi{10.1051/0004-6361/201322068}

\bibitem[{{Sandell} {et~al.}(1987){Sandell}, {Reipurth}, \&
  {Gahm}}]{1987A&A...181..283S}
{Sandell}, G., {Reipurth}, B., \& {Gahm}, G. 1987, \aap, 181, 283

\bibitem[{{Sanders} {et~al.}(1984){Sanders}, {Solomon}, \&
  {Scoville}}]{1984ApJ...276..182S}
{Sanders}, D.~B., {Solomon}, P.~M., \& {Scoville}, N.~Z. 1984, \apj, 276, 182,
  \dodoi{10.1086/161602}

\bibitem[{{Scoville}(2013)}]{2013seg..book..491S}
{Scoville}, N.~Z. 2013, {Evolution of star formation and gas}, ed.
  J.~{Falc{\'o}n-Barroso} \& J.~H. {Knapen}, 491

\bibitem[{{Shan} {et~al.}(2012){Shan}, {Yang}, {Shi}, {Yao}, {Zuo}, {Lin},
  {Chen}, {Zhang}, {Duan}, {Cao}, {Li}, {Li}, {Liu}, \& {Zhong}}]{6313968}
{Shan}, W., {Yang}, J., {Shi}, S., {et~al.} 2012, IEEE Transactions on
  Terahertz Science and Technology, 2, 593

\bibitem[{{Shimajiri} {et~al.}(2014){Shimajiri}, {Kitamura}, {Saito}, {Momose},
  {Nakamura}, {Dobashi}, {Shimoikura}, {Nishitani}, {Yamabi}, {Hara},
  {Katakura}, {Tsukagoshi}, {Tanaka}, \& {Kawabe}}]{2014A&A...564A..68S}
{Shimajiri}, Y., {Kitamura}, Y., {Saito}, M., {et~al.} 2014, \aap, 564, A68,
  \dodoi{10.1051/0004-6361/201322912}

\bibitem[{{Shore} {et~al.}(2006){Shore}, {Larosa}, {Chastain}, \&
  {Magnani}}]{2006A&A...457..197S}
{Shore}, S.~N., {Larosa}, T.~N., {Chastain}, R.~J., \& {Magnani}, L. 2006,
  \aap, 457, 197, \dodoi{10.1051/0004-6361:20053962}

\bibitem[{{Shore} {et~al.}(2007){Shore}, {Larosa}, {Magnani}, {Chastain}, \&
  {Costagliola}}]{2007IAUS..237...17S}
{Shore}, S.~N., {Larosa}, T.~N., {Magnani}, L., {Chastain}, R.~J., \&
  {Costagliola}, F. 2007, in Triggered Star Formation in a Turbulent ISM, ed.
  B.~G. {Elmegreen} \& J.~{Palous}, Vol. 237, 17--23,
  \dodoi{10.1017/S1743921307001160}

\bibitem[{{Shore} {et~al.}(2003){Shore}, {Magnani}, {LaRosa}, \&
  {McCarthy}}]{2003ApJ...593..413S}
{Shore}, S.~N., {Magnani}, L., {LaRosa}, T.~N., \& {McCarthy}, M.~N. 2003,
  \apj, 593, 413, \dodoi{10.1086/376355}

\bibitem[{{Shu} {et~al.}(1987){Shu}, {Adams}, \&
  {Lizano}}]{1987ARA&A..25...23S}
{Shu}, F.~H., {Adams}, F.~C., \& {Lizano}, S. 1987, \araa, 25, 23,
  \dodoi{10.1146/annurev.aa.25.090187.000323}

\bibitem[{{Stark}(1979)}]{1979PhDT.........9S}
{Stark}, A.~A. 1979, PhD thesis, Princeton Univ., NJ.

\bibitem[{{Tennekes} \& {Lumley}(1972)}]{1972fct..book.....T}
{Tennekes}, H., \& {Lumley}, J.~L. 1972, {First Course in Turbulence}

\bibitem[{{van Dishoeck} \& {Black}(1988)}]{1988ApJ...334..771V}
{van Dishoeck}, E.~F., \& {Black}, J.~H. 1988, \apj, 334, 771,
  \dodoi{10.1086/166877}

\bibitem[{{van Dishoeck} {et~al.}(1991){van Dishoeck}, {Black}, {Phillips}, \&
  {Gredel}}]{1991ApJ...366..141V}
{van Dishoeck}, E.~F., {Black}, J.~H., {Phillips}, T.~G., \& {Gredel}, R. 1991,
  \apj, 366, 141, \dodoi{10.1086/169547}

\bibitem[{{Vanden Bout} {et~al.}(2012){Vanden Bout}, {Davis}, \&
  {Loren}}]{2012JAHH...15..232V}
{Vanden Bout}, P.~A., {Davis}, J.~H., \& {Loren}, R.~B. 2012, Journal of
  Astronomical History and Heritage, 15, 232

\bibitem[{{Williams} {et~al.}(1994){Williams}, {de Geus}, \&
  {Blitz}}]{1994ApJ...428..693W}
{Williams}, J.~P., {de Geus}, E.~J., \& {Blitz}, L. 1994, \apj, 428, 693,
  \dodoi{10.1086/174279}

\bibitem[{{Wilson} \& {Rood}(1994)}]{1994ARA&A..32..191W}
{Wilson}, T.~L., \& {Rood}, R. 1994, \araa, 32, 191,
  \dodoi{10.1146/annurev.aa.32.090194.001203}

\bibitem[{{Wouterloot} {et~al.}(2000){Wouterloot}, {Heithausen}, {Schreiber},
  \& {Winnewisser}}]{2000A&AS..144..123W}
{Wouterloot}, J.~G.~A., {Heithausen}, A., {Schreiber}, W., \& {Winnewisser}, G.
  2000, \aaps, 144, 123, \dodoi{10.1051/aas:2000200}

\bibitem[{{Wu} \& {Evans}(1989)}]{1989ApJ...340..307W}
{Wu}, Y., \& {Evans}, Neal~J., I. 1989, \apj, 340, 307, \dodoi{10.1086/167393}

\bibitem[{{Wu} {et~al.}(2012){Wu}, {Liu}, {Meng}, {Li}, {Qin}, \&
  {Ju}}]{2012ApJ...756...76W}
{Wu}, Y., {Liu}, T., {Meng}, F., {et~al.} 2012, \apj, 756, 76,
  \dodoi{10.1088/0004-637X/756/1/76}

\bibitem[{{Yamamoto} {et~al.}(2003){Yamamoto}, {Onishi}, {Mizuno}, \&
  {Fukui}}]{2003ApJ...592..217Y}
{Yamamoto}, H., {Onishi}, T., {Mizuno}, A., \& {Fukui}, Y. 2003, \apj, 592,
  217, \dodoi{10.1086/375128}

\bibitem[{{Yang} {et~al.}(2017){Yang}, {Chen}, {Shen}, {Li}, {Wang}, {Jiang},
  {Li}, {Dong}, {Wu}, {Qiao}, \& {Ren}}]{2017ApJ...846..160Y}
{Yang}, K., {Chen}, X., {Shen}, Z.-Q., {et~al.} 2017, \apj, 846, 160,
  \dodoi{10.3847/1538-4357/aa8668}

\bibitem[{{Yuan} {et~al.}(2017){Yuan}, {Wu}, {Ellingsen}, {Evans}, {Henkel},
  {Wang}, {Liu}, {Liu}, {Li}, \& {Zavagno}}]{2017ApJS..231...11Y}
{Yuan}, J., {Wu}, Y., {Ellingsen}, S.~P., {et~al.} 2017, \apjs, 231, 11,
  \dodoi{10.3847/1538-4365/aa7204}

\bibitem[{{Zhang} {et~al.}(2020){Zhang}, {Wu}, {Liu}, {Qin}, {Liu}, {Yuan},
  {Li}, {Meng}, {Zhang}, {Tang}, {Yuan}, {Zhou}, {Esimbek}, {Zhou}, {Chen}, \&
  {Hu}}]{2020ApJS..247...29Z}
{Zhang}, C., {Wu}, Y., {Liu}, X., {et~al.} 2020, \apjs, 247, 29,
  \dodoi{10.3847/1538-4365/ab720b}

\bibitem[{Zhang {et~al.}(2016)Zhang, Wu, Liu, \& Meng}]{Zhang_2016}
Zhang, T., Wu, Y., Liu, T., \& Meng, F. 2016, The Astrophysical Journal
  Supplement Series, 224, 43, \dodoi{10.3847/0067-0049/224/2/43}

\end{thebibliography}

\begin{table*}[!htp]
\centering
\caption{CO Line Survey at High Galactic Latitude\label{tab:surveys}}
\begin{tabular}{ccccc}
\hline
\hline
Sample Selection\tablenotemarknew{a.} & Telescope &  Resolution & Sampling & Covering Range \\
\hline
POSS plates\tablenotemarknew{b.} & Texas 5-m\tablenotemarknew{c.} & 2\fmg3 & 10$^\prime$ & $|b|\ge20\degr$ \\
ESO plates\tablenotemarknew{d.} & South CU 1.2-m\tablenotemarknew{e.} & 8\fmg6 & 7\fmg5 & R.A.: 3-15$^\mathrm{h}$, Decl.$<-20^\circ$ \\
Infrared cirrus\tablenotemarknew{f.} & North CU 1.2-m\tablenotemarknew{g.} & 8\fmg6 & 8\fmg6 & $l:141\sim149\degr, b:34\sim42\degr$ \\
2nd quadrant HGal\tablenotemarknew{h.} & North CU 1.2-m & 8\fmg7 & 7\fmg5 & $l:117\sim160\degr, b:16\sim44\degr$ \\
Catalog1996\tablenotemarknew{i.} & & & & $|b|\ge25\degr$ \\
North\tablenotemarknew{j.} & North CU 1.2-m & 8\fmg4 & 0.3-1\degr\ \tablenotemarknew{k.} & $b\ge30\degr$ \\
South\tablenotemarknew{l.} & North CU 1.2-m & 8\fmg4 & 0.3-1\degr & $b\le-30\degr$ \\
\textsc{Hi} Filament\tablenotemarknew{m.} &  NANTEN 4-m\tablenotemarknew{n.} & 2\fmg6 & 4$^\prime(\times\cos b)$ \tablenotemarknew{o.} & $l:82\sim98\degr, b:-45\sim-29\degr$ \\
PGCC at HGal\tablenotemarknew{p.} & PMO 13.7-m\tablenotemarknew{q.} & $\sim$56$^{\prime\prime}$ & 30$^{\prime\prime}$ & $|b|\ge25\degr$ \\
\hline
\end{tabular}
\\
\tablenotetextnew{\textbi{a.}}{The criterion of the source selection in the survey.} 
\tablenotetextnew{\textbi{b.}}{Optical obscuration selected from POSS southern sky survey plates \citep{1985ApJ...295..402M}.} 
\tablenotetextnew{\textbi{c.}}{5-m Millimeter Wave Observatory in Texas.}
\tablenotetextnew{\textbi{d.}}{European Southern Observatory (ESO) southern sky survey plates were examined for regions of faint obscuration or reflection. 135 in 151 prints were searched for CO emission \citep{1986ApJ...304..466K}.} 
\tablenotetextnew{\textbi{e.}}{The Columbia University 1.2-m Millimeter-Wave Telescope in Cerro Tololo, Chile. The telescope was constructed at Columbia and shipped to Chile in 1982.} 
\tablenotetextnew{\textbi{f.}}{The patchy far-infrared emission (“cirrus”) in Ursa Major, discovered by \textit{IRAS} \citep{1984ApJ...278L..19L,1987ApJ...319..723D}.} 
\tablenotetextnew{\textbi{g.}}{The Columbia University 1.2-m Millimeter-Wave Telescope which was removed to the Center for Astrophysics (CfA) in 1986. One of the twin instruments was located in the Northern hemisphere, the other one is in Chile\tablenotemarknew{d}.}
\tablenotetextnew{\textbi{h.}}{Composite large-scale CO survey at high Galactic latitude in the second quadrant \citep{1993AA...268..265H}.}
\tablenotetextnew{\textbi{i.}}{The catalog of molecular gas at high Galactic latitude by 1996. Related references are included in \citet{1996ApJS..106..447M}.}
\tablenotetextnew{\textbi{j.}}{the northern Galactic hemisphere.}
\tablenotetextnew{\textbi{k.}}{The sampling interval is $\Delta l\times\cos b$.}
\tablenotetextnew{\textbi{l.}}{the sorthern Galactic hemisphere.}
\tablenotetextnew{\textbi{m.}}{The \textsc{Hi} filament that contains the HGal clouds complex of MBM 53, 54, and 55 complex which is one of the largest and most massive complexes at HGal \citep{2003ApJ...592..217Y}.}
\tablenotetextnew{\textbi{n.}}{NANTEN 4-m (sub)millimeter-wave telescope of Nagoya University at Las Campanas Observatory in Chile.}
\tablenotetextnew{\textbi{o.}}{To begin with, $^{12}$CO has been measured with 8$^\prime$ in latitude $b$ and 8$^\prime\times\cos b$ in longitued $l$; next observe $^{13}$CO with 4$^\prime$ in $b$ and 4$^\prime\times\cos b$ in $l$; finanly observe $^{13}$CO and C$^{18}$O with 2$^\prime$ in $b$ and 2$^\prime\times\cos b$ in $l$.}
\tablenotetextnew{\textbi{p.}}{This work: Planck Galactic Cold Clumps (PGCCs) at high Galactic latitude.}
\tablenotetextnew{\textbi{q.}}{Purple Mountain Observatory 13.7-m Millimeter Radio Telescope.}
\end{table*}

\begin{deluxetable*}{ccccccccccc}
\tablecaption{Catalogue of PGCCs at High Galactic Latitude 
\label{tab:samples}}
\tablehead{
\colhead{Name} & \colhead{Glon.} & \colhead{Glat.} & \colhead{R.A.(J2000)} &
\colhead{Decl.(J2000)} & \colhead{R.A.(B1950)} & \colhead{Decl.(B1950)} & \colhead{Distance} & \colhead{Elevation} & \colhead{OTF Map} & \colhead{Comment} \\
\colhead{} & \colhead{deg} & \colhead{deg} & \colhead{h:m:s} &
\colhead{d:m:s} & \colhead{h:m:s} & \colhead{d:m:s} & \colhead{pc} & \colhead{pc} & \colhead{} & \colhead{}
}
\decimalcolnumbers
\startdata
\hline
G004.15+35.77 & 4.1528 & 35.7772 & 15 53 29.82 & -04 38 52.39 & 15 50 51.56 & -04 40 30.16 & 145.0 & 85.0 & Y & L134\tablenotemarknew{a}, MBM 36\tablenotemarknew{b} \\
G004.17+36.67 & 4.1748 & 36.6789 & 15 50 42.66 & -04 04 20.84 & 15 48 05.00 & -03 55 20.10 & 130.0 & 77.0 & Y & CB 63\tablenotemarknew{c}, LBN\tablenotemarknew{d}\\
G004.54+36.74 & 4.5483 & 36.7487 & 15 51 14.27 & -03 47 40.14 & 15 48 36.88 & -03 38 41.35 & 130.0 & 78.0 & Y & L169, MBM 37 \\
G004.81+37.02 & 4.8120 & 37.0285 & 15 50 52.79 & -03 27 20.38 & 15 48 15.74 & -03 18 20.28 & 130.0 & 78.0 & Y & L169, MBM 37 \\
G005.69+36.84 & 5.6909 & 36.8419 & 15 53 11.88 & -03 00 56.50 & 15 50 35.24 & -02 52 04.98 & 122.0 & 73.0 & N & MBM 37 \\
G006.04+36.74 & 6.0424 & 36.7487 & 15 54 10.81 & -02 50 56.32 & 15 51 34.33 & -02 42 08.46 & 122.0 & 73.0 & Y & L183 \\
G008.43+36.35 & 8.4375 & 36.3540 & 16 00 02.55 & -01 32 25.18 & 15 57 27.34 & -01 23 59.32 & 130.0 & 77.0 & Y & MBM 38, CB 64 \\
G011.40+36.19 & 11.4038 & 36.1921 & 16 06 03.58 & +00 19 32.84 & 16 03 30.26 & +00 27 35.75 & 130.0 & 77.0 & Y & MBM 39, LBN \\
G037.51+44.57 & 37.5193 & 44.5732 & 16 10 54.35 & +21 45 37.26 & 16 08 44.35 & +21 53 20.74 & 122.0 & 86.0 & N & MBM 40, LBN \\
G089.03-41.28 & 89.0331 & -41.2869 & 23 08 43.43 & +14 43 35.71 & 23 06 13.66 & +14 27 19.66 & 183.0 & 121.0 & Y & MBM 55 \\
G091.29-38.15 & 91.2963 & -38.1584 & 23 08 26.74 & +18 18 20.84 & 23 05 58.02 & +18 02 05.05 & 218.0 & 134.0 & N & MBM 54, LBN\\
G101.62-28.84 & 101.6235 & -28.8436 & 23 24 47.63 & +30 25 04.98 & 23 22 20.08 & +30 08 35.36 & 244.0 & 118.0 & Y & New\tablenotemarknew{e} \\
G102.72-25.98 & 102.7221 & -25.9859 & 23 24 03.29 & +33 26 00.51 & 23 21 36.65 & +33 09 31.40 & 259.0 & 113.0 & Y & New \\
G118.25-52.70 & 118.2512 & -52.7054 & 00 39 55.42 & +10 03 41.94 & 00 37 19.73 & +09 47 13.89 & 194.0 & 154.0 & N & MBM 02\\
G126.65-71.43 & 126.6563 & -71.4392 & 00 56 13.82 & -08 36 07.67 & 00 53 42.53 & -08 52 21.24 & 183.0 & 174.0 & N & New \\
G127.31-70.09 & 127.3183 & -70.0954 & 00 57 27.24 & -07 16 29.84 & 00 54 55.63 & -07 32 42.13 & 145.0 & 137.0 & N & New \\
G131.35-45.73 & 131.3571 & -45.7346 & 01 15 57.51 & +16 44 08.07 & 01 13 17.42 & +16 28 18.47 & 259.0 & 185.0 & N & MBM 03, LBN \\
G133.72-45.31 & 133.7257 & -45.3151 & 01 23 04.84 & +16 53 32.68 & 01 20 24.08 & +16 37 53.49 & 231.0 & 164.0 & N & MBM 04, LBN \\
G145.08-39.30 & 145.0854 & -39.3061 & 02 04 00.04 & +20 25 25.62 & 02 01 13.71 & +20 11 03.42 & 163.0 & 103.0 & Y & MBM 06, LBN \\
G150.35-38.06 & 150.3588 & -38.0636 & 02 22 11.93 & +19 55 36.57 & 02 19 24.33 & +19 41 57.58 & 173.0 & 107.0 & N & MBM 07, LBN \\
G151.58-38.58 & 151.5893 & -38.5866 & 02 24 53.06 & +19 01 47.06 & 02 22 05.92 & +18 48 14.91 & 154.0 & 96.0 & Y & MBM 08, LBN \\
G153.74+35.91 & 153.7426 & 35.9152 & 08 36 34.59 & +62 26 27.09 & 08 32 20.45 & +62 36 52.58 & 365.0 & 214.0 & Y & HSVMT 27\tablenotemarknew{f} \\
G156.42+32.53 & 156.4233 & 32.5313 & 08 06 27.98 & +60 34 20.05 & 08 02 12.51 & +60 42 57.52 & 290.0 & 156.0 & Y & MBM 26, HSVMT 29 \\
G158.77-33.30 & 158.7744 & -33.3089 & 02 57 33.29 & +20 38 31.01 & 02 54 42.16 & +20 26 30.43 & 205.0 & 113.0 & Y & MBM 11--12, L1457-c\tablenotemarknew{g} \\
G158.88-34.18 & 158.8842 & -34.1837 & 02 55 48.23 & +19 51 50.97 & 02 52 57.95 & +19 39 45.12 & 231.0 & 129.0 & Y & MBM 11--12, L1457-c \\
G158.97-33.01 & 158.9721 & -33.0193 & 02 58 50.03 & +20 47 23.13 & 02 55 58.66 & +20 35 26.45 & 274.0 & 149.0 & Y & MBM 11--12, L1457-c \\
G159.23-34.49 & 159.2358 & -34.4999 & 02 56 04.75 & +19 26 23.35 & 02 53 14.84 & +19 14 18.34 & 244.0 & 138.0 & Y & MBM 11--12, L1457-c \\
G159.41-34.36 & 159.4116 & -34.3643 & 02 56 54.67 & +19 28 15.70 & 02 54 04.66 & +19 16 13.20 & 205.0 & 116.0 & Y & MBM 11--12, L1457-c \\
G159.58-32.84 & 159.5873 & -32.8415 & 03 01 05.17 & +20 38 43.23 & 02 58 13.76 & +20 26 53.47 & 290.0 & 157.0 & Y & MBM 11--12, L1457-c \\
G159.67-34.31 & 159.6752 & -34.3191 & 02 57 46.94 & +19 23 09.02 & 02 54 56.94 & +19 11 09.16 & 194.0 & 109.0 & Y & MBM 11--12, L1457-c \\
G160.64-35.04 & 160.6420 & -35.0449 & 02 58 48.02 & +18 20 11.95 & 02 55 58.90 & +18 08 15.22 & 173.0 & 99.0 & Y & MBM 11--12, L1457-c \\
G161.43-35.59 & 161.4330 & -35.5935 & 02 59 41.08 & +17 30 58.04 & 02 56 52.65 & +17 19 04.04 & 183.0 & 107.0 & Y & MBM 13, LBN \\
G161.85-35.75 & 161.8505 & -35.7542 & 03 00 26.79 & +17 11 16.08 & 02 57 38.60 & +16 59 24.43 & 163.0 & 95.0 & Y & MBM 13, LBN \\
G162.64-31.67 & 162.6415 & -31.6726 & 03 12 56.11 & +20 04 33.85 & 03 10 04.36 & +19 53 21.59 & 231.0 & 121.0 & Y & MBM 14, LBN \\
G165.91-44.02 & 165.9188 & -44.0278 & 02 50 21.82 & +08 41 43.64 & 02 47 41.31 & +08 29 21.83 & 173.0 & 120.0 & N & DIR 164-44\tablenotemarknew{h} \\
G173.32+31.27 & 173.3203 & 31.2789 & 08 03 20.34 & +46 11 45.20 & 07 59 46.38 & +46 20 12.22 & 274.0 & 142.0 & N & MBM 25, LBN \\
G173.91+31.16 & 173.9135 & 31.1699 & 08 03 07.66 & +45 40 41.82 & 07 59 34.74 & +45 49 08.07 & 290.0 & 150.0 & N & MBM 25, LBN \\
G182.54-25.34 & 182.5488 & -25.3444 & 04 23 04.85 & +12 13 03.09 & 04 20 18.04 & +12 06 06.23 & 365.0 & 156.0 & Y & New \\
G203.57-30.08 & 203.5766 & -30.0861 & 04 47 58.28 & -05 56 06.84 & 04 45 31.19 & -06 01 21.77 & 218.0 & 109.0 & N & New \\
G210.67-36.77 & 210.6738 & -36.7720 & 04 34 00.52 & -14 10 28.57 & 04 31 42.53 & -14 16 40.54 & 145.0 & 87.0 & N & MBM 20 \\
G210.89-36.53 & 210.8935 & -36.5395 & 04 35 10.69 & -14 14 37.73 & 04 32 52.81 & -14 20 44.94 & 145.0 & 87.0 & N & MBM 20 \\
\enddata
\tablecomments{Column(10) indicates, which observational modes are adopted. N for no OTF mapping, only single-point observation. Y for both OTF mapping and single-point observations.}
\tablenotetextnew{\textbf{a}}{Identifications with LDN prefix refers to optical dark nebula identified by \citet{1962ApJS....7....1L}}
\tablenotetextnew{\textbf{b}}{Identifications with MBM prefix refer to optically selected molecular clouds observed with CO by \citet*{1985ApJ...295..402M}}
\tablenotetextnew{\textbf{c}}{Identifications with CB prefix refer to optically selected molecular clouds \citep{1988ApJS...68..257C}.}
\tablenotetextnew{\textbf{d}}{Identifications with LBN prefix refer to bright nebula given in column[8] in \citet{1965ApJS...12..163L}.}
\tablenotetextnew{\textbf{e}}{New CO observation.}
\tablenotetextnew{\textbf{f}}{Identifications with HSVMT prefix refers to \citet{1993AA...268..265H}.}
\tablenotetextnew{\textbf{g}}{L1457 complex \citep{1997ApJ...475..642M}}
\tablenotetextnew{\textbf{h}}{Identifications with DIR prefix refers to the diffuse infrared clouds noted by \citet{1998ApJ...507..507R}.}
\end{deluxetable*}

\begin{deluxetable*}{cccccccccccc}
\renewcommand{\thetable}{\arabic{table}}
\centering
\tablecaption{Parameters of lines at core centers\label{tab:obs-para of cores}}
\tabletypesize{\scriptsize}
\tablehead{
\colhead{Core Name} & \colhead{RA. offset} & \colhead{Dec. offset} & \colhead{V$_{lsr}$($^{12}$CO)} & \colhead{${\rm \Delta V_{^{12}CO}}$} & \colhead{T$_b$($^{12}$CO)} & \colhead{ V$_{lsr}$($^{13}$CO)} & \colhead{${\rm \Delta V_{^{13}CO}}$} & \colhead{T$_b$($^{13}$CO)} & \colhead{ V$_{lsr}$(C$^{18}$O)} & \colhead{${\rm \Delta V_{C^{18}O}}$} & \colhead{T$_b$(C$^{18}$O)} \\
\colhead{} & \colhead{(arcsec)} & \colhead{(arcsec)} & \colhead{(km\ s$^{-1}$)} & \colhead{(km\ s$^{-1}$)} & \colhead{(K)} & \colhead{(km\ s$^{-1}$)} &
\colhead{(km\ s$^{-1}$)} & \colhead{(K)} & \colhead{(km\ s$^{-1}$)} & \colhead{(km\ s$^{-1}$)} & \colhead{(K)} 
}
\decimalcolnumbers
\startdata
\hline
G004.15+35.77a.C1 & 156.0 & -186.0 & 2.33(0.1) & 2.7(0.26) & 7.44(0.69) & 2.57(0.04) & 1.56(0.09) & 5.69(0.3) & 2.77(0.01) & 0.65(0.02) & 1.84(0.0) \\
G004.17+36.67a.C1 & 42.0 & 6.0 & 2.47(0.02) & 0.97(0.05) & 10.12(0.44) & 2.36(0.01) & 0.66(0.02) & 5.95(0.14) & 2.55(0.03) & 0.52(0.06) & 0.7(0.01) \\
G004.54+36.74a.C1 & -12.0 & -24.0 & 2.33(0.02) & 1.1(0.04) & 9.32(0.29) & 2.43(0.01) & 0.75(0.03) & 5.71(0.18) & 2.64(0.03) & 0.56(0.07) & 0.72(0.01) \\
G004.81+37.02a.C1 & 234.0 & -66.0 & 3.12(0.05) & 2.38(0.13) & 9.39(0.45) & 3.28(0.02) & 1.29(0.04) & 5.67(0.16) & 3.48(0.06) & 1.08(0.13) & 0.55(0.01) \\
G006.04+36.74a.C1 & -120.0 & -204.0 & 2.27(0.05) & 2.41(0.11) & 7.83(0.36) & 2.59(0.01) & 1.23(0.04) & 6.42(0.16) & 2.66(0.01) & 0.59(0.03) & 2.47(0.0) \\
G008.43+36.35a.C1 & 54.0 & -84.0 & 1.05(0.03) & 0.84(0.07) & 8.45(0.6) & 0.99(0.01) & 0.52(0.01) & 5.33(0.13) & \nodata & \nodata & \nodata \\
G011.40+36.19a.C1 & -90.0 & 42.0 & 2.37(0.03) & 1.19(0.06) & 8.02(0.38) & 2.42(0.03) & 0.92(0.08) & 3.68(0.28) & \nodata & \nodata & \nodata \\
G101.62-28.84a.C1 & 60.0 & 330.0 & -5.76(0.01) & 1.32(0.03) & 8.78(0.19) & -5.8(0.02) & 0.88(0.06) & 2.8(0.16) & \nodata & \nodata & \nodata \\
G101.62-28.84a.C2 & -78.0 & 144.0 & -5.68(0.01) & 1.33(0.03) & 6.99(0.14) & -5.87(0.01) & 0.71(0.03) & 2.43(0.08) & \nodata & \nodata & \nodata \\
G101.62-28.84a.C3 & -18.0 & 60.0 & -5.89(0.01) & 1.15(0.02) & 8.4(0.12) & -5.86(0.01) & 0.77(0.02) & 3.21(0.08) & \nodata & \nodata & \nodata \\
G101.62-28.84a.C4 & -48.0 & -126.0 & -5.9(0.01) & 1.38(0.03) & 6.22(0.11) & -5.98(0.02) & 0.81(0.04) & 2.28(0.09) & \nodata & \nodata & \nodata \\
G101.62-28.84b.C1 & 150.0 & -90.0 & -4.97(0.03) & 1.88(0.07) & 5.8(0.18) & -4.9(0.03) & 0.95(0.07) & 1.31(0.08) & \nodata & \nodata & \nodata \\
G102.72-25.98a.C1 & 126.0 & -42.0 & -7.33(0.01) & 2.41(0.03) & 8.1(0.09) & -7.32(0.03) & 1.82(0.06) & 2.42(0.07) & \nodata & \nodata & \nodata \\
G145.08-39.30a.C1 & -90.0 & -42.0 & 6.73(0.03) & 1.56(0.07) & 10.16(0.44) & 6.86(0.02) & 1.21(0.05) & 5.47(0.19) & \nodata & \nodata & \nodata \\
G151.58-38.58a.C1 & -114.0 & 18.0 & 4.34(0.03) & 1.01(0.07) & 7.8(0.49) & 4.35(0.02) & 0.67(0.06) & 1.99(0.14) & \nodata & \nodata & \nodata \\
G151.58-38.58b.C1 & -12.0 & 216.0 & 5.41(0.03) & 1.41(0.06) & 7.68(0.28) & 5.41(0.01) & 0.75(0.03) & 3.54(0.12) & \nodata & \nodata & \nodata \\
G151.58-38.58b.C2 & -366.0 & 0.0 & 4.75(0.04) & 1.6(0.09) & 7.47(0.37) & 4.88(0.12) & 1.64(0.28) & 1.49(0.23) & \nodata & \nodata & \nodata \\
G151.58-38.58b.C3 & -60.0 & -6.0 & 5.34(0.11) & 2.46(0.38) & 6.73(0.82) & 5.53(0.03) & 1.03(0.08) & 2.26(0.13) & \nodata & \nodata & \nodata \\
G151.58-38.58b.C4 & 240.0 & -366.0 & 5.15(0.02) & 1.2(0.04) & 8.48(0.26) & 5.05(0.04) & 0.92(0.12) & 2.16(0.21) & \nodata & \nodata & \nodata \\
G153.74+35.91a.C1 & -24.0 & 156.0 & -2.15(0.01) & 1.18(0.04) & 7.91(0.21) & -2.15(0.02) & 0.77(0.03) & 2.81(0.12) & \nodata & \nodata & \nodata \\
G156.42+32.53a.C1 & 18.0 & -12.0 & -0.29(0.04) & 2.19(0.1) & 3.87(0.58) & -0.3(0.06) & 1.49(0.12) & 1.02(0.23) & \nodata & \nodata & \nodata \\
G156.42+32.53a.C2 & 138.0 & -114.0 & -0.52(0.02) & 1.8(0.05) & 3.57(0.17) & -0.51(0.03) & 0.69(0.07) & 1.78(0.19) & \nodata & \nodata & \nodata \\
G156.42+32.53a.C3 & 228.0 & -120.0 & -0.7(0.04) & 1.58(0.09) & 4.49(0.33) & -0.78(0.03) & 0.74(0.07) & 1.49(0.14) & \nodata & \nodata & \nodata \\
G158.77-33.30a.C1 & -144.0 & -138.0 & -2.75(0.03) & 3.87(0.08) & 6.2(0.12) & -2.7(0.02) & 2.38(0.04) & 4.26(0.07) & -2.79(0.17) & 2.36(0.43) & 0.43(0.02) \\
G158.88-34.18a.C1 & -90.0 & 114.0 & -4.93(0.02) & 2.71(0.06) & 6.78(0.11) & -5.28(0.01) & 1.0(0.02) & 5.65(0.12) & \nodata & \nodata & \nodata \\
G158.88-34.18a.C2 & 120.0 & -108.0 & -5.5(0.02) & 3.74(0.07) & 8.04(0.12) & -5.5(0.01) & 1.62(0.03) & 4.5(0.07) & \nodata & \nodata & \nodata \\
G158.97-33.01a.C1 & -192.0 & 90.0 & -2.6(0.01) & 0.78(0.02) & 6.06(0.14) & -2.58(0.01) & 0.53(0.03) & 3.73(0.2) & \nodata & \nodata & \nodata \\
G158.97-33.01a.C2 & 30.0 & -6.0 & -2.75(0.05) & 1.24(0.1) & 5.25(0.4) & -2.76(0.05) & 1.19(0.13) & 2.39(0.22) & \nodata & \nodata & \nodata \\
G158.97-33.01a.C3 & -366.0 & -36.0 & -2.02(0.03) & 1.27(0.09) & 6.45(0.36) & -1.92(0.08) & 1.09(0.21) & 2.15(0.35) & \nodata & \nodata & \nodata \\
G158.97-33.01b.C1 & 114.0 & 324.0 & 0.43(0.04) & 1.44(0.12) & 6.98(0.81) & 0.28(0.02) & 0.82(0.06) & 3.64(0.37) & \nodata & \nodata & \nodata \\
G158.97-33.01b.C2 & -132.0 & -6.0 & 1.56(0.03) & 1.64(0.07) & 6.7(0.25) & 1.57(0.03) & 0.76(0.09) & 2.31(0.22) & \nodata & \nodata & \nodata \\
G158.97-33.01b.C3 & 336.0 & -354.0 & 1.96(0.03) & 1.87(0.06) & 7.21(0.24) & 1.96(0.03) & 1.0(0.05) & 2.7(0.14) & \nodata & \nodata & \nodata \\
G159.23-34.49a.C1 & -294.0 & 126.0 & -4.98(0.08) & 7.01(0.3) & 7.45(0.28) & -5.37(0.07) & 3.08(0.21) & 3.96(0.21) & \nodata & \nodata & \nodata \\
G159.23-34.49a.C2 & 108.0 & 48.0 & -4.17(0.09) & 4.51(0.22) & 8.25(0.39) & -5.01(0.14) & 2.67(0.45) & 5.1(0.53) & -4.89(0.11) & 1.46(0.23) & 0.97(0.02) \\
G159.23-34.49b.C1 & 6.0 & 66.0 & -3.07(0.07) & 2.5(0.17) & 7.62(0.47) & -2.83(0.05) & 1.85(0.12) & 2.24(0.13) & \nodata & \nodata & \nodata \\
G159.41-34.36a.C1 & -66.0 & -120.0 & -5.12(0.03) & 1.66(0.08) & 10.34(0.42) & -5.05(0.03) & 1.25(0.09) & 6.14(0.34) & \nodata & \nodata & \nodata \\
G159.58-32.84a.C1 & -324.0 & 222.0 & 3.08(0.04) & 1.0(0.08) & 2.91(0.23) & 3.1(0.1) & 0.79(0.21) & 0.8(0.2) & \nodata & \nodata & \nodata \\
G159.58-32.84a.C2 & 18.0 & 60.0 & 3.23(0.05) & 2.01(0.08) & 3.44(0.19) & 3.38(0.17) & 1.73(0.23) & 0.98(0.21) & \nodata & \nodata & \nodata \\
G159.58-32.84a.C3 & 60.0 & -72.0 & 4.72(0.03) & 3.99(0.07) & 4.07(2.69) & 2.03(0.08) & 0.84(0.26) & 0.18(0.04) & \nodata & \nodata & \nodata \\
G159.58-32.84b.C1 & 138.0 & 108.0 & 3.92(0.01) & 2.15(0.03) & 3.07(0.03) & 4.4(0.04) & 1.11(0.12) & 0.66(0.06) & \nodata & \nodata & \nodata \\
G159.58-32.84b.C2 & 6.0 & 42.0 & 3.53(0.01) & 2.29(0.03) & 3.92(0.04) & 3.66(0.02) & 1.83(0.06) & 0.96(0.03) & \nodata & \nodata & \nodata \\
G159.67-34.31a.C1 & 6.0 & 54.0 & -1.13(0.01) & 1.95(0.02) & 7.95(0.11) & -1.15(0.01) & 1.38(0.02) & 3.77(0.06) & \nodata & \nodata & \nodata \\
G160.64-35.04a.C1 & -66.0 & 270.0 & -5.07(0.01) & 0.91(0.02) & 7.39(0.13) & -5.07(0.01) & 0.69(0.02) & 3.5(0.11) & \nodata & \nodata & \nodata \\
G160.64-35.04a.C2 & -384.0 & 132.0 & -5.29(0.02) & 1.23(0.05) & 6.48(0.2) & -5.26(0.02) & 0.51(0.05) & 2.9(0.24) & \nodata & \nodata & \nodata \\
G160.64-35.04a.C3 & -246.0 & 30.0 & -4.82(0.02) & 1.38(0.06) & 5.94(0.22) & -4.86(0.03) & 0.88(0.08) & 2.51(0.17) & \nodata & \nodata & \nodata \\
G160.64-35.04a.C4 & -78.0 & -132.0 & -4.72(0.02) & 1.27(0.06) & 7.25(0.27) & -4.68(0.04) & 1.16(0.12) & 2.4(0.19) & \nodata & \nodata & \nodata \\
G160.64-35.04a.C5 & -312.0 & -264.0 & -4.47(0.02) & 1.05(0.04) & 7.5(0.23) & -4.31(0.04) & 0.55(0.1) & 2.22(0.36) & \nodata & \nodata & \nodata \\
G161.43-35.59a.C1 & 78.0 & 186.0 & -5.9(0.02) & 1.37(0.05) & 6.66(0.2) & -5.86(0.01) & 0.94(0.04) & 3.53(0.11) & \nodata & \nodata & \nodata \\
G161.43-35.59a.C2 & 36.0 & -42.0 & -5.9(0.01) & 0.74(0.02) & 10.43(0.22) & -5.85(0.0) & 0.57(0.01) & 7.19(0.13) & \nodata & \nodata & \nodata \\
G161.85-35.75a.C1 & -60.0 & 0.0 & -6.48(0.01) & 1.05(0.03) & 6.67(0.15) & -6.6(0.02) & 0.76(0.04) & 4.3(0.18) & -6.74(0.03) & 0.58(0.07) & 0.8(0.01) \\
G162.64-31.67a.C1 & -30.0 & 96.0 & -1.05(0.03) & 0.84(0.06) & 6.48(0.51) & -1.02(0.05) & 0.79(0.2) & 1.3(0.21) & \nodata & \nodata & \nodata \\
G162.64-31.67a.C2 & 18.0 & -0.0 & -1.06(0.03) & 0.9(0.06) & 8.13(0.58) & -1.02(0.02) & 0.59(0.05) & 1.71(0.12) & \nodata & \nodata & \nodata \\
G162.64-31.67a.C3 & -66.0 & -66.0 & -0.92(0.05) & 0.99(0.11) & 5.94(0.67) & -0.94(0.02) & 0.58(0.04) & 1.99(0.13) & \nodata & \nodata & \nodata \\
G182.54-25.34a.C1 & 24.0 & 48.0 & 0.72(0.04) & 2.54(0.09) & 7.08(0.43) & 0.67(0.05) & 1.74(0.11) & 2.33(0.2) & \nodata & \nodata & \nodata \\
\enddata
\tablecomments{All the cores are shown in the table.}
\end{deluxetable*}

\begin{deluxetable*}{ccccc cc}
\renewcommand{\thetable}{\arabic{table}}
\tablecaption{Parameters derived at the core center\label{tab:deri-para of cores}}
\tablewidth{0pt}
\tablehead{
\colhead{Core Name} & \colhead{T$_{\mathrm{ex}}$} & \colhead{$\tau_{13}$} &  \colhead{N$_{\mathrm{H}_2}$} &
\colhead{$\sigma_{\mathrm{Therm}}$} & \colhead{$\sigma_{\mathrm{NT}}$} & \colhead{$\sigma_{3\mathrm{D}}$}  \\
\colhead{} & \colhead{(K)} & \colhead{} &  \colhead{(10$^{21}$ cm$^{-2}$)} &
\colhead{(km\ s$^{-1}$)} & \colhead{(km\ s$^{-1}$)} & \colhead{(km\ s$^{-1}$)}
}
\decimalcolnumbers
\startdata
\hline
G004.15+35.77a.C1 & 10.8(0.7) & 1.45(0.35) & 15.9(2.48) & 0.2(0.01) & 0.66(0.04) & 1.19(0.06) \\
G004.17+36.67a.C1 & 13.5(0.4) & 0.89(0.07) & 6.11(0.31) & 0.22(0.0) & 0.27(0.01) & 0.61(0.01) \\
G004.54+36.74a.C1 & 12.7(0.3) & 0.95(0.07) & 6.66(0.4) & 0.21(0.0) & 0.31(0.01) & 0.66(0.02) \\
G004.81+37.02a.C1 & 12.8(0.5) & 0.93(0.08) & 11.3(0.65) & 0.21(0.0) & 0.54(0.02) & 1.01(0.03) \\
G006.04+36.74a.C1 & 11.2(0.4) & 1.71(0.24) & 15.8(1.53) & 0.2(0.0) & 0.52(0.02) & 0.96(0.03) \\
G008.43+36.35a.C1 & 11.8(0.6) & 1.0(0.13) & 4.28(0.28) & 0.2(0.01) & 0.21(0.0) & 0.51(0.01) \\
G011.40+36.19a.C1 & 11.4(0.4) & 0.61(0.08) & 4.37(0.53) & 0.2(0.0) & 0.39(0.03) & 0.76(0.05) \\
G101.62-28.84a.C1 & 12.2(0.2) & 0.38(0.03) & 2.93(0.26) & 0.21(0.0) & 0.37(0.03) & 0.73(0.04) \\
G101.62-28.84a.C2 & 10.4(0.1) & 0.43(0.02) & 1.99(0.11) & 0.19(0.0) & 0.3(0.01) & 0.61(0.02) \\
G101.62-28.84a.C3 & 11.8(0.1) & 0.48(0.02) & 3.04(0.11) & 0.2(0.0) & 0.32(0.01) & 0.66(0.01) \\
G101.62-28.84a.C4 & 9.6(0.1) & 0.46(0.02) & 2.12(0.14) & 0.18(0.0) & 0.34(0.02) & 0.67(0.03) \\
G101.62-28.84b.C1 & 9.1(0.2) & 0.26(0.02) & 1.29(0.12) & 0.18(0.0) & 0.4(0.03) & 0.76(0.05) \\
G102.72-25.98a.C1 & 11.5(0.1) & 0.36(0.01) & 5.06(0.23) & 0.2(0.0) & 0.77(0.03) & 1.38(0.04) \\
G145.08-39.30a.C1 & 13.6(0.4) & 0.77(0.06) & 9.81(0.61) & 0.22(0.0) & 0.51(0.02) & 0.96(0.03) \\
G151.58-38.58a.C1 & 11.2(0.5) & 0.3(0.03) & 1.48(0.17) & 0.2(0.0) & 0.28(0.03) & 0.59(0.04) \\
G151.58-38.58b.C1 & 11.1(0.3) & 0.62(0.04) & 3.4(0.19) & 0.2(0.0) & 0.31(0.01) & 0.64(0.02) \\
G151.58-38.58b.C2 & 10.8(0.4) & 0.22(0.04) & 2.59(0.6) & 0.2(0.0) & 0.69(0.12) & 1.25(0.2) \\
G151.58-38.58b.C3 & 10.1(0.8) & 0.41(0.07) & 2.65(0.27) & 0.19(0.01) & 0.43(0.03) & 0.82(0.05) \\
G151.58-38.58b.C4 & 11.9(0.3) & 0.29(0.03) & 2.24(0.37) & 0.2(0.0) & 0.39(0.05) & 0.76(0.08) \\
G153.74+35.91a.C1 & 11.3(0.2) & 0.44(0.03) & 2.57(0.15) & 0.2(0.0) & 0.32(0.01) & 0.66(0.02) \\
G156.42+32.53a.C1 & 7.1(0.6) & 0.31(0.1) & 1.62(0.4) & 0.16(0.01) & 0.63(0.05) & 1.13(0.09) \\
G156.42+32.53a.C2 & 6.8(0.2) & 0.69(0.12) & 1.58(0.25) & 0.16(0.0) & 0.29(0.03) & 0.57(0.05) \\
G156.42+32.53a.C3 & 7.8(0.3) & 0.4(0.06) & 1.22(0.17) & 0.17(0.0) & 0.31(0.03) & 0.61(0.05) \\
G158.77-33.30a.C1 & 9.5(0.1) & 1.16(0.06) & 15.8(0.52) & 0.18(0.0) & 1.01(0.02) & 1.78(0.03) \\
G158.88-34.18a.C1 & 10.1(0.1) & 1.79(0.13) & 11.3(0.64) & 0.19(0.0) & 0.42(0.01) & 0.8(0.01) \\
G158.88-34.18a.C2 & 11.4(0.1) & 0.82(0.03) & 10.3(0.28) & 0.2(0.0) & 0.69(0.01) & 1.24(0.02) \\
G158.97-33.01a.C1 & 9.4(0.1) & 0.96(0.09) & 2.82(0.25) & 0.18(0.0) & 0.22(0.01) & 0.49(0.02) \\
G158.97-33.01a.C2 & 8.6(0.4) & 0.61(0.1) & 3.45(0.52) & 0.17(0.0) & 0.5(0.06) & 0.92(0.09) \\
G158.97-33.01a.C3 & 9.8(0.4) & 0.4(0.09) & 2.64(0.68) & 0.19(0.0) & 0.46(0.09) & 0.86(0.14) \\
G158.97-33.01b.C1 & 10.3(0.8) & 0.74(0.17) & 3.96(0.58) & 0.19(0.01) & 0.34(0.03) & 0.68(0.04) \\
G158.97-33.01b.C2 & 10.1(0.3) & 0.42(0.05) & 2.01(0.31) & 0.19(0.0) & 0.32(0.04) & 0.64(0.06) \\
G158.97-33.01b.C3 & 10.6(0.2) & 0.47(0.04) & 3.19(0.24) & 0.19(0.0) & 0.42(0.02) & 0.8(0.03) \\
G159.23-34.49a.C1 & 10.8(0.3) & 0.76(0.07) & 16.5(1.53) & 0.2(0.0) & 1.31(0.09) & 2.29(0.15) \\
G159.23-34.49a.C2 & 11.6(0.4) & 0.96(0.18) & 20.6(4.39) & 0.2(0.0) & 1.13(0.19) & 1.99(0.33) \\
G159.23-34.49b.C1 & 11.0(0.5) & 0.35(0.04) & 4.68(0.42) & 0.2(0.0) & 0.78(0.05) & 1.4(0.09) \\
G159.41-34.36a.C1 & 13.8(0.4) & 0.9(0.1) & 12.1(1.23) & 0.22(0.0) & 0.53(0.04) & 0.99(0.06) \\
G159.58-32.84a.C1 & 6.1(0.2) & 0.32(0.1) & 0.71(0.26) & 0.15(0.0) & 0.33(0.09) & 0.63(0.14) \\
G159.58-32.84a.C2 & 6.7(0.2) & 0.34(0.09) & 1.86(0.48) & 0.15(0.0) & 0.73(0.1) & 1.3(0.17) \\
G159.58-32.84a.C3 & 7.3(2.8) & 0.05(0.03) & 0.14(0.05) & 0.16(0.03) & 0.35(0.11) & 0.67(0.18) \\
G159.58-32.84b.C1 & 6.3(0.0) & 0.24(0.02) & 0.79(0.11) & 0.15(0.0) & 0.47(0.05) & 0.85(0.08) \\
G159.58-32.84b.C2 & 7.2(0.0) & 0.28(0.01) & 1.85(0.08) & 0.16(0.0) & 0.78(0.03) & 1.37(0.04) \\
G159.67-34.31a.C1 & 11.3(0.1) & 0.64(0.02) & 6.79(0.16) & 0.2(0.0) & 0.58(0.01) & 1.07(0.01) \\
G160.64-35.04a.C1 & 10.8(0.1) & 0.64(0.03) & 3.1(0.14) & 0.2(0.0) & 0.29(0.01) & 0.6(0.01) \\
G160.64-35.04a.C2 & 9.8(0.2) & 0.59(0.07) & 1.82(0.24) & 0.19(0.0) & 0.21(0.02) & 0.49(0.03) \\
G160.64-35.04a.C3 & 9.3(0.2) & 0.55(0.06) & 2.63(0.31) & 0.18(0.0) & 0.37(0.03) & 0.71(0.05) \\
G160.64-35.04a.C4 & 10.6(0.3) & 0.4(0.04) & 3.19(0.42) & 0.19(0.0) & 0.49(0.05) & 0.91(0.08) \\
G160.64-35.04a.C5 & 10.9(0.2) & 0.35(0.07) & 1.38(0.34) & 0.2(0.0) & 0.23(0.04) & 0.52(0.06) \\
G161.43-35.59a.C1 & 10.0(0.2) & 0.76(0.05) & 4.4(0.25) & 0.19(0.0) & 0.4(0.02) & 0.76(0.03) \\
G161.43-35.59a.C2 & 13.9(0.2) & 1.17(0.06) & 7.24(0.26) & 0.22(0.0) & 0.23(0.0) & 0.56(0.01) \\
G161.85-35.75a.C1 & 10.0(0.2) & 1.03(0.09) & 4.88(0.37) & 0.19(0.0) & 0.32(0.02) & 0.64(0.03) \\
G162.64-31.67a.C1 & 9.8(0.5) & 0.22(0.05) & 1.06(0.32) & 0.19(0.0) & 0.33(0.09) & 0.66(0.13) \\
G162.64-31.67a.C2 & 11.5(0.6) & 0.24(0.03) & 1.1(0.12) & 0.2(0.01) & 0.24(0.02) & 0.55(0.03) \\
G162.64-31.67a.C3 & 9.3(0.7) & 0.41(0.07) & 1.29(0.13) & 0.18(0.01) & 0.24(0.02) & 0.52(0.03) \\
G182.54-25.34a.C1 & 10.4(0.4) & 0.4(0.05) & 4.62(0.51) & 0.19(0.0) & 0.74(0.05) & 1.32(0.08) \\
\enddata
\tablecomments{All of the cores are shown in the table.}
\end{deluxetable*}

\begin{deluxetable*}{cccccccccc}
\renewcommand{\thetable}{\arabic{table}}
\tablecaption{Emission Region Parameters\label{tab:region-para}}
\tablewidth{0pt}
\tablehead{
\colhead{Core Name} & \colhead{RA offset} &
\colhead{Dec offset} & \colhead{Distance} & \colhead{R} &  \colhead{n$_{vol}$}  &
\colhead{$M_{\mathrm{LTE}}$} & \colhead{$M_{\mathrm{Jeans}}$}  &
\colhead{$M_{\mathrm{vir}}$} & \colhead{$\alpha_{\mathrm{vir}}$}\\
\colhead{} & \colhead{(arcsec)} &
\colhead{(arcsec)} & \colhead{(kpc)} & \colhead{(pc)} & \colhead{(10$^3$ cm$^{-3}$)} &
\colhead{(M$_{\odot}$)} & \colhead{(M$_{\odot}$)}  &
\colhead{(M$_{\odot}$)} & \colhead{}
}
\decimalcolnumbers
\startdata
\hline
G004.15+35.77a.C1 & 156.0 & -186.0 & 0.145(0.02) & 0.12(0.02) & 21.7(4.55) & 8.85(2.86) & 11.8(2.25) & 65.7(65.7) & 7.42(1.75) \\
G004.17+36.67a.C1 & 42.0 & 6.0 & 0.13(0.022) & 0.04(0.01) & 24.5(4.42) & 0.39(0.14) & 1.46(0.16) & 5.76(5.76) & 14.8(2.74) \\
G004.54+36.74a.C1 & -12.0 & -24.0 & 0.13(0.019) & 0.05(0.01) & 23.8(3.76) & 0.53(0.16) & 1.86(0.22) & 7.52(7.52) & 14.1(2.37) \\
G004.81+37.02a.C1 & 234.0 & -66.0 & 0.13(0.019) & 0.09(0.01) & 20.3(3.19) & 3.58(1.07) & 7.43(0.84) & 35.8(35.8) & 10.0(1.66) \\
G006.04+36.74a.C1 & -120.0 & -204.0 & 0.122(0.018) & 0.1(0.01) & 26.2(4.61) & 5.95(1.85) & 5.64(0.7) & 35.3(35.3) & 5.92(1.1) \\
G008.43+36.35a.C1 & 54.0 & -84.0 & 0.13(0.022) & 0.07(0.01) & 9.64(1.78) & 0.86(0.3) & 1.39(0.15) & 7.3(7.3) & 8.44(1.58) \\
G011.40+36.19a.C1 & -90.0 & 42.0 & 0.13(0.019) & 0.04(0.01) & 19.2(3.65) & 0.23(0.07) & 3.16(0.73) & 8.12(8.12) & 35.1(8.27) \\
G101.62-28.84a.C1 & 60.0 & 330.0 & 0.244(0.064) & 0.04(0.01) & 11.3(3.12) & 0.2(0.11) & 3.78(0.8) & 8.74(8.74) & 43.4(12.8) \\
G101.62-28.84a.C2 & -78.0 & 144.0 & 0.244(0.064) & $<$0.03(0.01) & $>$11.8(3.13) & $<$0.06(0.03) & $<$2.15(0.35) & $<$3.97(3.97) & $>$68.0(18.6) \\
G101.62-28.84a.C3 & -18.0 & 60.0 & 0.244(0.064) & 0.04(0.01) & 13.8(3.62) & 0.15(0.08) & 2.5(0.36) & 6.04(6.04) & 39.8(10.6) \\
G101.62-28.84a.C4 & -48.0 & -126.0 & 0.244(0.064) & 0.03(0.01) & 10.3(2.77) & 0.09(0.05) & 3.02(0.54) & 5.79(5.79) & 62.9(17.6) \\
G101.62-28.84b.C1 & 150.0 & -90.0 & 0.244(0.064) & 0.04(0.01) & 5.61(1.56) & 0.07(0.04) & 5.98(1.39) & 8.36(8.36) & 119.0(36.1) \\
G102.72-25.98a.C1 & 126.0 & -42.0 & 0.259(0.045) & 0.11(0.02) & 7.49(1.34) & 2.37(0.83) & 31.0(4.0) & 80.8(80.8) & 34.1(6.47) \\
G145.08-39.30a.C1 & -90.0 & -42.0 & 0.163(0.034) & 0.05(0.01) & 35.3(7.69) & 0.78(0.33) & 4.83(0.74) & 16.1(16.1) & 20.7(4.75) \\
G151.58-38.58a.C1 & -114.0 & 18.0 & 0.154(0.054) & 0.02(0.01) & 9.93(3.64) & 0.03(0.02) & 2.14(0.56) & 3.29(3.29) & 98.2(38.0) \\
G151.58-38.58b.C1 & -12.0 & 216.0 & 0.154(0.054) & 0.02(0.01) & 27.5(9.68) & 0.05(0.04) & 1.63(0.32) & 3.2(3.2) & 60.0(21.4) \\
G151.58-38.58b.C2 & -366.0 & 0.0 & 0.154(0.054) & $<$0.02(0.01) & $>$25.3(10.6) & $<$0.03(0.02) & $<$12.5(6.52) & $<$10.0(10.0) & $>$360.0(189.0) \\
G151.58-38.58b.C3 & -60.0 & -6.0 & 0.154(0.054) & 0.04(0.01) & 11.3(4.1) & 0.15(0.1) & 5.29(1.43) & 9.88(9.88) & 66.5(25.7) \\
G151.58-38.58b.C4 & 240.0 & -366.0 & 0.154(0.054) & 0.02(0.01) & 16.1(6.2) & 0.04(0.03) & 3.49(1.28) & 5.01(5.01) & 113.0(49.2) \\
G153.74+35.91a.C1 & -24.0 & 156.0 & 0.365(0.032) & 0.1(0.01) & 4.25(0.46) & 0.96(0.18) & 4.43(0.45) & 16.4(16.4) & 17.0(2.07) \\
G156.42+32.53a.C1 & 18.0 & -12.0 & 0.29(0.044) & $<$0.02(0.0) & $>$10.6(3.05) & $<$0.04(0.01) & $<$14.2(3.82) & $<$12.2(12.2) & $>$315.0(102.0) \\
G156.42+32.53a.C2 & 138.0 & -114.0 & 0.29(0.044) & $<$0.03(0.0) & $>$10.1(2.18) & $<$0.04(0.01) & $<$1.88(0.5) & $<$3.19(3.19) & $>$80.1(21.6) \\
G156.42+32.53a.C3 & 228.0 & -120.0 & 0.29(0.044) & $<$0.02(0.0) & $>$8.66(1.75) & $<$0.02(0.01) & $<$2.49(0.62) & $<$3.29(3.29) & $>$133.0(33.5) \\
G158.77-33.30a.C1 & -144.0 & -138.0 & 0.205(0.051) & 0.11(0.03) & 23.4(5.87) & 7.42(3.7) & 37.4(5.04) & 134.0(134.0) & 18.1(4.57) \\
G158.88-34.18a.C1 & -90.0 & 114.0 & 0.231(0.048) & 0.06(0.01) & 28.6(6.21) & 1.84(0.78) & 3.09(0.37) & 16.0(16.0) & 8.7(1.91) \\
G158.88-34.18a.C2 & 120.0 & -108.0 & 0.231(0.048) & 0.09(0.02) & 18.1(3.84) & 3.43(1.44) & 14.4(1.69) & 54.7(54.7) & 16.0(3.42) \\
G158.97-33.01a.C1 & -192.0 & 90.0 & 0.274(0.076) & 0.04(0.01) & 13.0(3.81) & 0.14(0.08) & 1.08(0.2) & 3.32(3.32) & 24.3(7.33) \\
G158.97-33.01a.C2 & 30.0 & -6.0 & 0.274(0.076) & 0.06(0.02) & 9.98(3.16) & 0.42(0.24) & 8.0(2.68) & 18.5(18.5) & 43.5(16.2) \\
G158.97-33.01a.C3 & -366.0 & -36.0 & 0.274(0.076) & $<$0.03(0.01) & $>$15.2(5.77) & $<$0.08(0.05) & $<$5.25(2.82) & $<$8.04(8.04) & $>$98.6(49.9) \\
G158.97-33.01b.C1 & 114.0 & 324.0 & 0.274(0.076) & 0.05(0.01) & 13.1(4.14) & 0.37(0.21) & 2.82(0.66) & 8.79(8.79) & 23.9(8.02) \\
G158.97-33.01b.C2 & -132.0 & -6.0 & 0.274(0.076) & $<$0.03(0.01) & $>$10.0(3.2) & $<$0.08(0.05) & $<$2.68(0.84) & $<$5.16(5.16) & $>$62.7(23.0) \\
G158.97-33.01b.C3 & 336.0 & -354.0 & 0.274(0.076) & 0.05(0.01) & 10.8(3.11) & 0.29(0.16) & 5.09(0.98) & 12.0(12.0) & 41.7(12.5) \\
G159.23-34.49a.C1 & -294.0 & 126.0 & 0.244(0.064) & 0.07(0.02) & 37.8(10.4) & 3.24(1.71) & 62.9(15.3) & 144.0(144.0) & 44.4(13.6) \\
G159.23-34.49a.C2 & 108.0 & 48.0 & 0.244(0.064) & 0.11(0.03) & 30.0(10.1) & 9.96(5.6) & 46.6(24.2) & 171.0(171.0) & 17.2(8.06) \\
G159.23-34.49b.C1 & 6.0 & 66.0 & 0.244(0.064) & 0.08(0.02) & 10.1(2.77) & 1.04(0.55) & 27.9(6.41) & 57.3(57.3) & 55.0(16.6) \\
G159.41-34.36a.C1 & -66.0 & -120.0 & 0.205(0.068) & 0.08(0.03) & 23.1(7.97) & 3.39(2.26) & 6.51(1.65) & 32.2(32.2) & 9.48(3.47) \\
G159.58-32.84a.C1 & -324.0 & 222.0 & 0.29(0.091) & 0.06(0.02) & 2.09(1.01) & 0.08(0.06) & 5.58(4.02) & 8.47(8.47) & 101.0(66.6) \\
G159.58-32.84a.C2 & 18.0 & 60.0 & 0.29(0.091) & 0.05(0.02) & 6.15(2.49) & 0.18(0.12) & 28.4(12.3) & 32.0(32.0) & 183.0(87.7) \\
G159.58-32.84a.C3 & 60.0 & -72.0 & 0.29(0.091) & $<$0.03(0.01) & $>$0.67(0.34) & $<$0.01(0.0) & $<$12.0(9.91) & $<$5.97(5.97) & $>$935.0(676.0) \\
G159.58-32.84b.C1 & 138.0 & 108.0 & 0.29(0.091) & $<$0.02(0.01) & $>$5.14(1.77) & $<$0.02(0.01) & $<$8.84(3.03) & $<$6.98(6.98) & $>$372.0(148.0) \\
G159.58-32.84b.C2 & 6.0 & 42.0 & 0.29(0.091) & 0.06(0.02) & 4.64(1.47) & 0.3(0.19) & 38.7(7.14) & 47.1(47.1) & 156.0(50.5) \\
G159.67-34.31a.C1 & 6.0 & 54.0 & 0.194(0.056) & 0.07(0.02) & 15.2(4.38) & 1.38(0.79) & 10.1(1.5) & 31.9(31.9) & 23.1(6.65) \\
G160.64-35.04a.C1 & -66.0 & 270.0 & 0.173(0.032) & 0.03(0.01) & 14.6(2.78) & 0.14(0.05) & 1.85(0.21) & 4.84(4.84) & 33.6(6.55) \\
G160.64-35.04a.C2 & -384.0 & 132.0 & 0.173(0.032) & $<$0.02(0.0) & $>$15.3(3.49) & $<$0.03(0.01) & $<$0.95(0.2) & $<$1.76(1.76) & $>$67.2(17.2) \\
G160.64-35.04a.C3 & -246.0 & 30.0 & 0.173(0.032) & 0.04(0.01) & 12.1(2.65) & 0.13(0.05) & 3.37(0.84) & 6.94(6.94) & 54.5(14.4) \\
G160.64-35.04a.C4 & -78.0 & -132.0 & 0.173(0.032) & 0.04(0.01) & 12.3(2.8) & 0.22(0.09) & 6.98(2.06) & 13.6(13.6) & 61.4(17.8) \\
G160.64-35.04a.C5 & -312.0 & -264.0 & 0.173(0.032) & $<$0.01(0.0) & $>$16.8(5.18) & $<$0.01(0.0) & $<$1.1(0.4) & $<$1.39(1.39) & $>$147.0(55.5) \\
G161.43-35.59a.C1 & 78.0 & 186.0 & 0.183(0.042) & 0.04(0.01) & 18.5(4.42) & 0.26(0.12) & 3.28(0.52) & 8.61(8.61) & 33.6(8.39) \\
G161.43-35.59a.C2 & 36.0 & -42.0 & 0.183(0.042) & 0.06(0.01) & 19.0(4.47) & 1.08(0.5) & 1.28(0.16) & 7.44(7.44) & 6.9(1.63) \\
G161.85-35.75a.C1 & -60.0 & 0.0 & 0.163(0.036) & 0.04(0.01) & 20.2(4.73) & 0.29(0.13) & 1.89(0.32) & 6.22(6.22) & 21.3(5.26) \\
G162.64-31.67a.C1 & -30.0 & 96.0 & 0.231(0.048) & $<$0.02(0.0) & $>$9.08(3.31) & $<$0.01(0.01) & $<$3.06(1.89) & $<$3.19(3.19) & $>$213.0(115.0) \\
G162.64-31.67a.C2 & 18.0 & -0.0 & 0.231(0.048) & $<$0.02(0.0) & $>$9.21(2.16) & $<$0.02(0.01) & $<$1.76(0.35) & $<$2.25(2.25) & $>$141.0(36.4) \\
G162.64-31.67a.C3 & -66.0 & -66.0 & 0.231(0.048) & $<$0.02(0.0) & $>$9.35(2.14) & $<$0.03(0.01) & $<$1.5(0.28) & $<$2.36(2.36) & $>$93.7(23.2) \\
G182.54-25.34a.C1 & 24.0 & 48.0 & 0.365(0.106) & 0.05(0.01) & 15.5(4.84) & 0.42(0.25) & 18.8(4.45) & 32.5(32.5) & 77.4(25.8) \\
\enddata
\tablecomments{All of the cores are shown in the table.}
\end{deluxetable*}

\clearpage
\appendix
\section{Spectral Energy Distribution Fitting} \label{appendix:sed}
In order to be able to estimate physical properties from Herschel multiwavelength observations (160, 250, 350 and 500 $\mu m$) with different angular resolutions (12, 18, 25 and 36 arcsec), we convolved the images to a common angular resolution of $40\arcsec$ which is slightly larger than the poorest resolution among the four bands. The \textit{convolution} package of \textit{Astropy} \citep{2013Robitaille} was used with a Gaussian kernel of $\sqrt{\theta_0^2-\theta_\lambda^2}$ where $\theta_0 = 40\arcsec$ and $\theta_\lambda$ is the HPBW beam size for a given band. The convolved data were then regridded to be aligned pixel by pixel with a common pixel size of $10\arcsec$.

A modified blackbody model is applied to these data:
\begin{eqnarray}
	I_\nu = B_\nu(T)(1-e^{-\tau_\nu})
\end{eqnarray}
where optical depth is given by:
\begin{eqnarray}
	\tau_\nu = \mu_{\text{H}_2}m_{\text{H}}\kappa_\nu N_{\text{H}_2}/R_{\text{gd}}
\end{eqnarray}
\textbf{Here $\mu_{\text{H}_2} = 2.33$ is the mean molecular weight} adopted from \citet{2008A&A...487..993K}, $m_{\text{H}}$ is the mass of a hydrogen atom,  $N_{\text{H}_2}$ is the $\text{H}_2$ column density, and $R_{\text{gd}} = 100$ is the gas-to-dust ratio. The dust opacity $\kappa_\nu$ can be expressed as a power law in frequency \citep{1994A&A...291..943O},
\begin{eqnarray}
	\kappa_\nu = 3.33\left(\frac{\nu}{600\ (GHz)}\right)^\beta\ \text{cm}^2\text{g}^{-1}
\end{eqnarray}
where the power law index $\beta$ is assumed to be 2.0 in agreement with the standard value for cold dust emission \citep{1983QJRAS..24..267H}.
We filtrate out the pixel as long as any one of its four bands have non-positive value. Then the temperature and column density will be set null.

\begin{figure*}[!h]
\centering
\subfigure[]{
    \begin{minipage}[b]{0.4\textwidth}
    \includegraphics[width=1.0\textwidth]{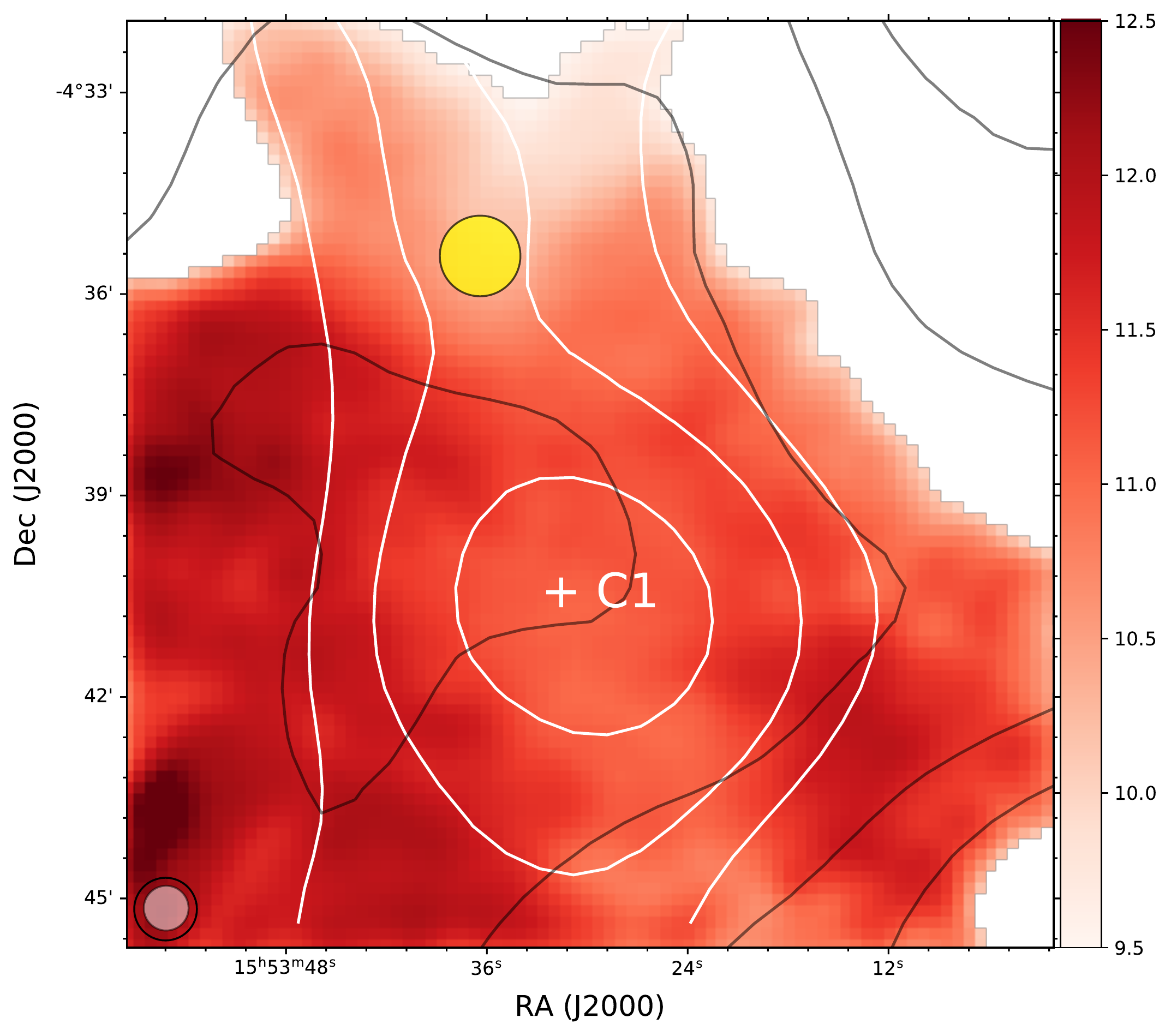}
    \end{minipage}
}%
\subfigure[]{
    \begin{minipage}[b]{0.4\textwidth}
    \includegraphics[width=1.0\textwidth]{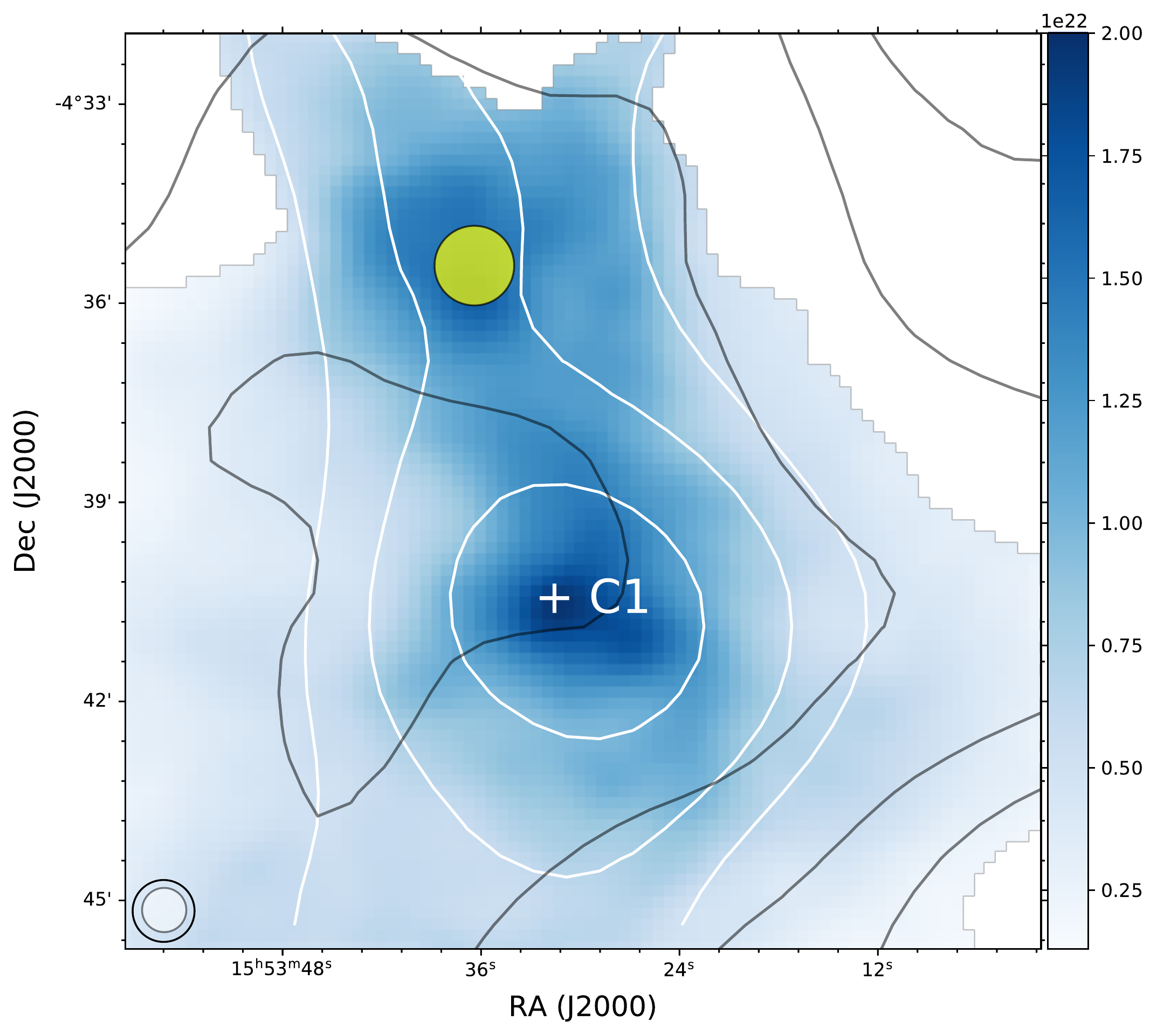}
    \end{minipage}
}%
\caption{(a) The dust temperature map of G004.15+35.77a is shown as background Reds colormap with linear scale between 9.5 and 12.5 K. The median value is 11.3 K. Part of null-value pixels are filled with pure white. (b) The column density map is shown as background blue colormap with linear scale between $1.3\times 10^{21}$ and $2.0\times 10^{22}\ \text{cm}^{-2}$. Both two maps contain the same group of contours and markers: 1) black contours: smoothed integrated $^{13}$CO intensity between velocity range of 0.22-0.40 km s$^{-1}$ with levels of 1.5, 3.1, 4.6, 6.2, 7.8 K km s$^{-1}$. 2) white contours: smoothed integrated C$^{18}$O intensity between velocity range of 0.22-0.40 km s$^{-1}$ with levels of 1.0, 1.3, 1.7, 2.0, 2.3, 2.6 K km s$^{-1}$. 3) Filled yellow circle: the center of the ammonia dense core (L134-A) identified by \citet{1983ApJ...264..517M} and the size of the circle gives the beamwidth $1\fdg2$ of NRAO 37-m Haystack Observatory. 4) White cross: the peak of C$^{18}$O emission, defined as the center of dense core G004.15+35.77a.C1 hereafter. 5) White filled circle: the beamwidth $40\arcsec$ of background dust map after the convolution of Herschel data. 6) black circle: the beamwidth $55\arcsec$ of PMO 13.7-m $^{13}$CO and C$^{18}$O observations.}
\label{fig:L134}
\end{figure*}

The column density and dust temperature of G004.15+35.77 are presented in Figure \ref{fig:L134}. A clear anti-correlation between the distributions of column density and temperature: the regions with lower temperature have higher column density. Along NE-SW direction, there is a low-temperature ($\leq$ 11 K ) valley where the column density reach its ridge ($\geq 10^{22}$ cm$^{-2}$). From Planck 857 GHz dust map \citep{2011A&A...536A..23P}, the clump G004.15+35.77 is at the tip of L134 and both the Northeast and Northwest part of the clump are on the edge of L134. The clump are exposed to Galactic starlight and the grains are heated not by local sources but by background emission instead \citep[also found in Ursa Major complex]{1987ApJ...319..723D}, which might explain the temperature gradient along West-East direction.


\clearpage
\section{Three CO-lines integrated intensity maps}\label{appendix:map}
\begin{figure*}[!h]
    \centering
    \includegraphics[width=1.0\linewidth,angle=0]{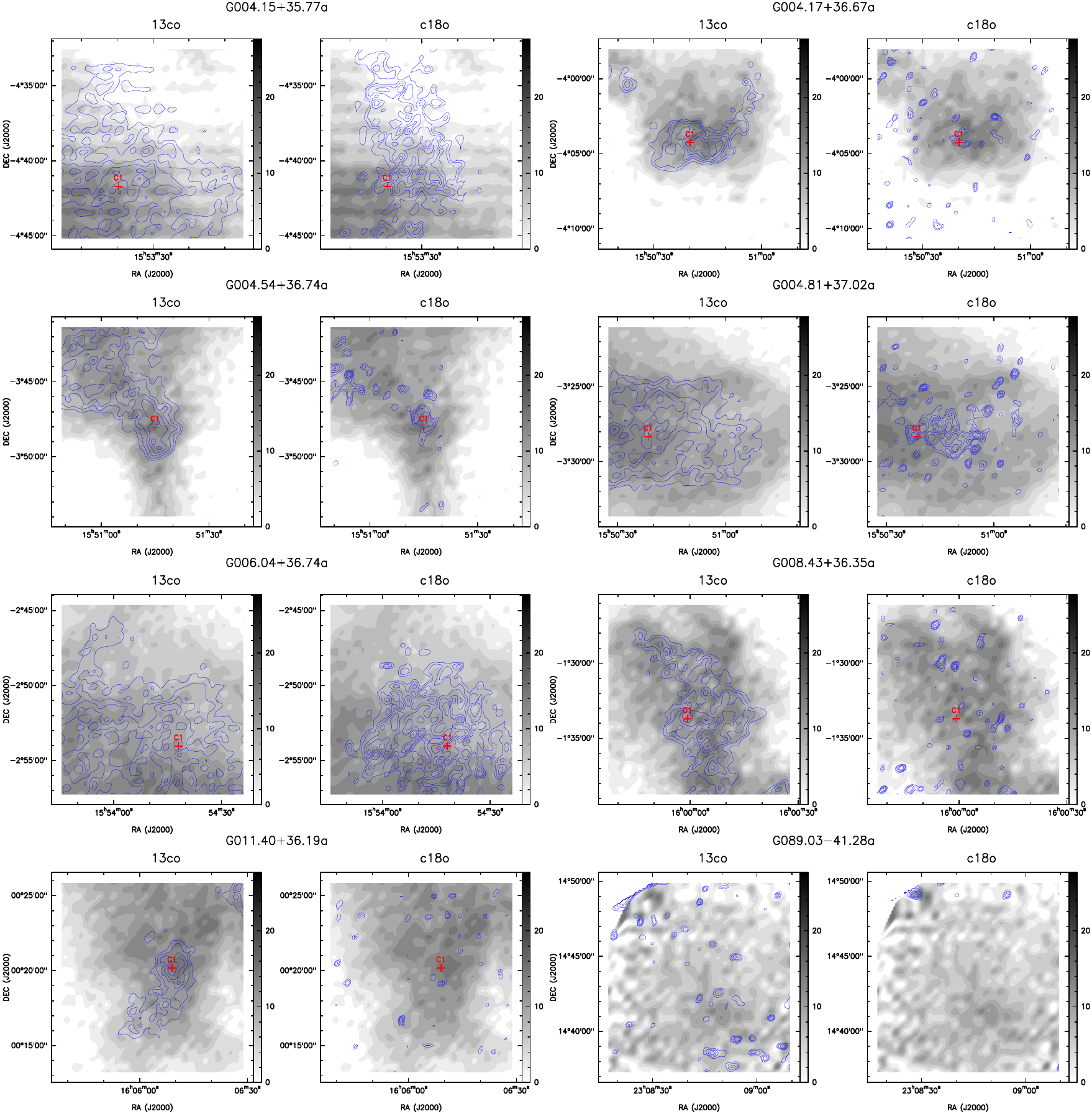}
	 \caption{CO integrated intensity maps for total 33 velocity-defined clumps. The background grey color maps show $^{12}$CO(1-0) emission and the colorbars show the value of W($^{12}$CO) in units of [K km s$^{-1}$]. The blue contours of $^{13}$CO(1-0) or C$^{18}$O(1-0) step from 50\% to 90\% by 10\% of the peak value. Defined cores are marked by red crosses and named `C1', `C2', and so on in the order of left-to-right and up-to-down. The names of CO emission lines are labeled on the top of each panel. Source names are labeled above each two panels.}
	\label{fig:map1}
\end{figure*}
\addtocounter{figure}{-1}

\begin{figure*}
    \centering
    \caption{continued}
    \includegraphics[width=1.0\linewidth,angle=0]{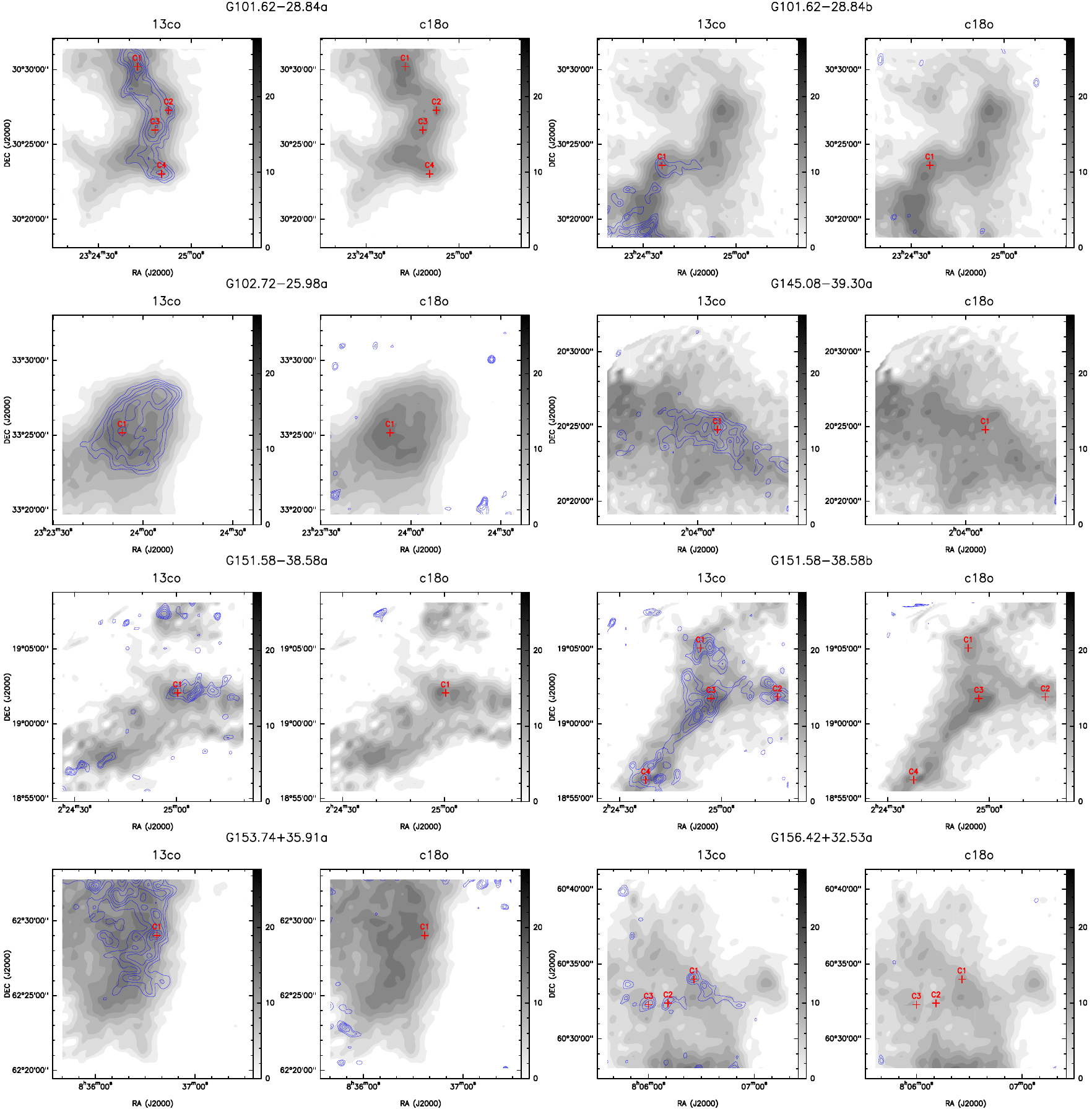}
	\label{fig:map2}
\end{figure*}
\addtocounter{figure}{-1}

\begin{figure*}
    \centering
    \caption{continued}
    \includegraphics[width=1.0\linewidth,angle=0]{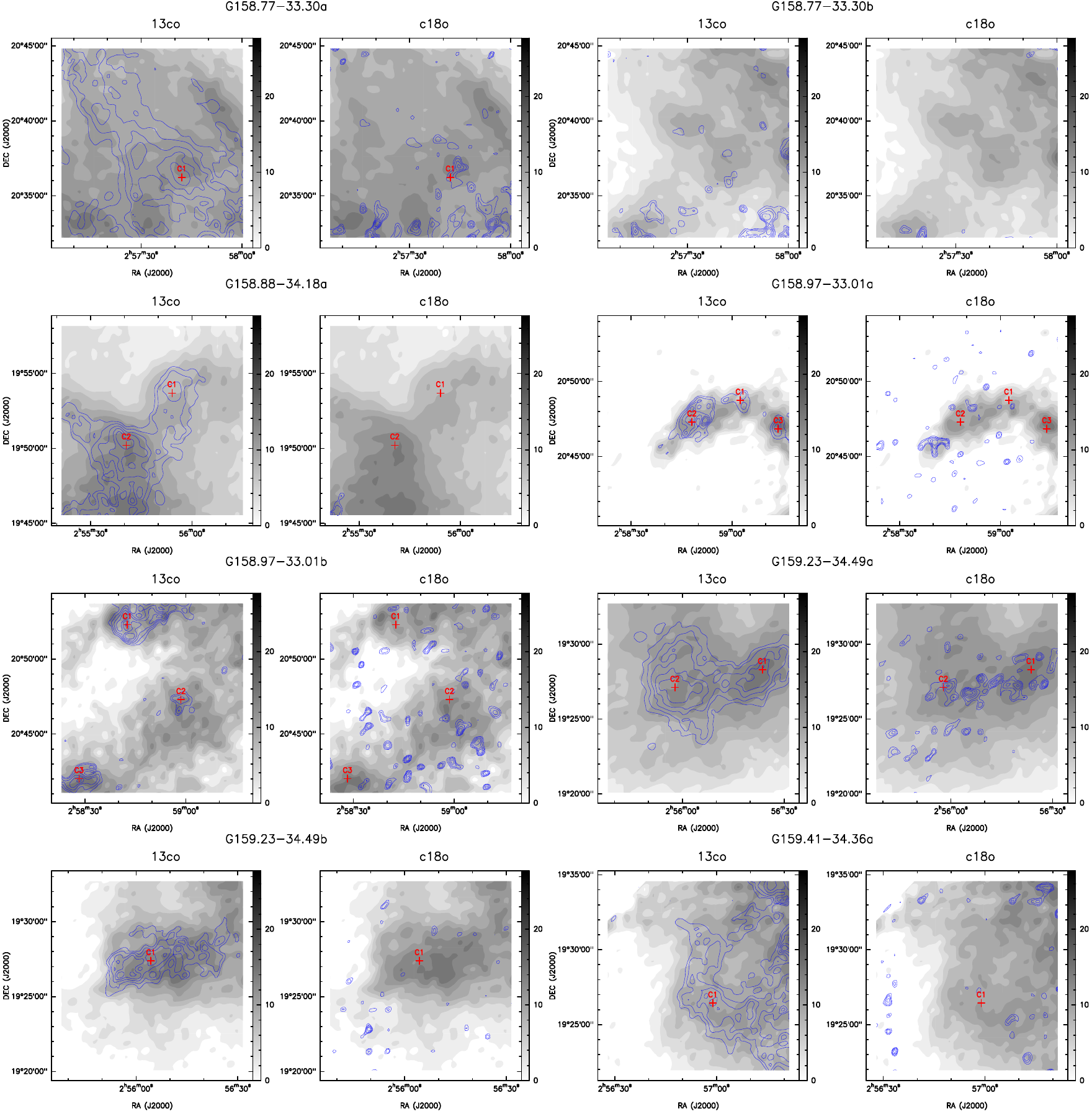}
	\label{fig:map3}
\end{figure*}
\addtocounter{figure}{-1}

\begin{figure*}
    \centering
    \caption{continued}
    \includegraphics[width=1.0\linewidth,angle=0]{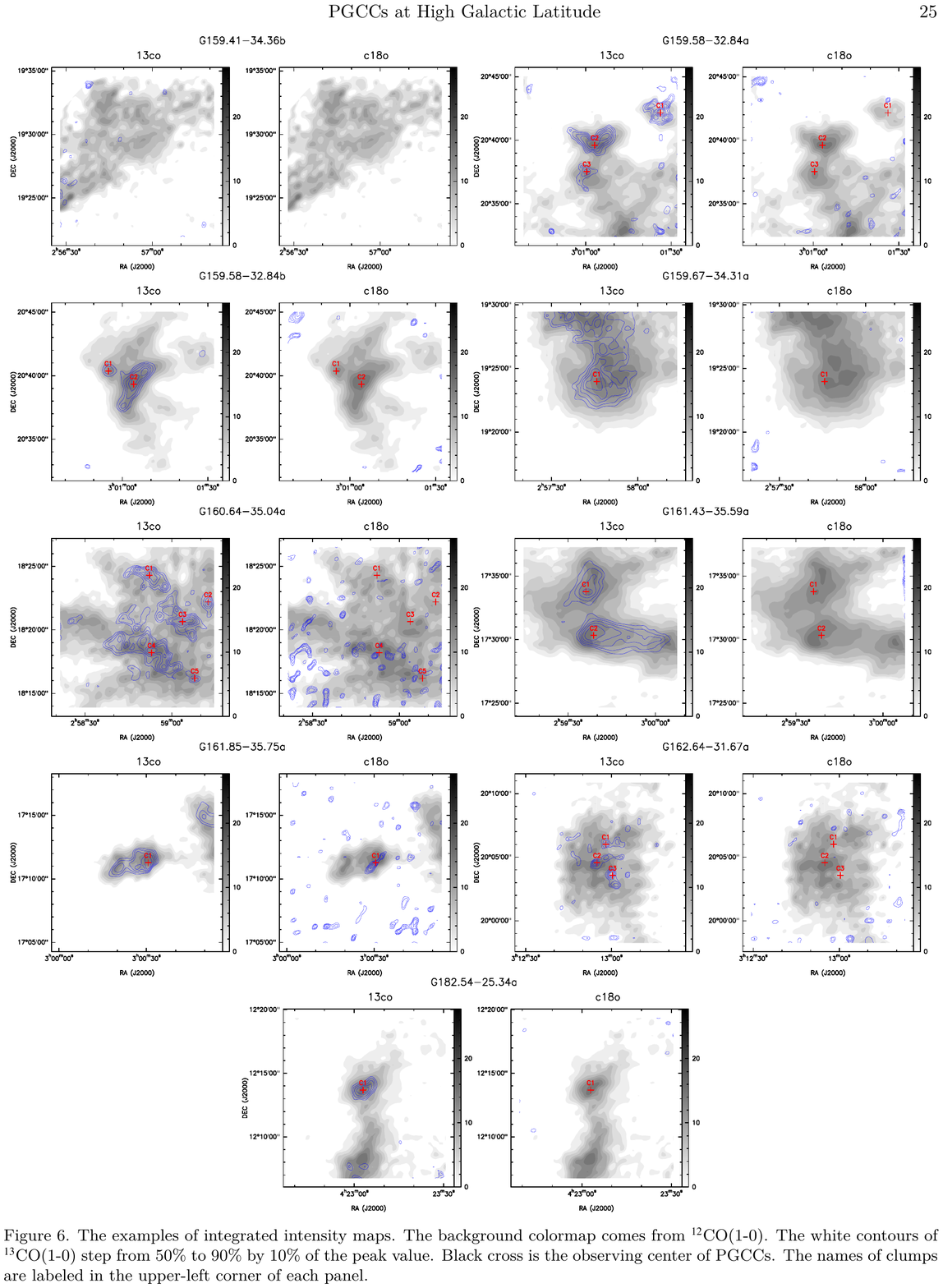}
	\label{fig:map4}
\end{figure*}
\addtocounter{figure}{-1}

\end{document}